\documentclass[a4paper,fleqn,usenatbib]{mnras}

\usepackage[T1]{fontenc}
\usepackage{ae,aecompl}

\usepackage{epsfig}
\usepackage{amsmath}
\usepackage{ulem}
\usepackage{color}
\usepackage{graphicx}
\usepackage{footmisc}
\usepackage{marginnote}
\usepackage{xcolor}
\usepackage{pdflscape}

\def\mnras#1{{MNRAS #1}}
\def\aj#1{{AJ #1}}
\def\apj#1{{ApJ #1}}
\def\apjl#1{{ApJL #1}}
\def\apjs#1{{ApJS #1}}
\def\aap#1{{Aap #1}}
\def\fcp#1{{fcp #1}}
\def\pasp#1{{PASP #1}}
\def\procspie#1{{procspie #1}}

\title[Nuclear stellar properties of active and inactive galaxies]{LLAMA: Nuclear stellar properties of Swift BAT AGN and matched inactive galaxies}
\author[M.-Y. Lin et al.]{Ming-Yi~Lin$^1$\thanks{E-mail: acdo2002@gmail.com},
R.I.~Davies$^1$,
E.K.S.~Hicks$^{2}$,
L.~Burtscher$^1$ 
A.~Contursi$^{1}$,
\newauthor
R.~Genzel$^{1}$,
M.~Koss$^{3}$,
D.~Lutz$^{1}$,
W.~Maciejewski$^{4}$,
F.~M\"uller-S\'anchez$^{5}$,
\newauthor
G.~Orban~de~Xivry$^{6}$,
C.~Ricci$^{7}$,
R.~Riffel$^{8}$,
R.A.~Riffel$^{9}$,
D.~Rosario$^{10}$,
\newauthor
M.~Schartmann$^{1,11,14}$,
A.~Schnorr-M\"uller$^{8}$,
T.~Shimizu$^{1}$,
A.~Sternberg$^{12}$,
\newauthor
E.~Sturm$^{1}$,
T.~Storchi-Bergmann$^{8}$,
L.~Tacconi$^{1}$ and
S.~Veilleux$^{13}$
\\
$^1$Max-Planck-Institut f\"ur extraterrestrische Physik, Postfach 1312, 85741, Garching, Germany\\
$^{2}$Department of Physics \& Astronomy, University of Alaska Anchorage, AK 99508-4664, USA\\
$^{3}$Institute for Astronomy, Department of Physics, ETH Zurich, Wolfgang-Pauli-Strasse 27, CH-8093 Zurich, Switzerland\\
$^{4}$Astrophysics Research Institute, Liverpool John Moores University, IC2 Liverpool Science Park, 146 Brownlow Hill, L3 5RF, UK\\
$^{5}$Center for Astrophysics and Space Astronomy, University of Colorado, Boulder, CO 80309-0389, USA\\
$^{6}$Space Sciences, Technologies, and Astrophysics Research Institute, Universit\'e de Li\`ege, 4000 Sart Tilman, Belgium\\
$^{7}$Instituto de Astrof\'isica, Facultad de F\'isica, Pontificia Universidad Cat\'olica de Chile, Casilla 306, Santiago 22, Chile\\
$^{8}$Departamento de Astronomia, Universidade Federal do Rio Grande do Sul, IF, CP 15051, 91501-970 Porto Alegre, RS, Brazil\\
$^{9}$Departamento de F\'isica, Centro de Ci\^encias Naturais e Exatas, Universidade Federal de Santa Maria, 97105-900 Santa Maria, RS, Brazil\\
$^{10}$Department of Physics, Durham University, South Road, Durham, DH1 3LE, UK\\
$^{11}$Centre for Astrophysics and Supercomputing, Swinburne University of Technology, Hawthorn, Victoria, 3122, Australia\\
$^{12}$Raymond and Beverly Sackler School of Physics \& Astronomy, Tel Aviv University, Ramat Aviv 69978, Israel\\
$^{13}$Department of Astronomy and Joint Space-Science Institute, University of Maryland, College Park, MD 20742-2421 USA \\
$^{14}$Universit\"{a}ts-Sternwarte M\"{u}nchen, Scheinerstrasse 1, D-81679 M\"{u}nchen, Germany
}

\date{Accepted 2017 October 05 . Received 2017 October 05 ; in original form 2017 June 10}

\pubyear{2017}

\begin{document}
\label{firstpage}
\pagerange{\pageref{firstpage}--\pageref{lastpage}}
\maketitle

\begin{abstract}
In a complete sample of local 14-195 keV selected AGNs and inactive galaxies, matched by their host galaxy properties, we study the spatially resolved stellar kinematics and luminosity distributions at near-infrared wavelengths on scales of 10-150 pc, using SINFONI on the VLT.
In this paper, we present the first half of the sample, which comprises 13 galaxies, 8 AGNs and 5 inactive galaxies.
The stellar velocity fields show a disk-like rotating pattern, for which the kinematic position angle is in agreement with the photometric position angle obtained from large scale images. 
For this set of galaxies, the stellar surface brightness of the inactive galaxy sample is generally comparable to the matched sample of AGN but extends to lower surface brightness. 
After removal of the bulge contribution, we find a nuclear stellar light excess with an extended nuclear disk structure, and which exhibits a size-luminosity relation.
While we expect the excess luminosity to be associated with a dynamically cooler young stellar population, we do not typically see a matching drop in dispersion. This may be because these galaxies have pseudo-bulges in which the intrinsic dispersion increases towards the centre. And although the young stars may have an impact in the observed kinematics, their fraction is too small to dominate over the bulge and compensate the increase in dispersion at small radii, so no dispersion drop is seen.
Finally, we find no evidence for a difference in the stellar kinematics and nuclear stellar luminosity excess between these active and inactive galaxies.
\end{abstract}


\section{Introduction} 
\label{sec:intro}
It is widely accepted that most galaxies harbor a supermassive black hole (SMBH). The most remarkable BH observations are of the Galactic centre where the individual stars can be spatially resolved and followed through their orbits, accurately constraining the SMBH mass (for a review see \citealp{Genzel2010}). Beyond the Milky Way, the most compelling evidence is the correlation between the mass of SMBH and stellar velocity dispersion of the bulge component of the host galaxy, which is interpreted as the signature of coevolution and regulation between the SMBH and the bulge (see \citealp{Kormendy2013} and the reference therein).
The SMBH grows via inflowing gas accretion, resulting in active galactic nuclei (AGN), which have been observed across different cosmic times. 
The host galaxy growth typically follows two plausible modes \citep{Shlosman2013}: (i) galaxy merger: angular momentum dissipation leads the gas infall forming compact young stars in the host galaxy \citep{Holtzman1992, Hopkins2009}, and furthermore efficiently drives some amount of gas to feed the central SMBH. Examples of this include the ultra-luminous infrared bright galaxies with disturbed host galaxy morphologies, which are usually accompanied by a QSO-like luminous AGN \citep{Bennert2008,Veilleux2009, Teng2010, Ricci2017}. (ii) Secular process of cold gas inflow: gas transfers from outer host galaxy to inner circumnuclear regions through disk and bar instabilities. If a bar drives the gas inflow, the associated inner Lindblad resonance (ILR) may terminate the inflowing gas and redistribute it in a disk inside the ILR radius (see \citealt{Combes2001} and references therein). However, \citet{Haan2009} studied gravitational torques and concluded that such dynamical barriers can be easily overcome by gas flows from other non-axisymmetric structures. The direct observations of inflows in an ionized or warm molecular gas phase on $\sim$ 100 pc scales have been confirmed in nearby Seyferts (e.g. \citealt{Storchi2007, Muller2009,Ric2014, Riffel2013,Storchi2014,SchnorrM17}).

The studies above focus on the question of the origin of inflowing gas transport (e.g. ex-situ gas). Once the gas accumulates in the nuclear regions, we further want to know whether any in-situ star formation occurs. Some observations indicate on-going star formation in the nuclear region \citep{Esquej2014,R2009} while others point out the galaxies prefer to have post-starburst populations \citep{Cid2004, Ric2007, Sani12, Lin2016}. \citet{Erin2013} also find no evidence that on-going star formation is happening in the central 100 pc. Observationally, it is unclear whether nuclear star formation plays a decisive role in triggering nuclear activity.
While it is understood that AGN flicker on and off on very rapid timescales, a recent analysis points to 10$^{5}$ years as one timescale \citep{Schawinski2015}; longer duty cycles of 10$^{7-9}$ years corresponding to the lifetime of a typical AGN phase are superimposed on top of that \citep{Marconi2004}.
This means that focussing on the circumnuclear regions of galaxies (e.g. \citealt{Dumas2007,Erin2013,Ric2014}) where dynamical timescales are of order 10$^{6}$ years and star formation timescales are 10$^{6}$-10$^8$ years, is an appropriate strategy to study the feeding mechanisms of gas flows associated with AGN activity.
To address this issue comprehensively, \citet{Ric2015} built a near complete volume limited sample of 20 nearby active galaxies, which was complemented by a matched sample of inactive galaxies, with the aim to obtain high spatial resoution near-infrared observations with SINFONI together with high spectral resolution observations taken with XSHOOTER. This is the LLAMA (Luminous Local AGN with Matched Analogues) survey which has been the focus of several other studies (\citealt{SchnorrM16,Davies17}, Rosario et al. submitted, Burtscher et al. in prep.).

In this paper, we present the SINFONI H+K observations probing radial scales of $\sim$ 150 parsec for the first half of the AGNs from \citet{Ric2015}, and their corresponding inactive galaxies which are matched in stellar mass, morphology, inclination, and the presence of a bar. This contains a total of 13 galaxies, 8 AGNs and 5 inactive galaxies, which provide 8 pairs (since some inactive galaxies can be well matched to more than one AGN). 
In this study, because the sample number is small, when comparing a difference between active and inactive sample for any physical quantity, we directly compare the mean value and standard deviation rather than giving a statistical test.
In Section~\ref{sec:sample}, we introduce the observations and data reduction of all the galaxies. Section~\ref{sec:methods} describes the methodology to extract the stellar kinematics and constrain the bulge S\'{e}rsic parameters. The nuclear stellar photometry is in Section~\ref{section:photometry}, while the nuclear stellar kinematics is presented in Section~\ref{sec:kinematics}.
We summarize our conclusions in Section~\ref{sec:conclusion}. Throughout this paper, we focus the discussion on the overall kinematic and photometric properties of the active and inactive samples. A detailed descriptions of individual objects with special (or extreme) properties will be discussed throughout this paper. We assume a standard $\Lambda$CDM model with H0 = 73\,km\,s$^{-1}$\,Mpc$^{-1}$, $\Omega_{\Lambda}$ = 0.73 and $\Omega_{M}$ = 0.27.

\section{Sample selection, observations and data reduction} 
\label{sec:sample}

\begin{table*}
  \caption{Galaxy Properties: 
  (1) Pair ID (a -- AGN, i -- inactive galaxy); (2) galaxy name; (3) AGN classification; (4) Hubble type of host galaxy; (5) Presence of large-scale bar (B indicates a bar, AB a weak bar); (6) Stellar mass estimated from total 2MASS H-band luminosity; (7) $m_{K}$(nucleus), apparent K-band magnitude measured from SINFONI data cube within 3$\arcsec$ aperture size; (8) Large scale axis ratio (from NED or \citealt{Koss2011}); (9) Inclination derived from axis ratio; (10) Distance (the median value of redshift-independent distance measurements from NED); (11) Physical scale of 1$\arcsec$ (from NED); (12) Observed 14-195 keV luminosity (70 months average) from \textit{Swift}-BAT catalog \citep{Baumgartner2013}.  \newline}
    \begin{tabular}{*{12}{c}} \hline     
     (1) & (2) & (3) & (4) & (5) & (6) & (7) & (8) & (9) & (10) & (11) & (12)    \\
     Pair ID & Galaxy name & AGN   & Hubble type & Bar & M$_{*}$ & $m_{K}$   & a/b & Incl. &Distance & scale & log(L$_{14-195keV}$) \\
      & & &  &  & (M\sun) & (mag) &  & ($\degr$) & (Mpc) & (pc/$\arcsec$) & (erg s$^{-1}$)\\ 
  \hline \hline
  1a & ESO 137-34 & Sey 2    & S0/a & AB & 10.4 & 11.9 & 0.79 & 40 & 33 & 185 & 42.62$^{\dagger}$ \\  
  6a & NGC 3783    & Sey 1.2 & Sab & B & 10.2 & 10.2 & 0.89 & 27 & 48 & 212 & 43.49 \\
  7a & NGC 4593    & Sey 1    & Sb & B & 10.5 & 11.1 & 0.74 & 42 & 32 & 200 & 43.16 \\
  4a & NGC 5728    & Sey 2 & Sb & B & 10.5 & 11.6 & 0.57 & 55 & 30 & 199 & 43.21$^{\dagger}$ \\
  8a & NGC 6814    & Sey 1.5 & Sbc & AB & 10.3 & 11.0 & 0.93 & 22 & 23 & 89 & 42.69 \\
  3a & NGC 7172    & Sey 2 & Sa &   & 10.4 & 10.1 & 0.56 & 60  & 34 & 153 & 43.45 \\
  2a & NGC 7213    & Sey 1 & Sa &   & 10.6 & 10.2 & 0.90 & 26 & 22 & 102 & 42.50 \\
  5a & NGC 7582    & Sey 2 & Sab & B & 10.3 & 9.7 & 0.42 & 65 & 22 & 88 & 42.67$^{\dagger}$\\ \hline
  6i & NGC 718 & -  & Sa & AB & 9.8 & 11.2 & 0.87 & 30 & 22 & 96 & - \\
  7i & NGC 3351 & -  & Sb & B & 10.0 & 11.6 & 0.93 & 22 & 10 & 74 & - \\
  3i, 5i & NGC 4224 & - & Sa &   & 10.4 & 11.8 & 0.35 & 70 & 45 & 193 & - \\
  8i & NGC 4254 & - & Sc &   & 10.2 & 12.0 & 0.87 & 30 & 16 & 75 & - \\
  1i, 2i, 4i & NGC 7727 & - & Sa & AB & 10.4 & 10.8 & 0.74 & 42 & 27 & 100 & - \\  \hline
    \end{tabular} \\
   \label{tab1} 
$^{\dagger}$: Heavily obscured AGNs with N$_{H}$(column density of neutral hydrogen) $\ge 10^{24}$ cm$^{-2}$, which is based on C. Ricci et al. (2017, in preparation) modelling 0.3-150 keV spectrum.\\   
\end{table*}

\subsection{Matched Seyfert and inactive galaxy sample}

\footnotetext[1]{https://ned.ipac.caltech.edu/ \label{fn:ned}}

The sample is drawn from the LLAMA (Local Luminous AGN with Matched Analogues) project, the selection details and the scientific rationale for which have been described in \citet{Ric2015}. 
We briefly address and discuss the key aspects of our target strategy: 
\begin{enumerate}
\item \emph{Select AGNs from the 58-month \textit{Swift}-BAT catalog:} 

The \textit{Swift} Burst Alert Telescope (BAT) all-sky hard X-ray survey is a robust tool for selecting AGN, because it is based on observations in the 14-195 keV band. Emission in this band is generated close to the SMBH and can penetrate through foreground obscuration. In contrast to optical/near-infrared AGN classification techniques, hard X-ray surveys suffer little contamination from non-nuclear emission.
However, it is still biased against extremely obscured Compton-thick sources (N$_{H} \ge 10^{25}$ cm$^{-2}$, \citealp{Ricci2015,Koss2016}) where the hard X-ray photon attenuation is due to Compton scattering on electrons rather than photoelectric absorption. In order to create a complete, volume-limited sample of nearby bright hard X-ray selected AGNs, the selection criteria were solely (i) 14-195 keV luminosities: log L$_{14-195}$ $>$ 42.5 (using redshift distance), (ii) redshift: z $<$ 0.01 (corresponds to a distance of $\le$40 Mpc), and (iii) observable from the VLT ($\delta <$ 15$\degr$). The total sample contains 20 AGNs covering Seyfert 1, Seyfert 2, and intermediate Seyfert types. Classifications are based on NED\footref{fn:ned}, with additional information from the presence of near-infrared broad lines or polarized broad line emission, as well as the first spectroscopic data from the LLAMA survey itself.

\item \emph{Finding a matched sample of inactive galaxies:}

Studying the difference between Seyferts and inactive galaxies can provide a direct comparison and give clues to understand what mechanisms can fuel a central BH and how the gas flows (inflows or outflows) interact with the interstellar medium.
However, it is important that the inactive galaxies are well matched.
To achieve this, the inactive galaxies in LLAMA are selected as specific pairs to the AGN.
The criteria to select an inactive galaxy for each AGN are based on: host galaxy morphology (Hubble type), inclination (axis ratio), and Two Micron All Sky Survey (2MASS) H-band luminosity (the proxy of stellar mass). Figure 3 in \citet{Ric2015} shows that there is no significant difference in the distribution of host galaxy properties between the \textit{Swift}-BAT AGN sample and the matched inactive sample, except the distance distribution, the active galaxies being slightly more distant than the inactive pairs. 
We also note that the presence of large scale bar is matched if possible, but is not strictly necessary. A large scale bar in the host galaxy is an efficient way to drive some gas into the central region \citep{Buta1996}.
However, numerous studies have found that the bar fraction in Seyfert and inactive galaxies is similar, suggesting that while large scale bars may assist in fuelling SMBH growth they are unlikely to be the sole mechanism regulating it \citep{Mulchaey1997,Cisternas2015}. Our total sample contains 19 matched inactive galaxies.

\end{enumerate}

This project includes observations from the high resolution spectrograph XSHOOTER covering 0.3-2.3$\micron$ (\citealt{SchnorrM16}, and Burtscher et al. in prep) and adaptive optics near-infrared integral field spectroscopy covering 1.8-2.4$\micron$ taken with SINFONI (this paper and Lin et al. in prep). The two independent sets of observations and analyses allow us to approach, from two different perspectives, one of the primary science goals of the overall project: looking for evidence of young or recent stellar populations (stellar age of a few to a few hundred Myr) related to AGN accretion. The properties of the sample galaxies analysed in this paper are listed in Table~\ref{tab1}. 

\subsection{Observations and standard data reduction}

We present the first part of near-infrared IFU data for the LLAMA project. Eight AGNs and five matched inactive galaxies have been observed with SINFONI between 2014 April and 2015 June from programme 093.B-0057. In total, this provides eight Seyfert-inactive galaxy pairs because, by relaxing slightly the matching criteria, some inactive galaxies provide a good match to several AGN. Specifically, NGC 7727 and NGC 4224 are the inactive pair of three and two AGNs respectively.
SINFONI, installed at the Cassegrain focus of VLT-UT4, consists of the Spectrometer for Infrared Faint Field Imaging (SPIFFI), a NIR cryogenic integral field spectrometer with a HAWAII 2RG (2k$\times$2k) detector and an adaptive optics (AO) module, Multi-Application Curvature Adaptive Optics (MACAO) \citep{Eisenhauer2003,Bonnet2004}. 

We observe each galaxy with the H+K grating at a spectral resolution of R $\sim$ 1500 for each 0$\arcsec$.05$\times$0$\arcsec$.1 spatial pixel, leading to a total field of view (FOV) on the sky of 3$\arcsec \times$3$\arcsec$. All scientific objects were observed in AO mode, either using a natural guide star (NGS) or an artificial sodium laser guide star (LGS). For our observations to achieve the best correction with a NGS, it should be brighter than R $\sim$ 15 mag and within a distance of 10$\arcsec$ from the scientific object. During such observations, the typical Strehl ratio achieved was $\sim$ 20\%. A standard near-infrared nodding technique with an observation sequence of Object-Sky-Object was applied. In each observing block (OB), a total of three sky and six on-source exposures of 300 sec each were obtained, the on-source frames being dithered by 0.3$\arcsec$ and the sky frames offset by 60-100$\arcsec$. Data from several OBs were combined to make the data cube. Telluric stars were observed at similar airmass, either before or after each observing block to make sure they sample similar atmospheric conditions. The data were reduced using the SINFONI custom reduction package SPRED \citep{Abuter2006}, which includes the typical reduction steps used for near-infrared spectra with the additional routines to reconstruct the data cube. The standard reduction procedure comprises flat fielding, identifying bad/hot pixels, finding slit position, correcting optical distortion and wavelength calibration. The night-sky OH airglow emission have been removed by using the methods described by \citet{Ric2007}. The telluric and flux calibrations for the scientific data were carried out with B-type stars. In our observing strategy, there are at least two data sets for each standard star. We apply the same data processing procedure to standard star observations, and use these to make a single flux calibration to each science data set. 
The final flux calibration for both the H-band and K-band is accurate to $\pm$0.05 mag.  

\subsection{Differential atmospheric refraction}
To improve the image quality in SINFONI data cubes, we quantify the displacements induced by differential atmospheric refraction (DAR), which appears as a spatial offset of the observed object as a function of wavelength. The refraction is due to the Earth's atmosphere causing the light to deviate from its original trajectory and appear closer to the zenith by an amount that is dependent on wavelength. The DAR will be more important for observations with larger zenith distance, i.e. higher air mass. In principle, the DAR effect is stronger in the optical and, for seeing limited observations, can usually be ignored in the near-infrared.
However, with adaptive optics on large telescopes, the impact of DAR relative to the spatial resolution is more significant.
We correct DAR using standard analytical expressions based on a simple model of the atmosphere. And we include a description here of our method because it differs from the empirical method outlined by \citet{Menezes2015}.

An object at an actual zenith distance of $z$, has an apparent zenith distance $\zeta$ after the light has passed through the atmosphere.
The refraction angle is $R = z - \zeta$, and can be written as:
 \begin{equation}
 R = 206 265 \times (n - 1) \times \tan \zeta
 \label{eq:R}
 \end{equation}
where R is in arcsec and $n$ is the refraction index close to Earth's surface.

Assuming standard atmospheric conditions, a temperature $T = 20\degr$C, an atmospheric pressure $P = 10^{5}$ Pa, and CO$_{2}$ fraction of 0.0004 with low humidity, the refraction index $n_{20,10^{5}}$ is expressed as a function of wavelength \citep{Bonsch1998,Filippenko1982} 
 \begin{equation}
 (n_{20,10^{5}}(\lambda) - 1)\times 10^{8}=8091.37 + \frac{2333983}{130-(\frac{1}{\lambda})^{2}} + \frac{15518}{38.9-(\frac{1}{\lambda})^{2}}
\label{eq:n}
 \end{equation}
 where $\lambda$ is wavelength in $\micron$. To take into account the variation of T and P between the various observations, the refraction index $n_{T,P}$($\lambda$) can be written as
\begin{eqnarray}
 (n_{T,P}(\lambda) - 1) = (n_{10^{5},20}(\lambda) - 1) \times  \nonumber
\end{eqnarray} 
\begin{eqnarray}
 \frac{P\times[1+(0.5953-0.009876\times T)\times 10^{8}\times P]}{93214.6\times (1+0.003661\times T)} 
\label{eq:nTP}
\end{eqnarray}
If including the vapour pressure of water, the equation above is reduced by a factor of $f$:
 \begin{equation}
 n_{T,P,f}(\lambda) = n_{T,P}(\lambda) - f \times (3.8020 - \frac{0.0384}{\lambda ^{2}}) \times 10^{-10}
 \label{eq:nTPf}
 \end{equation}
where $f$ is measured in Pa (the empirical relation between the change of refractive index and water vapour pressure refer to Figure 4 of \citet{Bonsch1998}):
\begin{displaymath}
f = \exp(20.386 - \frac{5132}{273+T}) \times 133.32
\end{displaymath}
Combing the equation(\ref{eq:R}) and (\ref{eq:nTPf}) allows one to find the wavelength dependent differential refraction at constant $\zeta$:
 \begin{equation}
 \Delta R = 206265 \times (n_{T,P,f}(\Delta \lambda)) \times \tan \zeta
 \end{equation}
In most cases, the observed shift of the image with wavelength was reasonably well approximated by this analytical expression, although there were a few cases where the match was not so good.
Figure 2 in \citet{Menezes2015} shows some examples in which there are additional offsets that cannot be interpreted as DAR. 
For example, foreground obscurations (dust filaments or dust lanes crossing the nuclear regions) can cause the peak in the H-band image slightly deviated from the nucleus position of K-band image \citep{Mezcua2016}. In order to keep this intrinsic measurement, in this work, we correct only the displacements induced by DAR, the residual offset is typically small, the average offset in K-band being only $\sim$0.5 pixels relative to H-band.
 
\section{Analysis methods}
\label{sec:methods}

\subsection{Stellar distribution and kinematics}
\begin{figure}
\begin{center}
      \hspace*{-1.0cm}
      \includegraphics[width=95mm]{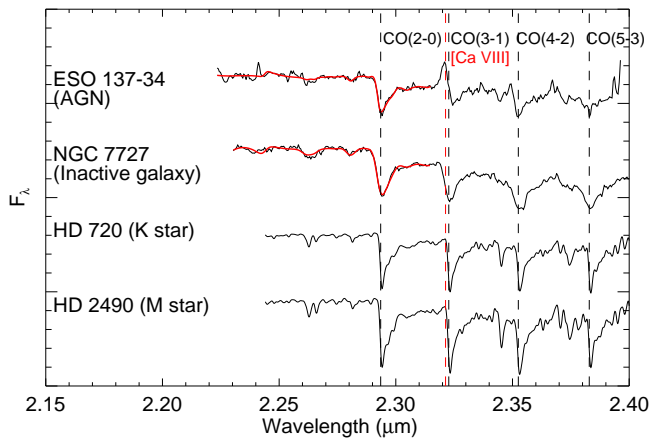}      
         \caption{The CO absorption features of one active and one inactive galaxy observed with SINFONI/VLT (top two spectra), and stellar templates from GNIRS/Gemini convolved to the resolution of SINFONI (bottom two spectra). Each spectrum has been corrected for systematic velocity and is shown at rest-frame wavelength. The top spectrum is the sum within the AGN-dominated region, and the non-thermal continuum has been removed by a polynomial function. The coronal line [Ca VIII] 2.3213$\micron$ is blended with CO(3-1) 2.3227$\micron$. The red solid lines for ESO 137-34 (AGN) and NGC 7727 (inactive galaxy) are the example ppxf kinematic fits of the CO(2-0) absorption.}
\label{fig:plot_sample}
\end{center} 
\end{figure}

In all 13 galaxies, the first two CO absorption bandheads are well detected. 
The stellar kinematics are extracted by fitting the first of these, the CO(2-0) absorption at 2.2935$\micron$, which has better signal-to-noise ratio (S/N) and less contamination by other absorption and emission lines. The second CO(3-1) absorption bandhead at 2.3227$\micron$ is excluded from the fit since it is contaminated by the coronal line [Ca VIII] 2.3213$\micron$ in AGN. In order to ensure that the active and inactive galaxies have a consistent analysis, we fit the kinematics using only the CO(2-0) bandhead.
To improve the S/N of the K-band continuum to 50, and simultaneously preserve the two-dimension (2D) kinematics, we have smoothed each slice of the IFU data cube. This is done by convolving with a point spread function (PSF) that has a FWHM of 3 pixels, matching that achieved on the brightest Seyfert nucleus of NGC 3783 in the same run.
To fit each galaxy spectrum in the resulting data cubes, we use the direct pixel fitting code Penalized Pixel-Fitting (pPXF) developed by \citet{CE04}. We choose the GNIRS sample of Gemini NIR late-type stellar library \citep{Winge2009}, which contains 30 stars with stellar type ranging from F7III to M2III and spectral resolution of 3.4{{\AA}} ($\sigma \sim$19\,km\,s$^{-1}$). To have the same spectral resolution between the stellar library and SINFONI, the stellar templates have been convolved with the instrument's line spread function of 70\,km\,s$^{-1}$, which is measured from the OH sky lines in K-band. To obtain the line-of-sight velocity distribution (LOSVD) of each galaxy spectrum, the stellar templates are shifted to the systematic velocity and convolved with a Gaussian broadening function. A polynomial function of fourth order is added to take into account the power-law continuum from AGN. Unlike the stellar absorption in the optical wavelengths which often have a higher S/N $\ge$ 50 (e.g. Ca II triplet lines measured by \citealt{Riffel2015}), the CO(2-0) absorption in the near-infrared has lower S/N of $\sim$ 10 per spatial element. We thus do not fit higher-order moments of the Gauss-Hermite series, the h3 and h4 terms, which indicate asymmetric deviation and peakiness of the profile respectively \citep{van1993,Bender1994}. Examples of fits and the smoothed stellar templates are shown in Figure~\ref{fig:plot_sample}. We apply this fitting procedure to the whole sample across the whole FOV to extract the 2D kinematics. The results will be discussed in Section~\ref{sec:kinematics}.

\begin{figure*}
\begin{center}
      \includegraphics[width=135mm]{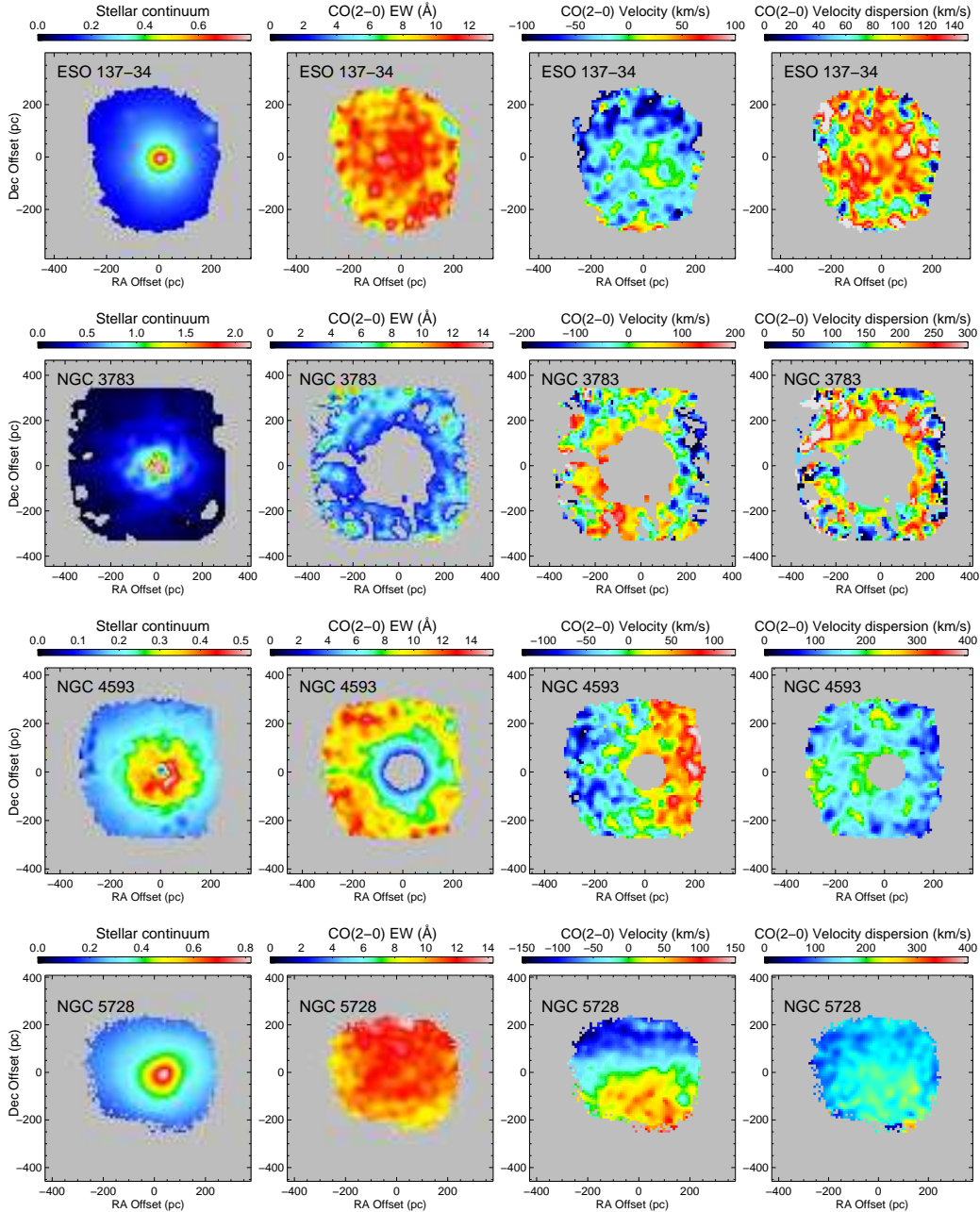}
         \caption{Active galaxy sample. Maps are labeled from left to right: stellar continuum flux, CO(2-0) equivalent width (EW), stellar velocity, and stellar velocity dispersion. The stellar continuum flux has been corrected for the contribution from non-stellar emission. In CO EW map, the central vacant hole is the region dominated by non-stellar emission that the ppxf program returns a unreliable kinematic measurement. We truncate it for display purposes. Because there is not non-stellar continuum dilution on stellar features for ESO 137-34 and NGC 5728, we do not outline any vacant hole in the centre. In all maps, north is up and east is to the left, the coordinate offset has been converted into the physical scale in parsec.}
\label{fig:plot_agn1}
\end{center} 
\end{figure*}

\begin{figure*}
\begin{center}
      \includegraphics[width=135mm]{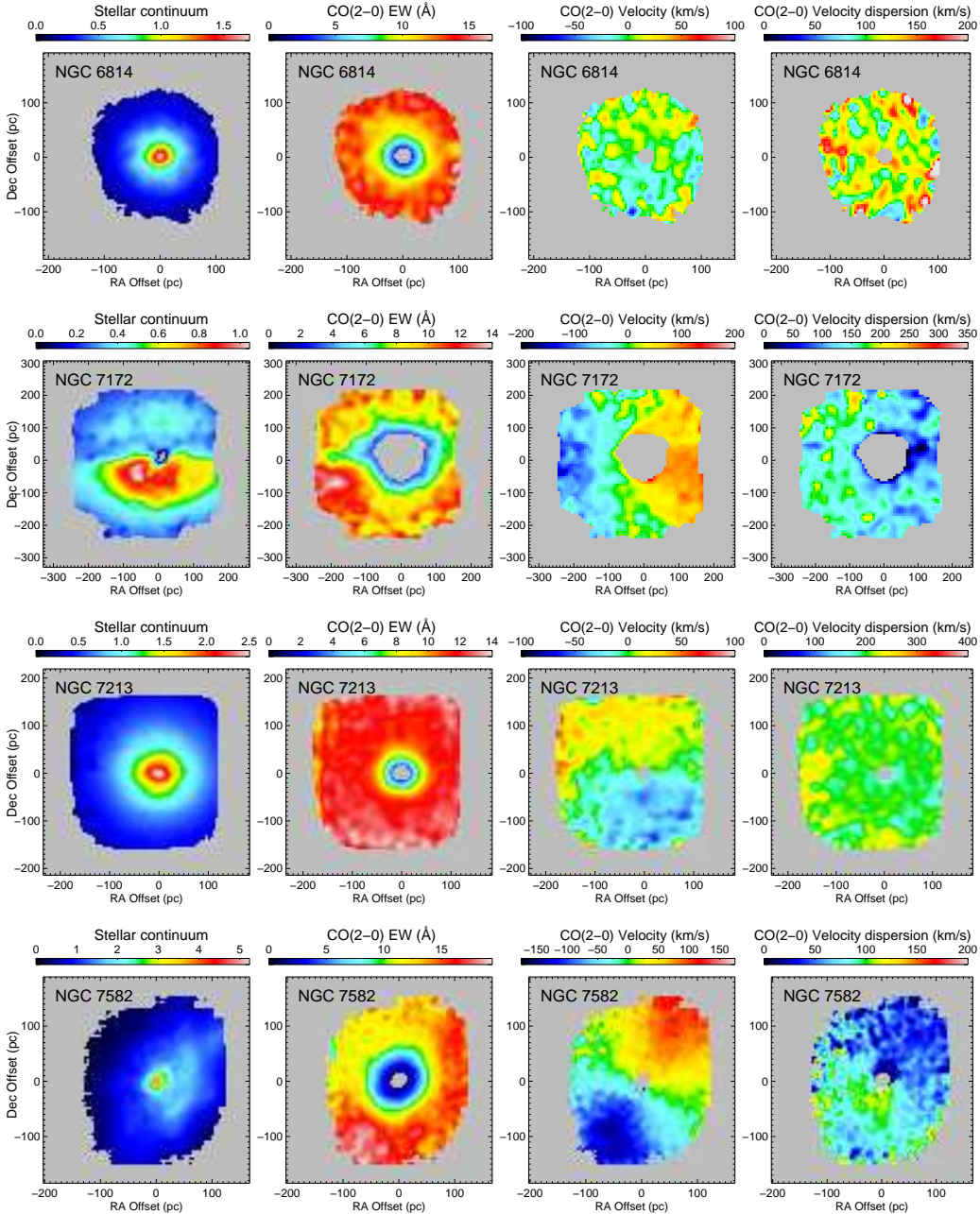}      
         \caption{$continued.$ The extreme irregularity in stellar continuum for NGC 7172 and NGC 7582 is due to the asymmetric reddening induced by dust lane, which passes through their nuclear region. The dust lane extinction does not influence the kinematic measurement. Although NGC 6814 is a face-on system, the weak rotation still is apparent from the map, consistent with the stellar velocity map of \citet{Ric2014}.  }
\label{fig:plot_agn2}
\end{center} 
\end{figure*}

\begin{figure*}
\begin{center}
      \includegraphics[width=135mm]{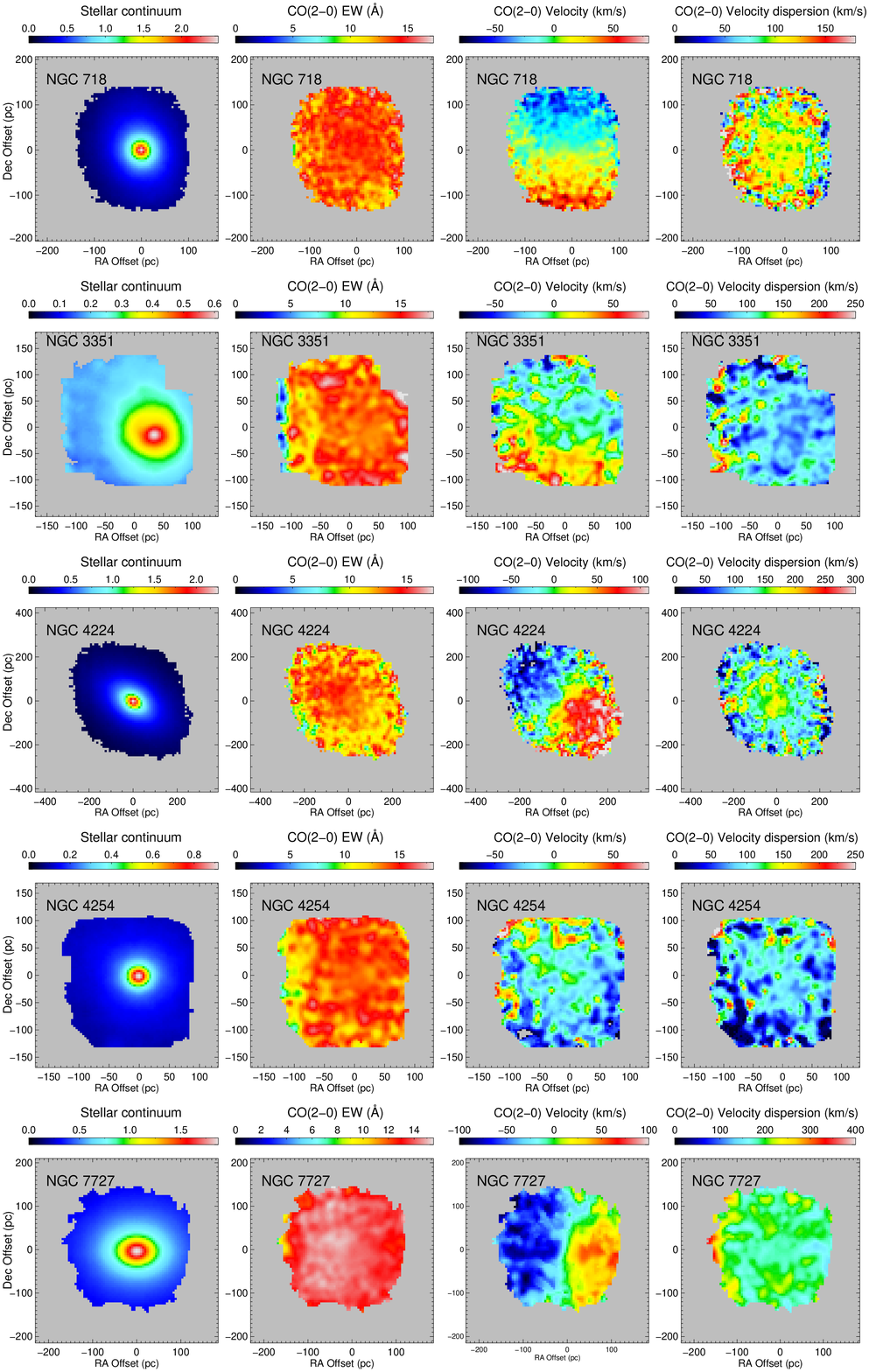}      
         \caption{Matched inactive galaxy sample. Maps are labeled from left to right: stellar continuum flux, CO(2-0) equivalent width (EW), stellar velocity, and stellar velocity dispersion. There is no non-stellar continuum to dilute stellar absorption features, thus the kinematics can be simply extracted. NGC 4254 has no clear velocity gradient because the system is very close to face-on. In all maps, north is up and east is to the left, the coordinate offset has been converted into the physical scale in parsec.}
\label{fig:plot_inactive}
\end{center} 
\end{figure*}

The resulting maps showing the flux distribution of the stellar continuum, CO(2-0) equivalent width (EW), stellar velocity, and stellar velocity dispersion are shown for the AGNs and inactive galaxies in Figure~\ref{fig:plot_agn1}-\ref{fig:plot_agn2} and Figure~\ref{fig:plot_inactive} respectively. 

\subsection{Continuum luminosity profile}
\label{section:constrain_bulge}
The goal of this study is searching for any nuclear excess stellar flux, which could indicate a young stellar population, and may be associated with a stellar velocity dispersion drop. 
The nuclear excess stellar flux in this paper is defined as additional light in the innermost regions that does not follow the bulge S\'{e}rsic profile.
Our FOV is only 3$\arcsec$ which, for the nearby galaxies in our sample, is well within the galactic bulge.
Thus an important step is to understand the larger scale bulge contribution in which these data reside. Once the bulge S\'{e}rsic profile has been derived, then we can check whether all the nuclear stellar flux follows the larger-scale bulge light profile.

To constrain the bulge S\'{e}rsic profile properly and systematically, we use the two-dimensional profile fitting algorithm GALFIT \citep{Peng2002,Peng2010} to decompose bulge and disk on scales of 2$\arcsec$-100$\arcsec$ with 2MASS Ks-band data, which trace the stellar light with less bias to extinction or stellar age than optical data. GALFIT requires the sky background and a PSF image. The sky background is set as a fixed parameter and obtained by measuring the mean value in the blank field of the same image.
The PSF, which allows us to correct the seeing, is generated as a Gaussian with FWHM of 2.5$\arcsec$. This provided a better residual map than when we used a bright star obtained from the 2MASS image.
Since 2MASS does not provide a pixel noise map, we do not include it in the calculation. Ordered lists of pixel coordinates have been used as a bad pixel mask, if needed to block bright stars close to the galaxy. For each galaxy, we iteratively fit two S\'{e}rsic profiles: one with a variable index, and one with the index fixed to an exponential profile. These components aim to model the large scale bulge and the disk respectively. Initial parameters are estimated by visual inspection, e.g. position angle (PA), ellipticity, and effective radius, etc. However, we note that the best fit value derived in the literatures also can be regarded as an initial guess for each parameter. By slightly changing the initial guess of each model component, we find that the choice of initial guesses does not influence the final result significantly.
We also note that while the GALFIT output provides a formal $\chi^{2}$ showing the difference between model and data, this is a purely quantitative assessment, and does not fully reflect whether a fit is good. It is more important to judge the quality of the fit from the residual maps. There are two steps in our fitting procedure:

(1) Fitting the large scale disk and bar:
We keep the number of free parameters to a minimum by fixing the S\'{e}rsic index n$_{disk}$ = 1 and leaving the effective radius r$_{\,e; disk}$ as a free parameter. In addition, if a large scale bar has been identified in the host galaxy, we fit it using an additional S\'{e}rsic profile. Initially, we allow n$_{bar}$ and r$_{\,e; bar}$ to be varied; however, if the GALFIT output value for n$_{bar}$ is too small (i.e. n$_{bar}$ $\le$ 0.1) or is too similar to the disk (i.e. n$_{bar}$ $\sim$ 1), we then fix n$_{bar}$ = 0.5, which remains a fairly constant value across different Hubble type within a limited range of M$_{\star}$ around $\sim$ 10$^{10}$ M$_{\odot}$ \citep{Weinzirl2009}. 

(2) Fitting the large scale bulge:
There is no constraint on n$_{bulge}$ and r$_{\,e; bulge}$ when we fit the bulge component. The PA and ellipticity are set to be in a reasonable range of quantitative agreement with the observed image. 
For the active sample, we also include a central point source to account for the AGN and avoid n$_{bulge}$ growing unrealistically large.
Note that any structures inside the bulge -- for example a nuclear disk, circumnuclear ring, or nuclear bar -- are not considered during the fitting, since they could be a part of the young stellar population which we expect to find.

\begin{figure}
\begin{center}
     \hspace*{-1.0cm}
      \includegraphics[width=95mm]{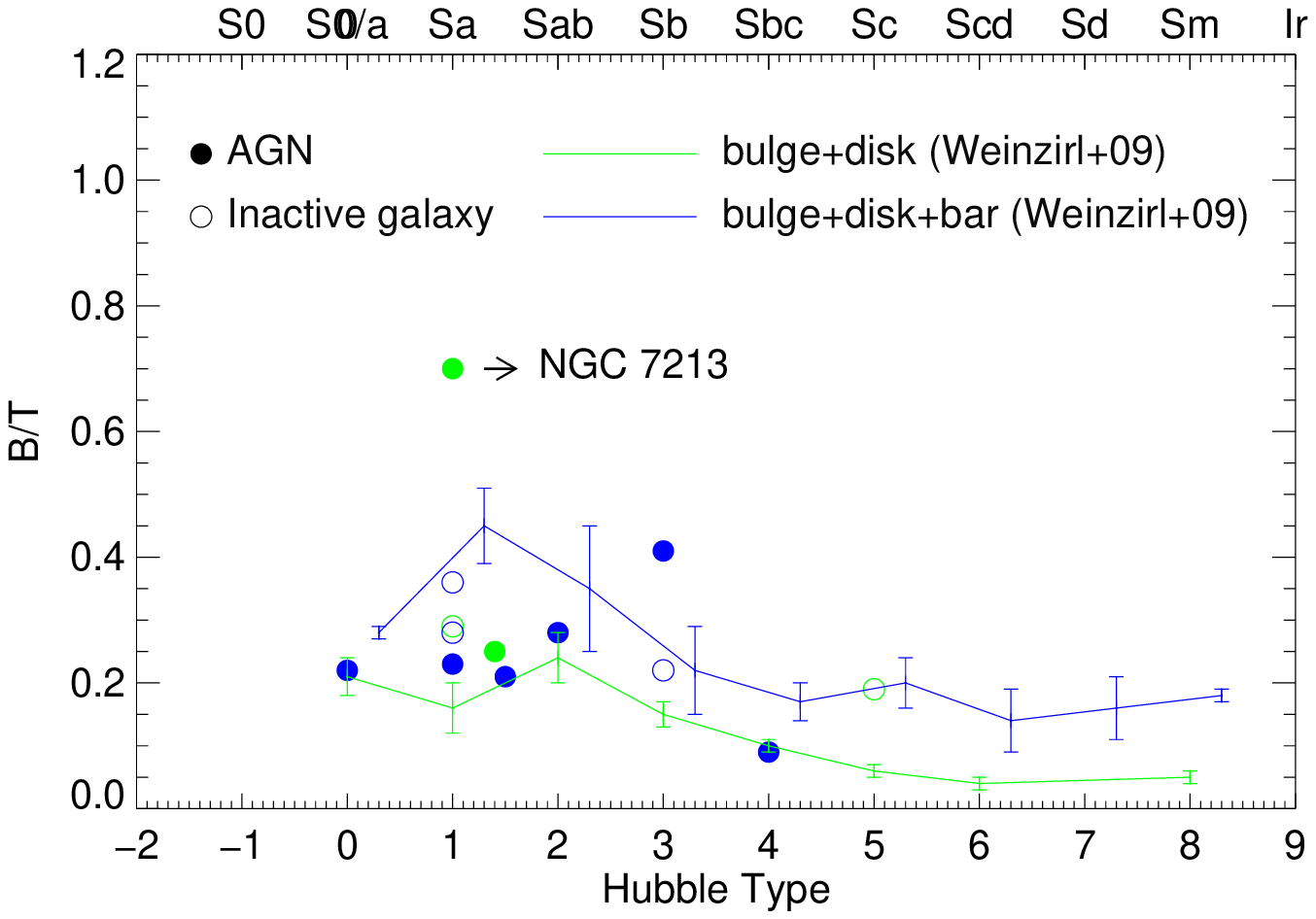}      
         \caption{Individual bulge to total luminosity ratio (B/T) as a function of Hubble type. With and without bar component in the two-dimensional decomposition fitting is shown in blue and green colour labelled both in lines and symbols. The filled circles represent Seyfert galaxies, while the open circles are inactive galaxies. The lines are the mean and standard deviation of B/T as a function of Hubble type from \citet{Weinzirl2009}.}
\label{fig:plot_B/T}
\end{center} 
\end{figure}

Studies with large samples of galaxies show that the bulge-to-total luminosity ratio (B/T) increases from late-type galaxies to early-type galaxies, i.e. as a function of Hubble type \citep{Weinzirl2009}. The results of our fitting show a similar trend in our small sample as presented in Figure~\ref{fig:plot_B/T}, confirming that the fits are reasonable.
We also note that there is no difference in B/T between AGN and inactive galaxies, consistent with our selection strategy of matching the active galaxies to the AGN based on host galaxy morphology and stellar mass. We find NGC 7213 deviates from the relation between the B/T and Hubble type. Its bulge parameter coupling problem has been discussed and tested in Section 4.4 of \citet{Weinzirl2009}. Given the Sa morphology, a B/T of $\sim$ 0.3 only occurs when we fix n$_{bulge}$ = 1, which cannot be distinguished from the outer disk, thus we decide to adopt the solution of n$_{bulge}$ = 2.57 with r$_{\,e; bulge}$ = 13.7$\arcsec$. The details of the S\'{e}rsic parameters for the bulge, bar, and disk are listed in Appendix~\ref{appendix:sec:1}. Most galaxies in our sample tend to have small n$_{bulge}$ $\sim$ 1--2.5, which is likely to be a disky bulge (pseudobulge) instead of a classical bulge with n = 4.

The next step is to match the flux scaling between the 2MASS profile fit and the SINFONI data. In order to compare the radial gradient between the observation scales, we extract the 1-D flux profile along the major-axis direction both from the 2MASS Ks-band image and the SINFONI stellar continuum image. The stellar continuum is measured from the CO(2-0) absorption bandhead after correcting the non-stellar AGN contribution \footref{fn:dilution} \citep{Ric2007-b,Leo2015}. 
The major-axis PA is the mean value measured from Spitzer near-infrared photometry \citep{Sheth2010} and SINFONI stellar kinematics. Once the bulge S\'{e}rsic index and effective radius have been obtained from the GALFIT decomposition, and assuming the SINFONI outer region $\sim$ 1.0$\arcsec$-2.0$\arcsec$ is bulge dominated, we extrapolate the bulge 1-D flux profile to a radius $<$ 1.5$\arcsec$ and look directly at the residual, to check whether the central few parsecs follow the outer bulge S\'{e}rsic profile. If there is an HST F160W archive image on scales of 0.2$\arcsec$-15$\arcsec$, we use it to reinforce the connection between SINFONI and 2MASS. However, the HST data have smaller pixel scales (0.075$\arcsec$ for NICMOS and 0.038$\arcsec$ for WFC3) and higher spatial resolution that better resolves the circumnuclear structures, and these can induce a slight inconsistency in the radial gradient between HST and 2MASS. We carefully minimize this effect due to fine structures when matching the profiles by extending the normalised region to include the large scale disk. 
Near-infrared wavelengths are less sensitive to extinction than the optical (the V-band to Ks-band extinction ratio is 1:0.062, \citealt{Nishiyama2008}), thus we did not correct for any near-infrared extinction in this study. The results are presented in Section~\ref{section:photometry}.

\footnotetext[2]{The non-stellar AGN light is estimated from the equivalent width (EW) of dilution CO(2-0) bandhead with a given intrinsic EW, which is expected to be a constant value over a wide range of star formation histories and ages (i.e. $L_{AGN} = L_{obs} \times f_{agn}$ where $f_{agn} = 1 - (EW_{obs, diluted}/EW_{intrinsic}$)). Note that $f_{agn}$ close to 1 corresponds to $\sim$ 100\% AGN contribution. \label{fn:dilution}}

\section{Nuclear Stellar Continuum Excess}
\label{section:photometry}

In these sections, we present the one-dimensional radial stellar continuum profile for our objects, in order to assess whether there is any photometric difference between the AGN and the matched inactive galaxy sample. We address this issue from two perspectives: the stellar surface luminosity distribution and the central excess light.

\subsection{Radial distribution of stellar luminosity}
\label{subsection:total}

\begin{figure*}
\begin{center}
      \includegraphics[width=180mm]{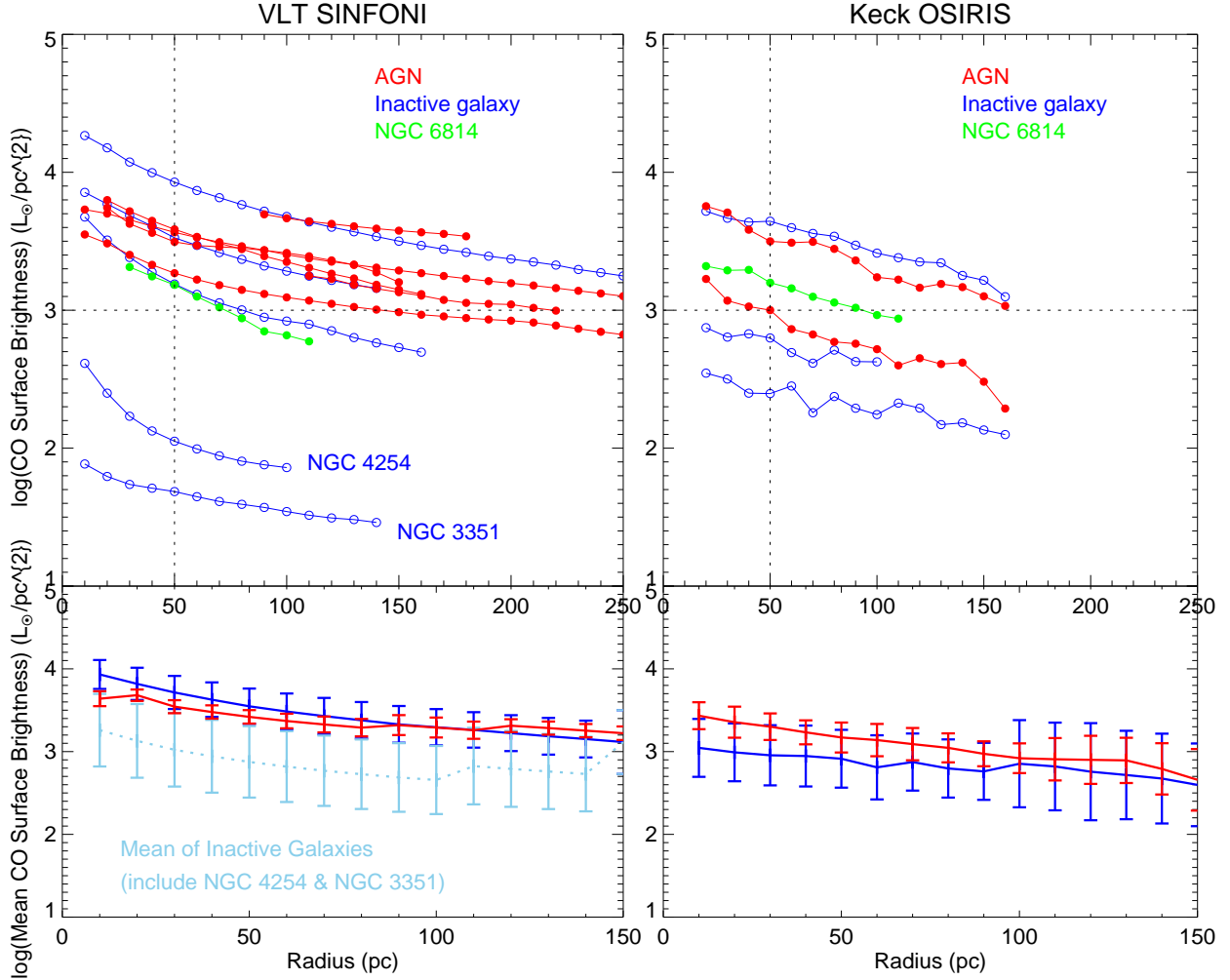}      
         \caption{The radial stellar light distribution. Top left panel: the stellar surface brightness as a function of radius for LLAMA sample. All of them were observed with VLT-SINFONI. Top Right panel: the stellar surface brightness as a function of radius for three AGNs and three inactive galaxies, where were observed with Keck-OSIRIS. AGNs and inactive galaxies are labelled as red and blue lines respectively. Both observations have covered NGC 6814 (AGN), which we highlight it as green colour. Bottom panels: the mean value at that radius and the radial bin size, the error bars are the standard deviation of measurements within each radial bin. The mean value of AGN and inactive galaxy sample are illustrated as bold red and blue lines. The light blue colour is the mean of all inactive galaxies, while the dark blue colour is the mean of inactive galaxies except for these two outliers (NGC 3351 and NGC 4254).} 
\label{fig:plot_stellar-conti}
\end{center} 
\end{figure*}

Figure~\ref{fig:plot_stellar-conti} shows the stellar surface brightness of AGNs and inactive galaxies, drawn as red and blue lines respectively. The top left panel of Figure~\ref{fig:plot_stellar-conti} is the directly observed stellar surface brightness from VLT-SINFONI. 
An observational caveat is that there are three AGNs (NGC 3783, NGC 4593 and NGC 7172) with strong non-stellar continuum contribution in the centre; therefore it is difficult to extract the stellar surface brightness at radii below 50 pc. 
Obviously, there are two inactive galaxies, which have lower stellar surface brightness: NGC 3351 and NGC 4254. The reason is that they have about 10 times lower K-band luminosity within our SINFONI field and 2 times closer distance, resulting in $\sim$1.5 dex lower surface brightness than other galaxies. Except for these two outliers, the surface brightness for other galaxies which obtained with SINFONI H+K grating are about 10$^{3-4}$ \,L$\sun$\,pc$^{-2}$ and the radial surface brightness distributions are generally similar for both active and inactive samples (refer to the bottom left panel of Figure~\ref{fig:plot_stellar-conti}).

A comparison of the stellar surface brightness between AGNs and matched inactive galaxies has been presented in Figure 15 in \citet{Erin2013}.
They showed that at radii greater than 150 pc, the Seyferts in their sample had a lower surface brightness than the inactive galaxies. However, the luminosity profile was steeper for the AGN, which led to similar, or in some cases higher, surface brightnesses at small radii.
The reason leading to the differing results of our LLAMA sample and \citet{Erin2013} will be discussed together with kinematic comparison in Section~\ref{subsection:avdkin}.

In addition, we are aware that it is difficult to compare our work directly to the previous \citet{Erin2013} study because the spatial pixel scales are different.
Unlike \citet{Erin2013} where observations cover a radial range of 50--250 pc, most of our observations cover a smaller range of 10--150 pc.
Thus we compare our work to recent Keck OSIRIS data (Hicks et al. in prep) of a similarly matched sample at spatial scale comparable to VLT SINFONI.
The preliminary results have been presented in the right panels of Figure~\ref{fig:plot_stellar-conti}, where the AGN sample is plotted in red while the matched inactive sample is plotted in blue. NGC 6814 (AGN) has similar surface brightness in both observations, which is highlighted in green. 
Based on the data currently available, our small sample shows that inactive galaxies cover a wider range of surface brightness in radial range of 10--150 pc, although this tentative conclusion is due to 2 inactive galaxies that are at leat one dex lower in surface brightness than the rest of the sample.
We note that some inactive galaxies have stellar surface brightness comparable to that of AGNs, which may suggest that the timescale of AGN switching on and off is shorter than the timescale to form nuclear stars, and the nucleus of those inactive galaxies may be just in a quiescent phase. Neither data set shows any AGN with stellar surface brightness below 10$^{3}$ \,L$\sun$\,pc$^{-2}$ within the central 50 pc, implying the mechanism to trigger AGN happens more effectively in galaxies which have higher stellar surface brightness. These conclusions will be revisited once the full sample is available.

\subsection{Central excess of stellar light}
\label{subsection:total2}

\begin{figure}
\begin{center}
      \hspace*{-1.0cm}
      \includegraphics[width=90mm]{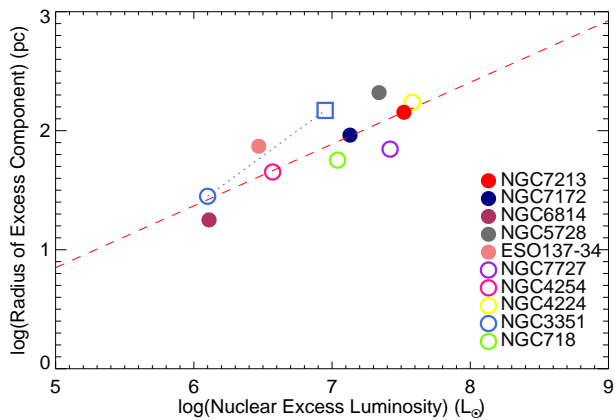}      
         \caption{The size-luminosity relation of excess nuclear star light (i.e. excluding the objects with central deficit star light). The excess star light is defined as a region, where the radial stellar light distribution does not follow the prediction of outer fitted bulge S\'{e}rsic profile. 
The filled circles represent Seyfert galaxies, while the open circles are inactive galaxies. Circle with different colour is given for each object. NGC 3351 has two measurements, the open square is the entire excess star light within the entire SINFONI FOV, while the open circle is an excess cusp appearing in the innermost region. Since the strong non-stellar dilution surrounds entire SINFONI FOV in NGC 3783, it has been excluded in this plot.}
\label{fig:plot_LR}
\end{center} 
\end{figure}

In this section we look at whether there is an excess or deficit in the stellar continuum compared to the bulge contribution that was extrapolated from the fit at larger scales.
In Section~\ref{section:constrain_bulge}, we described the method we used to measure the S\'{e}rsic index and effective radius of the bulge component from large scale 2MASS Ks band data.
By inward extrapolation of the fitted bulge S\'{e}rsic profile to the SINFONI FOV, we find that the stellar light at $<$ 1$\arcsec$ does not always follow the bulge profile. We classify the central stellar light profile as an excess (i.e. cusp) or deficit (see middle left panel in Appendix~\ref{appendix:sec:2}).
A similar dichotomy has been found in early-type galaxies in Virgo and Fornax clusters \citep{Cote2007}. Although our study focuses on Seyferts and inactive galaxies, many of which are late-type, we adopt a similar concept to parameterize the inner stellar profile.
We introduce a parameter, $\Delta_{L}$, as a ratio of the observed discrepancy in the SINFONI K-band luminosity (i.e. the difference between the stellar luminosity and the extrapolated bulge contribution) to the total SINFONI K-band luminosity in a 3$\arcsec$ aperture. This is estimated from the 1-D flux profile extracted along the major axis PA and then, assuming their radial distribution is symmetric, $\Delta_{L}$ is a radial integration until the radius where there is no significant excess light (i.e. R$_{excess}$ in Table~\ref{tab2}). Note that this 1-D flux extraction method of the central profile has the advantage that it is less susceptible to the stellar light asymmetry induced by foreground dust lane extinction, which can often create a large discrepancy along the minor axis. 
Galaxies with a central stellar light deficit then have $\Delta_{L} <$ 0, while those with excess have $\Delta_{L} >$ 0.

Within our sample, most galaxies have a central stellar light excess ranging between 1-12\%. 
Regarding NGC 6814, the central excess is small in physical size. However, we still consider this source to be a robust central excess detection. While the size is on 0.2$\arcsec$ in radius, it is still more than double the typical PSF FWHM of 0.1$\arcsec$. Further, the inner radial slope corresponding to the excess differs from the slope in the outermost regions (0.4-2$\arcsec$).
The $\Delta_{L}$ looks small (1\% for NGC 6814) because it is measured as the ratio of the additional luminosity to the total K-band luminosity within a 3$\arcsec$ aperture. We use this large aperture size because it can be easily compared to other public catalog, e.g. 2MASS.
On the other hand, if we used a small aperture, which is matched to the size of the nuclear excess component in each galaxy (refer to column (9) in Table 2), we would find the numbers in the range of 10-50\%, rather than 1-12\%. Thus, even though the $\Delta_{L}$ is small, the central excess is still significant.
But notably there are two AGNs which have a central stellar light deficit: NGC 4593 and NGC 7582. 
In order to confirm these light deficit features, we plot the HST/NICMOS/F160W radial flux along the major axis and find that it matches well with our SINFONI radial flux profile. The deficit could be due to either the intrinsic central behaviour or to foreground dust extinction. NGC 7582 is an example of the latter case, where the foreground dust lane across the circumnuclear region causes the stellar light asymmetry. However we cannot rule out that its stellar light deficit is intrinsic. 
The central stellar light deficit in NGC 4593 is likely to be intrinsic, and is unambiguously observed in both SINFONI and NICMOS radial flux profiles. Furthermore, we try to measure the bulge S\'{e}rsic profile solely based on HST/NICMOS/F160W image for these two galaxies by using GALFIT algorithm. NGC 7582 has strong asymmetric photometry, so that we cannot constrain the bulge S\'{e}rsic profile properly. On the other hand, for NGC 4593, although the radial S\'{e}rsic profile of bulge component is shallower (i.e. the amount of deficit light decreases), we still do not find any significant stellar excess toward the centre; the detailed results are presented in the top right panel of Figure~\ref{fig:appendix-galfit-4593}. 

For those galaxies with central stellar light excess, we fit a Gaussian function to characterize the size of the excess component (a similar method has been used in dwarf elliptical galaxies by \citealt{Graham2003}). 
The radius encloses 99\% of the Gaussian profile (i.e. 3$\sigma$ away from the centre).
NGC 3351 is a special case. Within the SINFONI FOV, the nuclear stellar light is entirely above the extrapolated bulge profile, and the integrated luminosity is 32\% higher than expected. 
In addition, to match the radial profile of the excess, we included a second Gaussian to fit the central cusp at a radius $<$ 0.5$\arcsec$, the slope of which is distinct from that of the 0.5-2$\arcsec$ outermost regions.
Interestingly, a similar situation in which there appear to be two components to the nuclear stellar excess was reported for another nearby Seyfert 1 galaxy NGC 3227 by \citet{Ric2006} (their Figure 6 and 7).
That work shows there is a clear excess starting at a radius of 0.5$\arcsec$, and the central cusp appears within a radius of 0.1$\arcsec$.
In our sample, NGC 3783 is a difficult case because the non-stellar contribution is strong across the whole SINFONI FOV. Thus, while we do measure an excess in the nuclear stellar luminosity, the scale is very uncertain and so we exclude it from our comparison of size versus luminosity.
Making use of the 10 objects with central excess stellar light (see Table~\ref{tab2}), we find the nuclear excess luminosity is proportional to the size of the excess component as shown in Figure~\ref{fig:plot_LR}. There is no significant difference in light excess between AGNs and inactive galaxies. This size-luminosity relation can be written as:
 \begin{equation}
  log (R_{excess}) = (0.52\pm 0.10) \times log (L_{excess}) - (1.74\pm 0.73)
 \label{eq:LR}
 \end{equation} 
We find that Spearman's rank correlation coefficient is $\rho$ $\sim$ 0.80 indicating a 98\% significance for the correlation (noting that $\rho$ = 1 corresponds to two variables being monotonically related). 

A central stellar light excess has been found in different types of galaxies, and is usually considered to be a nuclear star cluster (NSC). \citet{Boker2004} investigated the nearby late-type face-on spiral galaxies and found the optical $i$-band luminosity of NSCs (mean $\sim$ 10$^{6.4}$ L$\sun$) strongly correlates to its host galaxy luminosity, but the size-luminosity correlation of NSCs is weak. On the other hand, \citet{Cote2006} studied the compact central nuclei of early-type galaxies in the Virgo cluster in both g-band and z-band images and found that the size-luminosity relation of NSCs is $r \propto L^{0.5}$, where the mean luminosity of an NSC is $\sim$ 10$^{7.7}$ L$\sun$. Such a relation can be understood in terms of a merger model: the radius of the nucleus increases with increasing total luminosity as globular clusters merge \citep{Antonini2013}. In our sample, the mean K-band luminosity for nuclear excess stellar light is $\sim$ 10$^{7}$ L$\sun$, which is comparable to those NSCs found in the nuclei of early-type or late-type galaxies, if we assume NSCs mass of $\sim$ 10$^{7}$ L$\sun$ with M/L$_{H}$ of 0.6 \citep{Seth2010,Antonini2013}. 
However the size of the excess nuclear stellar light in our study is not matched to those of NSCs.
The typical size of NSCs is 5 pc, although a few of them can extend to 20-30 pc.
In contrast, for our sample, we find a mean size of $\sim$ 80 pc, and the size of individual nuclei ranges from 200 pc down to 10 pc. These are more likely to be an extended nuclear stellar disk rather than NSCs \citep{Balcells2007}. 
The reason that we cannot observe NSCs is due to the distance of our sample and the corresponding physical pixel scale is at least 10 pc, thus the NSCs cannot be spatially resolved.
The extended size of the nuclear disks gives clues, that such objects might have a different nature and structure than either compact NSCs or the bulge. The size-luminosity relation of the nuclear disks suggests their formation may nevertheless have some similarities to that of NSCs, in terms of the merging of sub-units.

\begin{table*}
  \caption{Nuclear properties of each galaxy: (1) Galaxy name, upper rows are AGNs and lower rows are inactive galaxies; (2) Systematic velocity derived from stellar kinematics; (3) Kinematic position angle from stellar velocity field; (4) Mean velocity dispersion of bulge ($\sigma$ at a radius $>$ R$_{excess}$); (5) The trend of velocity dispersion at a radius $<$ R$_{excess}$; (6) Stellar luminosity from SINFONI data cube within approximately 3$\arcsec$ aperture size (depend on how far of good pixels we can achieve); (7) Stellar luminosity of nuclear excess light; (8) $\Delta_{L}$; (9) Size of the nuclear excess light with $^{\dagger}$ (without taking into account the PSF of 0.1" radius)
  \newline}
    \begin{tabular}{*{9}{c}} \hline     
     (1) & (2) & (3) & (4) & (5) & (6) & (7) & (8) & (9)     \\
     Galaxy name & v$_{sys}$ & PA$_{kin}$   & $\sigma$ & $\sigma$ trend & log(L$_{K}$) & log(L$_{excess}$) & $\Delta_{L}$ & R$_{excess}$ \\
      & (\,km\,s$^{-1}$) & ($\degr$) & (\,km\,s$^{-1}$) & & (L\sun) & (L\sun) & (\%) & ($\arcsec$) \\ 
  \hline \hline
  ESO 137-34  & 2791.11 & 37 & 105 $\pm$ 5 & flat & 8.19 & 6.47 & 1.94 & 0.40  \\  
   NGC 3783$^{a}$  &  3044.28  & 137 & 154 $\pm$ 4 & drop & 8.73 & 7.95 & 16.67 & 0.80 \\
   NGC 4593     &  2553.97 & 106 & 149 $\pm$ 3 & increase & 8.15 & -- & -- & --\\
   NGC 5728     & 2834.28 & 12 & 164 $\pm$ 4 & flat & 8.29 & 7.34 & 12.15 & 1.05  \\
   NGC 6814     & 1612.63 & 37 & 115 $\pm$ 2 & flat & 8.10 & 6.11 & 1.00 & 0.20 \\
   NGC 7172    & 2591.40 & 93 & 103 $\pm$ 4 & drop & 8.52 & 7.13  & 4.32 & 0.60  \\
   NGC 7213    & 1876.60  & 31 & 201 $\pm$ 3 & flat & 8.52 & 7.52 & 11.00 & 1.40 \\
   NGC 7582    & 1651.05  & 155 & 68 $\pm$ 4 & flat & 8.79 & -- & -- & --\\ \hline
   NGC 718   & 1775.06 & 13  & 100 $\pm$ 2 & flat & 8.3 & 7.04  & 6.03 & 0.59 \\
   NGC 3351   & 867.93  & 174 & 84 $\pm$ 2 & drop & 7.57 & 6.95 [6.10]$^{b}$ & 32.11[4.52] & 2 [0.38]$^{b}$ \\
   NGC 4224   & 2651.79 & 56 & 153 $\pm$ 2 & increase & 8.76 & 7.58 & 6.95 & 0.91 \\
   NGC 4254 & 2514.77 & 87& 92 $\pm$ 2  & increase & 7.53 & 6.57 & 11.55 & 0.60 \\
   NGC 7727 & 1885.17 & 50 & 187 $\pm$ 3 & flat & 8.53 & 7.42 & 8.40 & 0.70 \\  \hline
    \end{tabular} \\
   \label{tab2} 
$^{\dagger}$: We assume the nuclear excess light followed the Gaussian profile, the radius encloses 99\% of Gaussian profile. \\
$^{a}$: Strong non-stellar continuum do exist across whole SINFONI FOV that nuclear stellar excess does not take into account in our analysis. \\
$^{b}$: The luminosity and radius of excess component in the central cusp. \\ 
\end{table*}

\section{Nuclear stellar kinematics}
\label{sec:kinematics}

In terms of stellar kinematics, the inactive galaxy sample is relatively simple to analyze, while the situation for the AGN sample is more complicated. This is because there is non-stellar hot dust contamination in the near-infrared which causes dilution of the stellar features, making it challenging to extract the kinematics. This issue will be discussed in Section~\ref{subsection:dilution}.
Looking at the stellar continuum maps, there are relatively noisy structures around the nuclear region in the AGN sample. This is because they are not direct measurements, but their K-band continuum includes the non-stellar continuum from AGN and stellar continuum, the latter of which can be extracted via CO EW (note that ESO 137-34 and NGC 5728 do not show any CO dilution and hence have no AGN hot dust observable in the K-band).
Overall, most galaxies (11/13) in our sample show nearly symmetric stellar continuum maps with regular elliptical isophotes. The two exceptions are NGC 7172 and NGC 7582, for both of which the maps exhibit extreme irregularities coinciding with known dust lanes \citep{Bianchi2007,Smajic2012}. These pass across their nuclei and are clearly visible in optical HST F606W images \citep{Malkan1998}. It is difficult to quantify the Ks-band extinction and correct it with present data. Fortunately, such extinction does not have a major influence on the extraction of stellar kinematics. Although the stellar continuum asymmetry runs along the minor axis, there is no significant feature in the kinematics along the same direction. 
NGC 3783 suffers strong dilution from non-stellar light, making it difficult to extract the kinematics and measure the PA. As a result the kinematics maps for this AGN are very noisy.
For the galaxies NGC 6814 and NGC 4254 (Pair 8), the observed rotating velocity pattern is weak because they are both nearly face-on. This leads also to larger uncertainties in estimating the kinematic PA on small scales. In the following sections, we discuss CO dilution and present the analysis of the two-dimensional stellar kinematic maps.

\subsection{Nuclear dilution by non-stellar light}
\label{subsection:dilution}

\begin{figure}
\begin{center}
      \hspace*{-1.0cm}
      \includegraphics[width=95mm]{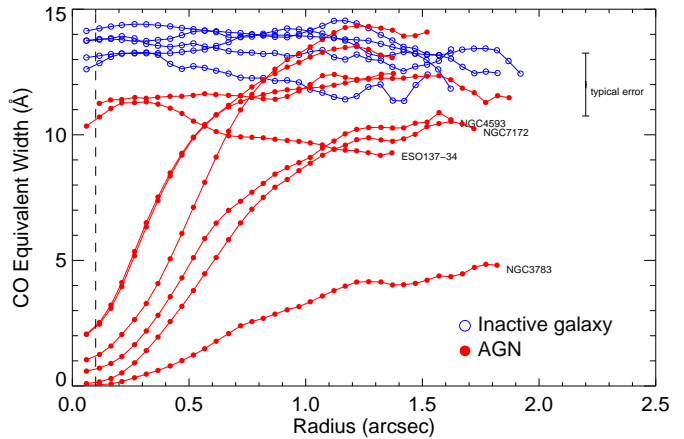}      
         \caption{Radial averaged CO EW. Red filled circles and blue open circles represent the AGN sample and the matched inactive galaxy sample. The decreased CO EW trends toward the centre in AGNs suggest the increasing contribution of non-stellar continuum. The FWHM radius of AO-corrected PSF is 0.1\arcsec, presented in black dashed line.} 
\label{fig:plot_COEW}
\end{center} 
\end{figure}

\begin{figure}
\begin{center}
      \hspace*{-1.0cm}
      \vspace*{-0.3cm}
      \includegraphics[width=95mm]{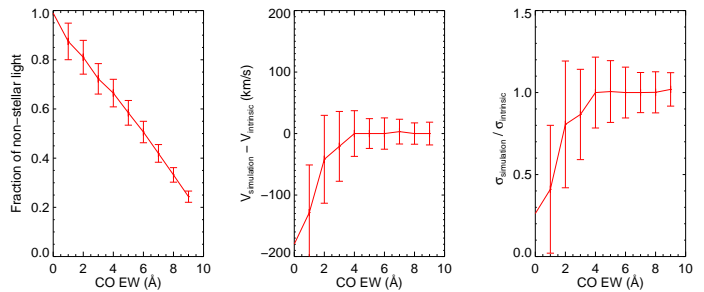}      
         \caption{Simulation of the impact of non-stellar continuum contribution on CO EW and stellar kinematics from 100 synthetic spectra. Red lines are the median value at each CO EW bin, the red error bars represent the 1$\sigma$ uncertainties. Left panel: The fraction of non-stellar continuum as a function of CO EW. Middle panel: The velocity offset from the intrinsic stellar velocity measurement. Right panel: The velocity dispersion ratio of the simulation to the intrinsic measurement. Stellar velocity and velocity dispersion start to deviate from the intrinsic measurements at CO EW $\le$ 3-4\AA, suggesting any kinematic measurement below this CO EW range is uncertain.} 
\label{fig:plot_simudilu}
\end{center} 
\end{figure}

For all the inactive galaxies, the stellar CO(2-0) EW has a relatively uniform distribution. In contrast, most AGNs (6/8), with the exception of ESO137-34 and NGC 5728, have a decreasing CO EW toward the centre.
The reason for this is that the stellar absorption features are diluted by the strong non-stellar continuum which is linked to hot dust associated with the AGN. 
Figure~\ref{fig:plot_COEW} shows the radial CO EW gradient.
The average intrinsic CO EW is 10-15\AA, and for the inactive galaxies is slightly higher than typical value of 11\AA\ reported by \cite{Leo2015} but within the range expected. 
We find there are four AGNs, for which the CO EW at 1.5\arcsec\ is lower than the CO EW of inactive galaxies and other AGNs at the same radii. Their names are labeled in  Figure~\ref{fig:plot_COEW}.
For NGC 3783, NGC 4593, and NGC 7172, they have higher $L_{14-195}$ and less obscuration with respect to other AGNs, suggesting, even at 1.5$\arcsec$ outer regions, the strong non-stellar continuum may marginally dilute the stellar absorption features.
We note here, in the case of NGC 3783 (the bright Seyfert nucleus in our sample), the CO bandhead appears to be very diluted everywhere within the 3$\arcsec$ field of view.
We therefore analyse also a SINFONI datacube with a FOV of 8$\arcsec$ and find that the CO EW in outer regions is $\sim$ 8\AA. We adopt this value as the intrinsic CO EW for this galaxy, despite it being lower than the mean value we found in other galaxies. In contrast, there is no CO dilution in ESO 137-34, and the CO EW of $\sim$10\AA\ across the whole FOV is likely to be intrinsic, implying the age of stars could be different from those galaxies with CO EW of 12-15\AA\ \citep{Ric2007-b}.
For the purpose of assessing the impact of the non-stellar continuum on the kinematics extraction, we produce a simple test to simulate the observed CO EW dilution. We start with the best-fit template spectrum of NGC 7727 (an inactive galaxy without any non-stellar contamination) and add a pure blackbody emitter with temperature of 1000K, representing the non-stellar continuum \citep{R2009,Leo2015}:
\begin{equation}
 F_{synthetic} = (F_{template} \times (1-c)) + (F_{blackbody} \times c)
 \label{eq:hot_dust}
\end{equation}
where c is the fraction of blackbody emitter, increasing from 0\% to 100\%, in the steps of 10\%:
Noise has been included in the synthetic spectrum to ensure that the S/N reaches $\sim$ 10 as the real data. The synthetic spectrum is then treated with the same analysis procedure as for the real data, measuring the kinematics and the CO EW. We repeat this process 100 times in order to make the experiment statistically robust.
The results are illustrated in Figure~\ref{fig:plot_simudilu}. 
The left panel shows how much the non-stellar continuum dilutes the CO EW. 
Once the CO EW becomes small, it is difficult to extract kinematics reliably because noise overwhelms the stellar absorption features. The middle and right panel of Figure~\ref{fig:plot_simudilu} show the velocity offsets and the velocity dispersion ratios to the intrinsic stellar kinematic measurement.
The velocity offsets and the velocity dispersion ratios deviate significantly when CO EW $\le$ 3-4\AA. Below this CO EW threshold, the velocity offsets and the velocity dispersion ratios are dominated by the randomly generated noise, indicating that the CO EW threshold is relatively sensitive to the data quality rather than just to the non-stellar continuum contribution fraction. Notably, NGC 7582 has a small CO EW in the centre but the velocity and velocity dispersion can still be measured reliably even close to the AGN.
The reason is that it is the brightest galaxy in our sample in the K-band, and the noise in the data is small enough that we can extend the CO EW dilution limit to 2\AA. 
For other active galaxies, we are limited to a radius corresponding to a CO EW threshold of 3-4\AA, and exclude any kinematic measurement within this radius. We have therefore truncated this region in Figures~\ref{fig:plot_agn1}-\ref{fig:plot_agn2}.

\subsection{Kinematic PA versus photometric PA}
\label{subsection:PAs}

\begin{figure}
\begin{center}
      \hspace*{-0.5cm}
      \includegraphics[width=85mm]{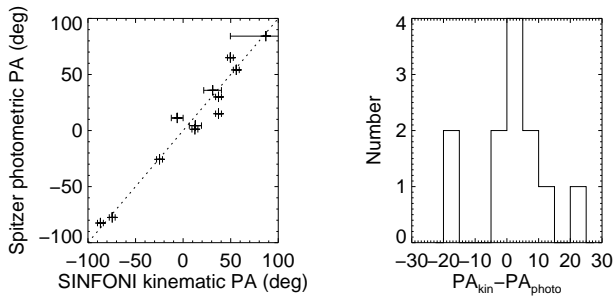}      
         \caption{Left panel: The Spitzer photometric PA plotted against the SINFONI kinematic PA. The photometric PA is redrawn from S4G catalog. The kinematic PA is measured by SINFONI stellar velocity field (see the third rows in Figure~\ref{fig:plot_agn1}-\ref{fig:plot_inactive}). The black dotted line is the 1:1 relation. Right panel: Histogram showing the distribution of PA difference measured with two independent methods.}
\label{fig:plot_PA}
\end{center} 
\end{figure}

In this section we compare the PA from the fits to the small scale kinematics (PA$_{kin}$) with the large scale photometric data (PA$_{photo}$). The kinematic PA is derived from the SINFONI stellar velocity map using the software developed by the SAURON team \citep{Cappellari2007,Krajnovic2011} \footref{fn:PA}. The kinematic PA is listed in Table~\ref{tab2}.
We obtain the photometric PA from the catalog for the Spitzer Survey of Stellar Structure in Galaxies (S4G), which observed numerous nearby galaxies at 3.6 and 4.5 $\micron$ with the Infrared Array Camera (IRAC). This catalog has the advantages that the seeing has been excluded when fitting the photometric ellipses and the observations cover the extension of the large scale disk.
NGC 3783 has been excluded in our comparison because there is no PA$_{photo}$ measurement from the catalog, which may be due to its bright Seyfert 1.5 nucleus preventing a measurement.
Figure~\ref{fig:plot_PA} shows that the PA$_{kin}$ is in good agreement with PA$_{photo}$, the differences between two measurements being in the range 8 $\pm$ 6$\degr$. In the following study, we simply adopt the mean of these two PAs and set $\pm$5$\degr$ as the uncertainty.
For comparison, the difference between our PA$_{kin}$ and PA$_{photo}$ is a factor two larger than \citet{Barnes2003} found, but it is robust enough for our analysis presented below. 
A similar result has been found in 16 nearby Seyferts with using Gemini NIFS data, both the large-scale photometric and small-scale kinematic axis are well aligned (Riffel et al. 2017 submitted).
We note that NGC 4254 has the largest uncertainty when calculating the kinematic PA, and NGC 6814 has the largest inconsistency in the measurements. It is because they are both face-on galaxies.

\footnotetext[3]{http://www-astro.physics.ox.ac.uk/$\sim$mxc/software/ \label{fn:PA}}

\subsection{Radial average kinematics}
\label{subsection:avdkin}

\begin{figure*}
\begin{center}
      \includegraphics[width=170mm]{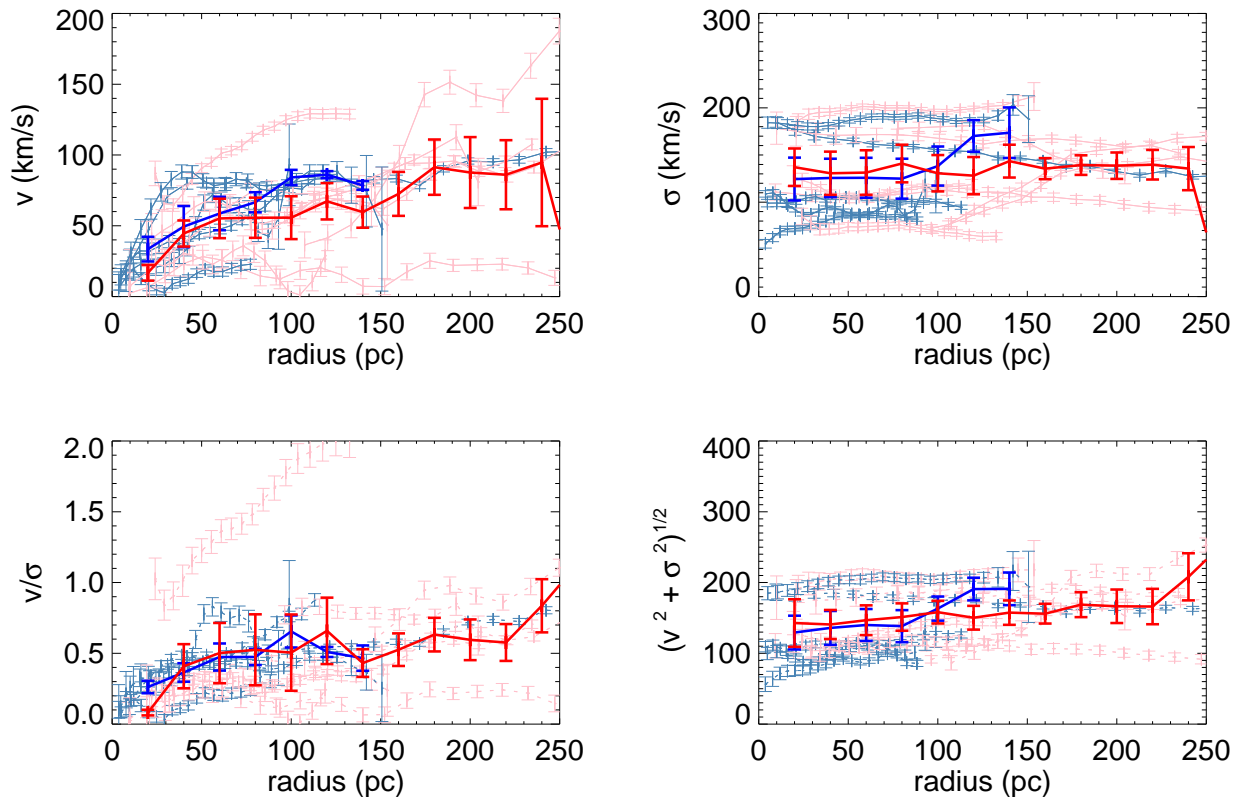}      
         \caption{Radial average kinematic properties of the CO(2-0) stellar absorption: AGNs and inactive galaxies are represented in pink and navy blue points. The mean value (uniformly weighted) of AGN and inactive galaxy sample are illustrated as bold red and blue lines, respectively. Beyond 150 pc, there is only one inactive galaxy NGC 4224 that has measurements, others inactive galaxies being so close that their outer regions exceed our SINFONI FOV. On the other hand, below 20 pc, there is only two AGNs, ESO 137-34 and NGC 5728, that have measurements, others have non-stellar continuum that prevent us to extract the stellar kinematics. The typical error bars of the lines are the standard deviation at that radius and the radial bin size. We show the stellar kinematic properties with four quantities: inclination-corrected rotational velocity (top left), velocity dispersion (top right), v/$\sigma$ (bottom left), and $\sqrt{v^{2}+\sigma^{2}}$ (bottom right). The velocities have been corrected for inclination based on the large-scale host galaxy inclination, which is listed in Table~\ref{tab1}.}
\label{fig:radial_kin}
\end{center} 
\end{figure*}

We use the IDL routine $kinemetry$ to extract radial distributions for the velocity and velocity dispersion from the 2D kinematic maps. The details of the method are described by \citet{Krajnovic2006}. Here we briefly summarise the key equations, which provide an important insight to quantify the physical meaning. 
By using Fourier analysis to characterize the periodicity, the observed kinematic moments can be divided into a series of elliptical rings, each of which can be written as a finite sum of harmonic terms:
 \begin{equation}
 K({a,\psi}) = A_{0}(a) + \sum_{n=1}^{N} A_{n}(a)\ sin(n\psi) + B_{n}(a)\ cos(n\psi)
 \label{eq:harmonic}
 \end{equation} 
where $\psi$ is the azimuthal angle measured from the projected major axis in the plane of the
galaxy, and $a$ is the length of semi-major axis of the elliptical ring. The parameters $A_{n}$ and $B_{n}$ can be represented by the amplitude coefficient $K_{n} = \sqrt{(A_{n}^2+B_{n}^2)}$.
With respect to the shape of elliptical rings for our analysis, the PA is set to be the mean PA obtained in Section~\ref{subsection:PAs}, the flattening (q) is the axial ratio listed in Table~\ref{tab1}, and the central position is fixed at the location given by the centre of the brightest continuum. For an ideal rotating disk, the rotational velocity and velocity dispersion fields are dominated by K$_{1}$ and A$_{0}$. 
The rotational velocity has been corrected for the inclination listed in Table~\ref{tab1}.

For the purpose here, to study the differences in stellar properties between AGNs and inactive galaxies, we plot in Figure~\ref{fig:radial_kin} four quantities related to the stellar kinematics as a function of radius: velocity, velocity dispersion, v/$\sigma$, and $\sqrt{v^{2}+\sigma^{2}}$. The velocities used throughout this paper have been corrected for inclination based on the large-scale host galaxy inclination, which is listed in Table~\ref{tab1}.
The approximate dynamics of a stellar system can be established by the velocity and velocity dispersion, which reflect the contributions of rotation and random motion. The relative contribution between these two motions is written as v/$\sigma$ and indicates how disky or spheroidal the system is. To access the enclosed dynamical mass in the nuclear region, we use the quantity $\sqrt{v^{2}+\sigma^{2}}$. The radius over which we can compare AGNs and inactive galaxies is in the range of 20-150 pc: most AGNs are affected by non-stellar dilution meaning the kinematic information cannot be extracted from the inner 20 pc, and most inactive galaxies are so close to us that the field of view does not extend over regions beyond 150 pc.

In Figure~\ref{fig:radial_kin}, we find that all the quantities are comparable between AGNs and inactive galaxies across 20-150 pc, suggesting that the nuclear stellar kinematics of the AGN sample are similar to those of the matched inactive galaxy sample. In an examination of the eight individual pairs, no specific trend toward the AGN or inactive galaxy sample is shown by any of the stellar kinematic quantities.
In contrast, within a radius of 200 pc \citet{Erin2013} found a large difference in $\sigma$ and a smaller difference in $\sqrt{v^{2}+\sigma^{2}}$, both of these quantities being smaller in the AGN sample compared to their inactive sample.
The authors interpreted this to indicate there is a dynamically cold nuclear structure composed of a relatively young stellar population in the AGN sample.

There could be several reasons for these differing results.
One possibility is the AGN selection method: we use solely the 14-195 keV luminosity to select AGN while \citet{Erin2013} used Seyfert galaxies that meet the Revised Shapely-Ames (RSA) catalog magnitude requirement ($B <$ 13.4 mag).
It is worth noting also that log L$_{14-195}$ of the AGNs in our sample is 43.0$\pm$0.4, which is a factor of a few brighter than those in the \citet{Erin2013} sample where log L$_{14-195}$ is 42.4$\pm$0.4 (and there is one AGN without a \textit{Swift}-BAT detection). 
A second possibility is the limited sample size and the selection variation. In this study we use 8 AGN and 5 inactive galaxies while \citet{Erin2013} use 5 AGN and 4 inactive galaxies in their kinematic comparison. As such, simple scatter in the individual properties at small scales could lead to the discrepancy. If all the objects in the LLAMA sample are observed, we will be able to resolve this issue.
A third possibility is the selection of the matched inactive galaxies. In both cases, this was done on parameters including galaxy integrated luminosity, Hubble type, inclination, bar, and distance. However, a difference is that we selected based on H-band luminosity while \citet{Erin2013} used matched samples from \cite{mar03} that had been selected on B-band luminosity. Due to the reduced impact of extinction and stellar age on the luminosity in the near-infrared, the H-band is a more reliable proxy for stellar mass. \citet{Erin2013} discussed this issue, and found that the H-band luminosity for their AGN was 24\% less than in their inactive sample. If H-band luminosity does linearly trace mass as we expect, this could then lead to a $\sim$ 12\% difference in kinematic tracers. This would reduce the difference in $\sqrt{v^{2}+\sigma^{2}}$ to a level where it is insignificant, but could not account for all the difference in $\sigma$.
Nevertheless, it underlines the importance of careful matching of the control sample.

\subsection{Central velocity dispersion}
\label{subsec:discussion}

When looking at the nuclear stellar kinematics, some AGNs show a significant velocity dispersion drop, suggesting the presence of a dynamically cold nuclear component \citep{Emsellem2001,Greene2014}. Self-consistent $N$-body simulations interpret these stellar velocity dispersion drops in the nuclear region as a consequence of young stars born from the dynamically cold gas reservoir coming in from larger scales.
As such, nuclear disk formation requires in-situ star formation, and it is kinematically cooler than the surroundings \citep{Cole2014}.
The simulations further suggest the young stellar population (less than 0.9 Gyr) should be brighter than the old underlying population at near-infrared wavelengths \citep{Wozniak2003}.
Adaptive optics observations with SINFONI have shown two AGN where this is seen (NGC 1097 and NGC 1068, \citealt{Ric2007-b}). 
And our new SINFONI observations can be used to test this interpretation further, as well as to assess whether a young stellar population commonly occurs in AGNs.

The nuclear stellar light excesses we have discussed in Section~\ref{section:photometry} exhibit an extended structure and a velocity field that indicates rotation, suggesting that they are nuclear disks. 
In particular, our data for the inactive galaxy NGC 3351 are consistent with the above mentioned simulations: it exhibits significant excess stellar light across the whole field of view, which is accompanied by an obvious central velocity dispersion drop to about $\sim$ 50\,km\,s$^{-1}$.
We can compare this to recent hydrodynamic simulations of the evolution of star formation as a result of gravitational instabilities in a nuclear gas disk (Schartmann et al. MNRAS accepted).
These show that, because of the large number of interactions between gas clumps and as a result of scattering between stars and gas clumps, the stellar disk undergoes significant gravitational heating before it relaxes in the global potential of the bulge.
These authors show that, in their simulation, the velocity dispersion can attain a constant value of $\sim$ 40\,km\,s$^{-1}$.

\begin{figure}
\begin{center}
      \includegraphics[width=85mm]{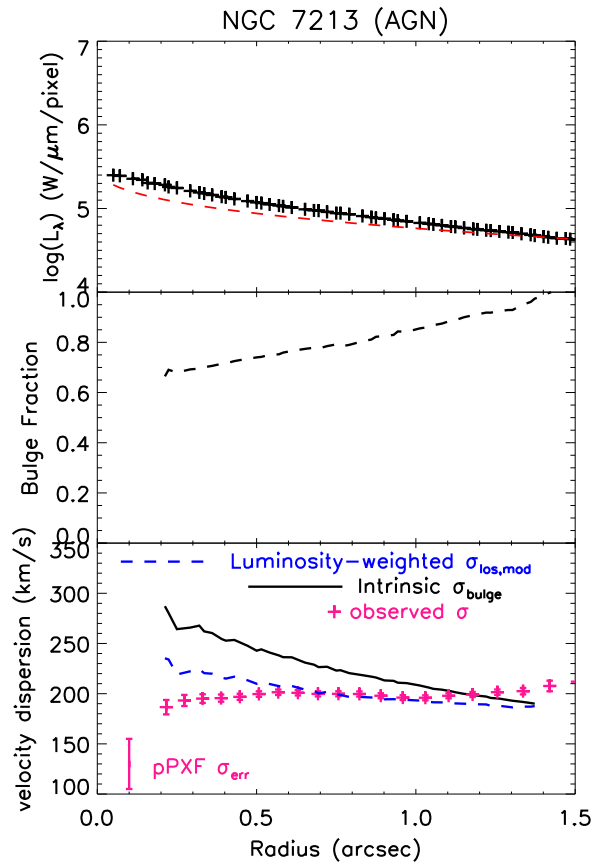}      
         \caption{The best sample of simple toy model to explain the trend of radial stellar velocity dispersion. Top panel: The radial luminosity for observed data point (black pluses) and bulge component (red dashed line). Middle panel: The bulge fraction as a function of radius. Bottom panel: The radial stellar velocity dispersion with radius. The black solid line represents the intrinsic bulge velocity dispersion, which is calculated based on the bulge surface brightness profile. Considering the contributions come from both a dynamically hot bulge population (typical is several hundred\,km\,s$^{-1}$) and a dynamically cold young star population (we assume 40-50\,km\,s$^{-1}$), the luminosity-weighted velocity dispersion is presented in blue dashed line. Pink pluses are the observed data with the error bar representing the 1-$\sigma$ error with respect to the velocity dispersion inside the 2-D elliptical rings at specific radius. We attached the pPXF return mean stellar velocity dispersion error (25\,km\,s$^{-1}$ in average) in the bottom left corner.}
\label{fig:model_disp}
\end{center} 
\end{figure}

In contrast to the case of NGC 3351, looking at the overall results of our sample, we surprisingly find the excess stellar light is not generally accompanied by the drop of stellar velocity dispersion toward the centre. These phenomena do not appear to correlate with each other, and the stellar velocity dispersion across the observable FOV is generally rather uniform. 
This observational evidence is quite distinct from the \citet{Wozniak2003} simulations, Figure 9 and 10 of their paper present that the stellar velocity dispersion drops with the increasing contribution of young (dynamically cold) stellar population. Once the mass density of the young population significantly overtakes the old population, the drops of stellar velocity dispersion can be easily seen. Looking at our data, the reason why we do not see the clear velocity dispersion drops in the central hundred pc scales, is probably because the contribution of old stars dominates over the young stars. To test this scenario, we describe a simple toy model to interpret the LOS velocity dispersion.

Based on the hydrodynamical disk models, $\sigma_{los}$ (line-of-sight velocity dispersion) can be projected into $\sigma_{R}$ (azimuthally average radial dispersion), $\sigma_{\theta}$ (tangential average radial dispersion), and $\sigma_{z}$ (vertical average radial dispersion) in cylindrical coordinates. The $\sigma_{R}$ and $\sigma_{\theta}$ are parallel to the disk plane, their contribution is marginal when the disk close to face-on (inclination $\sim$ 0$\degr$, i.e. the angle between line-of-sight and the disk is 90$\degr$; see Equations (28) and (29), and Figures (7) and (8) in \citealt{Tempel2006}). For simplicity, we assume that the simple toy models are all close to face-on.
Then $\sigma_{los}^2$ can be seen as $\sim \sigma_{z}^2$, in which the vertical velocity dispersion ($\sigma_{z}$(r)) can be derived from the surface density $\Sigma$(r) in the disk:
\begin{equation}
\sigma_{z}^2(r) = 2\pi G\Sigma(r)h_{z}
\label{eq:sigma_z}
\end{equation} 
where h$_{z}$ is the scale-height of the disk. Although this may vary with radius, we assume the disk models have a constant h$_{z}$ because in Section \ref{section:constrain_bulge}, we find that the bulges in our galaxies typically have n$_{bulge}$ $\sim$ 2, and hence are likely to be pseudo-bulges instead of classical bulges with n = 4. The bulge decomposition provides the bulge surface density profile which allows us to calculate the intrinsic velocity dispersion of bulge ($\sigma_{bulge}$). In addition, we set a second velocity dispersion ($\sigma_{new}$) to represent the young stellar population with a lower velocity dispersion of 40-50\,km\,s$^{-1}$. 
Adopting Equation (1) from \citet{Wozniak2003}, the line-of-sight velocity dispersion predicted by the toy model is weighted by the luminosity of two components (i.e. bulge and nuclear excess). It can be expressed as:
\begin{equation}
\sigma_{los, mod}^2(r) = \sigma_{z,bulge}^2(r) \times (\frac{L_{bulge}(r)}{L_{total}(r)})  
+ \sigma_{z,new}^2(r) \times (1 - \frac{L_{bulge}(r)}{L_{total}(r)})
\label{eq:sigma_los}
\end{equation} 
We apply this toy model to all objects and find that in its simplest form above, it can already explain the observed trend of velocity dispersion for NGC 7213 (AGN), which is presented Figure~\ref{fig:model_disp}. 
This galaxy has strong nuclear excess stellar luminosity. By checking its galactic plane inclination, we find NGC 7213 is close to face-on (26$\degr$), which is consistent with one of the assumptions in our simple toy model. Inspection of Figure~\ref{fig:model_disp} shows that the bulge contribution still dominates within the whole FOV, with the result that the kinematic signature of the young stars is dominated by that of the old stars, and hence the lower velocity dispersion from the young population is hidden by the dynamically hot galactic bulge. 
The increase in velocity dispersion at smaller radii is reduced, but not enough to be manifested as a drop in velocity dispersion.
Thus we can argue that we would not expect to see a clear stellar velocity dispersion drop in the centre. But, because we do not know the true intrinsic radial dispersion profile of the (pseudo-)bulge, we also cannot claim that the excess stellar luminosity must be associated with a dynamically cooler stellar population.

There is another explanation that most of the stars embedded in nuclear disk have not formed recently, and this stellar component is not dynamically cold anymore. \citet{Sarzi2016} find that the age of hot nuclear stellar disk in elliptical galaxy NGC 4458 is at least 6 Gyr old, while \citet{Corsini2016} measured the age of the nuclear stellar disk in the SB0 galaxy NGC 1023 is about 2 Gyr.
On the other hand, spectral synthesis fits to the detailed XSHOOTER spectra of the LLAMA sample (Burtscher et al. in prep.) show that while an old population dominates the optical continuum, a younger component with an age of 0.1--1\,Gyr is nearly always present at a level of a few to 20\%.

In addition to the question of whether the young stellar population is dynamically cold, a second issue concerns whether there is a difference between active and inactive galaxies.
In this study, we find there is no significant difference in stellar surface brightness and stellar kinematics between AGNs and most inactive galaxies; although we note that there are 2 inactive galaxies with surface brightness substantially lower than the rest of the sample.
Similarly photometric and kinematic characteristics between AGN and inactive galaxies suggest that the nuclear stellar properties are generally comparable.
This suggests that the timescale of switching AGN activity on and off, especially X-ray activity, is shorter than the timescale to form nuclear stars.
On the other hand, focusing on individual galaxies, we find that NGC 3351 has two components of nuclear stellar excess, together with a significant velocity dispersion drop toward the centre, indicating recent star formation around the galactic nucleus. These properties are similar to those reported in previous studies about recent star formation around AGN in Seyfert galaxies (e.g. NGC 3227 in \citealt{Ric2006}). It implies that, although they may be indications of gas inflow which triggers star formation and then feeds the AGN, there could still be a discrepancy between the timescales of these phenomena, with the AGN switching on and off multiple times during the period in which the nuclear stars can still be seen.

\section{Conclusions}
\label{sec:conclusion}
We present SINFONI data for the first half of a complete volume limited sample of bright local Seyferts, selected from the 14-195 keV \textit{Swift}-BAT catalog. These AGNs have been assigned a matched sample of inactive galaxies, in which the host galaxy properties (stellar mass, Hubble type, inclination, presence of a bar) share a similar distribution. In this paper, we present an analysis of the spatially resolved stellar luminosity and kinematics for a sample of 8 pairs of matched active and inactive galaxies, covering their central few hundred parsecs. 
Observations on this scale enable us to approach the galactic nucleus and search for a young stellar population, which simulations suggest should be brighter at near-infrared wavelengths and dynamically colder than the old population in the large-scale bulge. 
Based on this small set of galaxies, the main findings from the kinematic and photometric perspectives are as follows:
\begin{enumerate}
\item An apparent bimodality has been observed in the nuclear equivalent width of the stellar CO feature (CO EW) in nearby galaxies. Inactive galaxies show almost a constant CO EW across the entire SINFONI 3$\arcsec$ field of view, while most AGNs suffer from a strong non-stellar continuum contribution toward the centre, that dilutes the stellar light. For two AGNs (ESO 137-34 \& NGC 5728) the CO EW is nearly constant across the whole field, suggesting there is no non-stellar contamination in the nuclear region.\\

\item We find that the central (within the inner 1.5$\arcsec$ radius) stellar light distribution typically does not follow the S\'{e}rsic profile fitted to the larger scale bulge, and this difference can be classified as excess or deficit behaviour. Most of our galaxies show an excess corresponding to a few percent of the total stellar luminosity within a 3$\arcsec$ aperture; and the excess components show a clear relation of L $\sim$ R$^{0.5}$ which suggests surface brightness of the excess is constant across different galaxies. \\

\item The nuclear stellar photometry indicates that, except for the NGC 4254 and NGC 3351, the mean stellar surface luminosity of AGN is generally similar to the matched sample of inactive galaxies. We do not see any AGN with stellar surface brightness below 10$^{3}$ \,L$\sun$\,pc$^{-2}$ in the central 50 pc, while in contrast the matched sample of inactive galaxies includes lower surface brightness objects. \\

\item The stellar kinematics of the AGN and inactive galaxy samples show regular rotation patterns like typical disk-like systems, with a kinematic position angle that is in agreement with the photometric one fitted from large-scale near-infrared images. 
There is no direct evidence of a dynamically cold component (which can be seen as the young stellar population) in the stellar population; but we also show that one would not necessarily expect to see a central drop in the velocity dispersion.
There is no indication, either when looking at the whole sample or individual pairs, for any difference in stellar kinematics between the AGN sample and the matched sample of inactive galaxies within a radius of 150 pc.\\


\end{enumerate}


\section*{Acknowledgements}
We thank to the referee for a careful reading and giving useful suggestions to improve this paper.
The authors are grateful to K. Dodds-Eden for her major contribution to selecting the matched inactive sample. MY would like to thank Prof. Dr. Eric Emsellem for useful discussions in the Ringberg 2016 Meeting: on In-situ View of Galaxy Evolution.
E.K.S.H. acknowledges support from the NSF Astronomy and Astrophysics Research Grant under award AST-1008042. C.R. acknowledges financial support from the CONICYT-Chile "EMBIGGEN" Anillo (grant ACT1101). M.S. acknowledges financial support from the Deutsche Forschungsgemeinschaft (BU 842/25-1). R.R. thanks CNPq for financial support. M.K. acknowledges support from the Swiss National Science Foundation (SNSF) through the Ambizione fellowship grant PZ00P2 154799/1. 
This research is based on observations collected at the European Organisation for Astronomical Research in the Southern Hemisphere under ESO programmes: 093.B-0057(A) \& 093.B-0057(B). 
It also makes use of data products from the 2MASS, which is a joint project of the University of Massachusetts and the Infrared Processing and Analysis Center/California Institute of Technology, funded by the National Aeronautics and Space Administration and the National Science Foundation.
This research has made use of the NASA/IPAC Extragalactic Database (NED) which is operated by the Jet Propulsion Laboratory, California Institute of Technology, under contract with the National Aeronautics and Space Administration.

\appendix
\section{Disk, bar, and bulge decomposition}
\label{appendix:sec:1}

We present the GALFIT \citep{Peng2002,Peng2010} 2-dimensional decomposition results of bulge, bar, and disk from 2MASS Ks band image. The disk S\'{e}rsic index has been fixed to an exponential profile (e.g. n$_{disk}$ = 1). The detailed fitting procedure refers to Section~\ref{section:constrain_bulge}.

\begin{landscape}
\begin{table*}
  \caption{(1) Galaxy name, upper rows are AGNs and lower rows are inactive galaxies; (2) Bulge S\'{e}rsic index; (3) Bulge effective radius; (4) Bulge position angle; (5) Bulge axis ratio; (6) Bulge to total ratio; (7) Disk effective radius; (8) Disk position angle; (9) Disk axis ratio; (10) Bar S\'{e}rsic index; (11) Bar effective radius; (12) Bar position angle; (13) Bar axis ratio.
  \newline}
    \begin{tabular}{*{13}{c}} \hline     
     (1) & (2) & (3) & (4) & (5) & (6) & (7)  & (8) & (9) & (10) & (11) & (12) & (13)   \\
     Galaxy name & n$_{bulge}$   & r$_{\,e; bulge}$ & PA$_{bulge}$ & $\epsilon_{bulge}$ &  B/T & r$_{\,e; disk}$ &  PA$_{disk}$ & $\epsilon_{disk}$ & n$_{bar}$ & r$_{\,e; bar}$ & PA$_{bar}$ & $\epsilon_{bar}$  \\
      &  & ($\arcsec$) & ($\degr$) & &  & ($\arcsec$) & ($\degr$) & &  & ($\arcsec$) & ($\degr$)  \\ 
  \hline \hline
  ESO 137-34  & 2.13 & 6.94 & -45.0 & 0.75 & 0.22 & 27.15 & [35.0] & 0.91 & 0.31 & 8.02 & [-16.0] & [0.3]   \\  
   NGC 3783$^{a}$    & 1.24  & 1.45 & [-20.0] & 0.95 & 0.21 & 19.53 & -21.0 & 0.80 & 0.5 & 12.27 & [-20.0] & 0.26 \\
   NGC 4593 (2MASS)    &  2.73 & 3.66 & -85.0 & 0.83 & 0.41 & 38.64 & 71.0 & 0.57 & [0.5] & 37.65 & 55.0 & 0.29 \\
   NGC 4594 (HST)    &  1.67 & 8.1 & [-85.0] & 0.80 & - & - & - & - & - & - & - & -\\
   NGC 5728     & 1.1 & 4.02 & 3.7 & 0.92 & 0.23 & 48.85 & 32.0 & 0.37 & [0.5] & 42.5 & 34.0 & [0.1]  \\
   NGC 6814     & 1.08 & 1.07 & 27.0 & 0.94 & 0.09 & 30.53 & 56.0 & 0.96 & [0.5] & 5.43 & 26.0 & 0.66  \\
   NGC 7172    & 1.16 & 3.55 & -89.0 & 0.64 & 0.25 & 27.21 & -84.0 & 0.50 & - & - & - & -  \\
   NGC 7213    & 2.57  & 13.7 & -16.0 & 0.96 &  0.7 & 39.72 & 80.0 & 0.86 & - & - & - & -  \\
   NGC 7582    & 2.72  & 1.99 & -35.0 & 0.68 & 0.28  & 50.21 & -25.0 & 0.38 & 0.27 & 54.71 & [-22.0] & [0.15] \\ \hline
   NGC 718   & 1.32 & 2.09  & -5.0 & 0.92 & 0.28 & 16.25 & 0.0 & 1.0 & 0.73 & 15.73 & -27.0 & 0.42 \\
   NGC 3351   & 0.8  & 6.95 & 12.0 & 0.77 & 0.22  & 61.64 & 0.0 & 0.89 & [0.5] & 38.42 & [-70.0] & 0.3  \\
   NGC 4224   & 2.53 & 5.01 & [55.0] & 0.72 & 0.29  & 28.42 & [55.0] & 0.43 & - & - & - & -  \\
   NGC 4254 & 1.99 & 12.79 & 65.0 & 0.74 & 0.19  & 41.44 & 69.0 & 0.88 & - & - & - & - \\
   NGC 7727 & 1.68 & 5.07 & -80.0 & 0.72 & 0.36 & 22.86 & 10.0 & 0.95 & [0.5] & 12.73 & [-90.0] & 0.15  \\  \hline
    \end{tabular} \\
   \label{tab:appendix-decomposition} 
$^{a}$: Adding PSF to take into account the central bright nucleus. \\
Brackets: We hold the components fixed to the values in order to optimally perform galaxy fitting. \\
\end{table*} 
\end{landscape}

\section{Radial flux and kinematics of each galaxy}
\label{appendix:sec:2}
We show the radial flux along the major-axis PA (top and middle rows) and the average kinematics (bottom row) for each galaxy.

\begin{figure*}
\begin{center}
      \includegraphics[width=170mm]{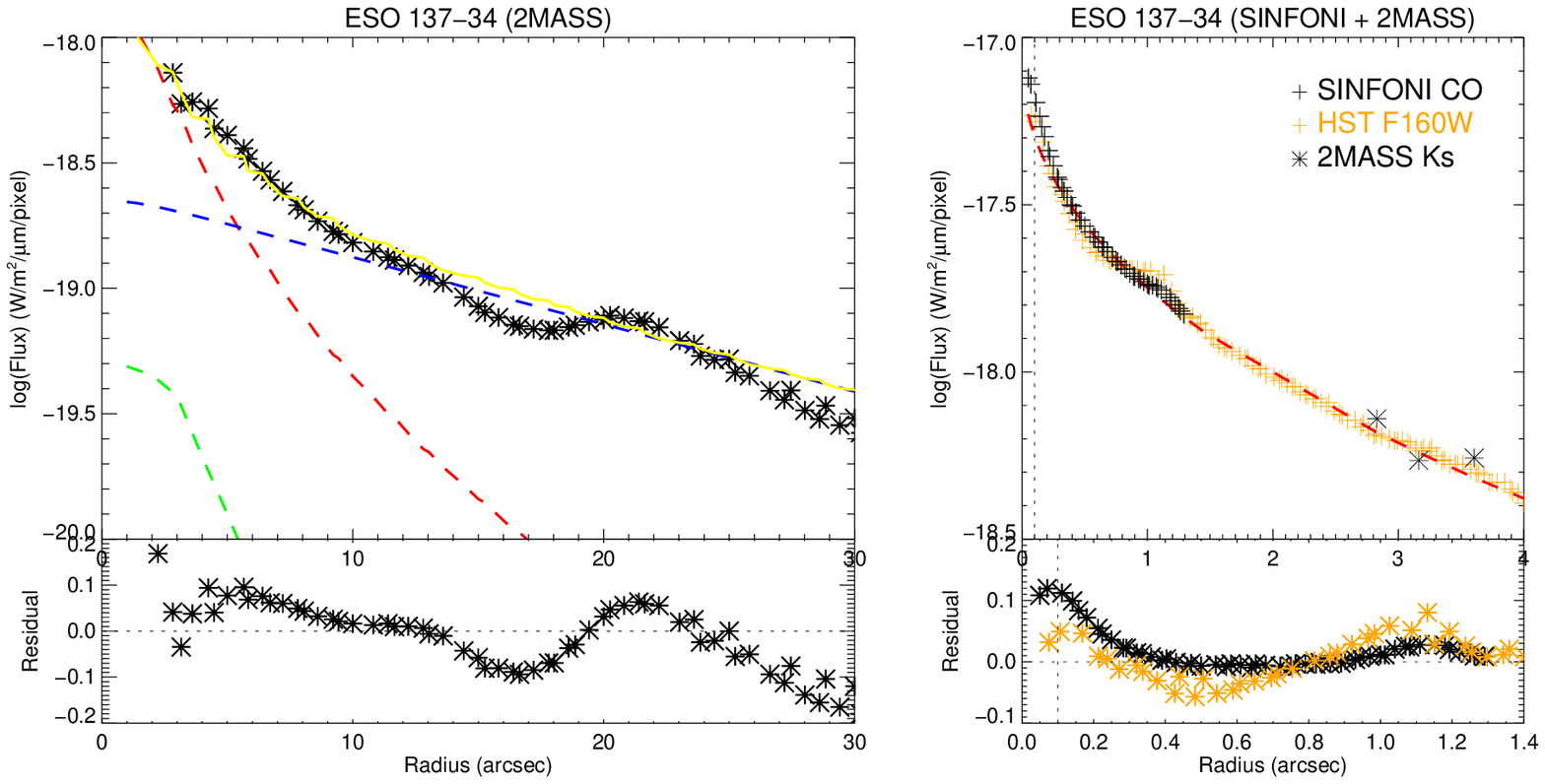}             
       \includegraphics[width=160mm]{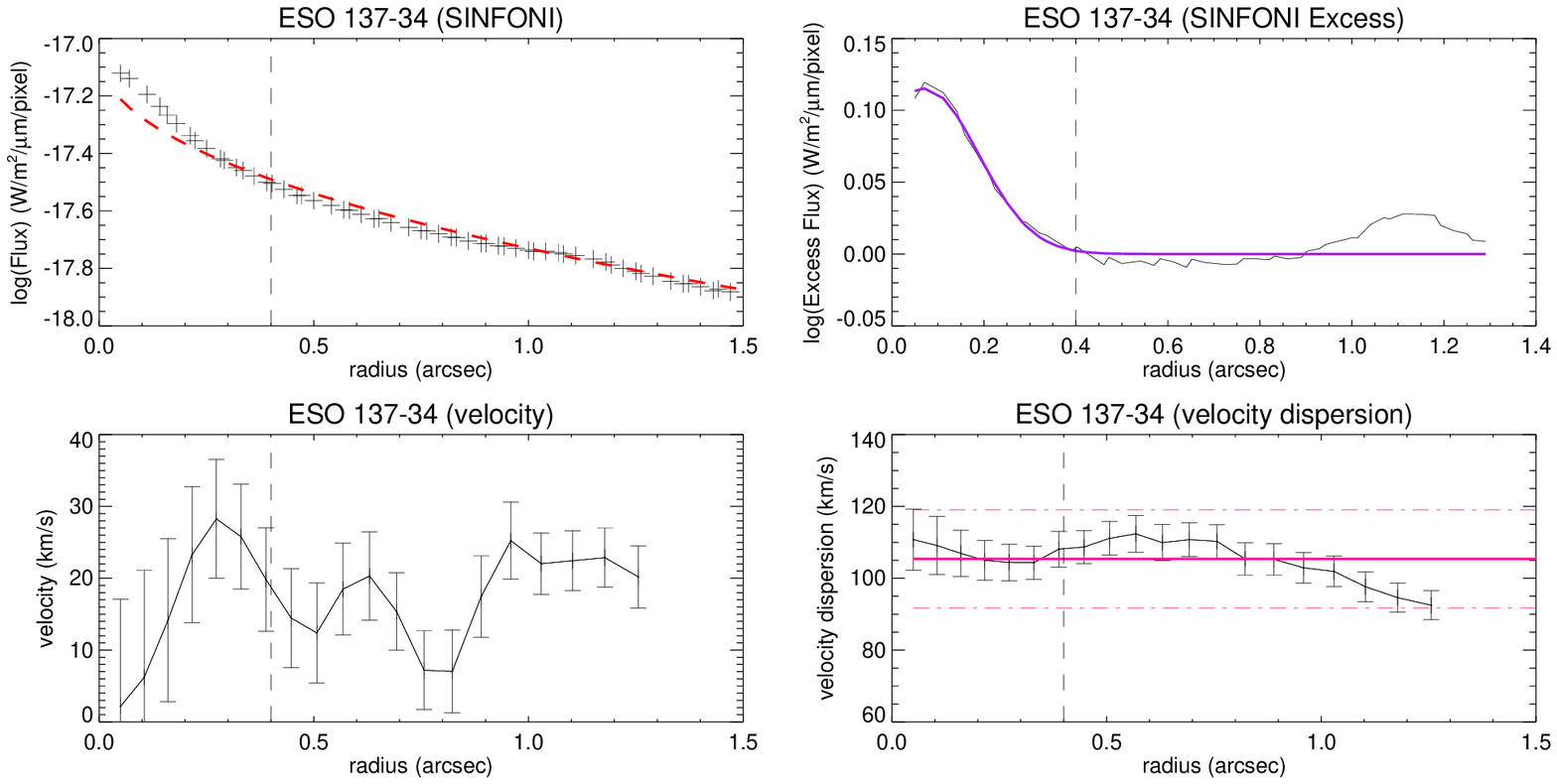}    
         \caption{ESO 137-34 (Active galaxy in Pair 1). \textit{Top left panel}: The radial flux profile along the major axis of 2MASS Ks band image (black asterisk signs). The decomposed disk, bar, and bulge component are presented in blue dashed line, green dashed line, and red dashed line, respectively. The yellow solid line indicates the integrated radial flux profile. This plot together with 2MASS radial residual enables us to identify the non-fitted structures (e.g. spiral arms). The horizontal black dotted line represents a residual value equal to zero. \textit{Top right panel:} The combined radial flux profiles along the major axis from the large scale 2MASS Ks image (black asterisk signs; at a radius $\ge$ 2$\arcsec$) and small scale SINFONI stellar continuum image (black plus signs; at a radius $\le$ 1.5$\arcsec$). The HST F160W radial flux profile (orange plus signs) is added to reinforce the scaling factor between two different scales. The red dashed line represents the bulge S\'{e}rsic profile. The vertical black dotted line is the SINFONI AO mode PSF, FWHM radius of $\sim$ 0.1$\arcsec$. This plot together with SINFONI (and HST) radial residual enables us to identify whether there is an excess toward the centre. The horizontal black dotted line represents a residual value equal to zero. \textit{Middle left panel}: The radial flux profile of SINFONI stellar lights (black plus signs). The red dashed line indicates the inward extrapolation of the fitted bulge S\'{e}rsic profile to SINFONI FOV. The dashed line encloses the extended radius of nuclear stellar excess component. \textit{Middle right panel:} The radial flux profile of nuclear stellar excess (the inward extrapolation of the fitted bulge S\'{e}rsic profile has been subtracted) and fit a simple gaussian profile that is plotted as purple solid line. \textit{Bottom left panel:} The circular velocity of stars as a function of radius to the SINFONI FOV, which is extracted by using the IDL routine $kinemetry$. The inclination has been corrected based on the large-scale host galaxy inclination. Black dashed line is the size of nuclear excess component. \textit{Bottom right panel:} The LOS velocity dispersion of stars as a function of radius to the SINFONI FOV, which is extracted by using the IDL routine $kinemetry$. Black dashed line is the size of nuclear excess component. Pink solid line is the mean LOS stellar velocity dispersion at the radius larger than black dashed line, pink dashed lines are the $\pm$3$\sigma$ errors of mean LOS stellar velocity dispersion, which can be a criterion to check whether there is a significant variation in the centre (refer to column (5) in Table~\ref{tab2}). }
\label{fig:appendix-galfit-137-34}
\end{center} 
\end{figure*}

\begin{figure*}
\begin{center}
      \includegraphics[width=170mm]{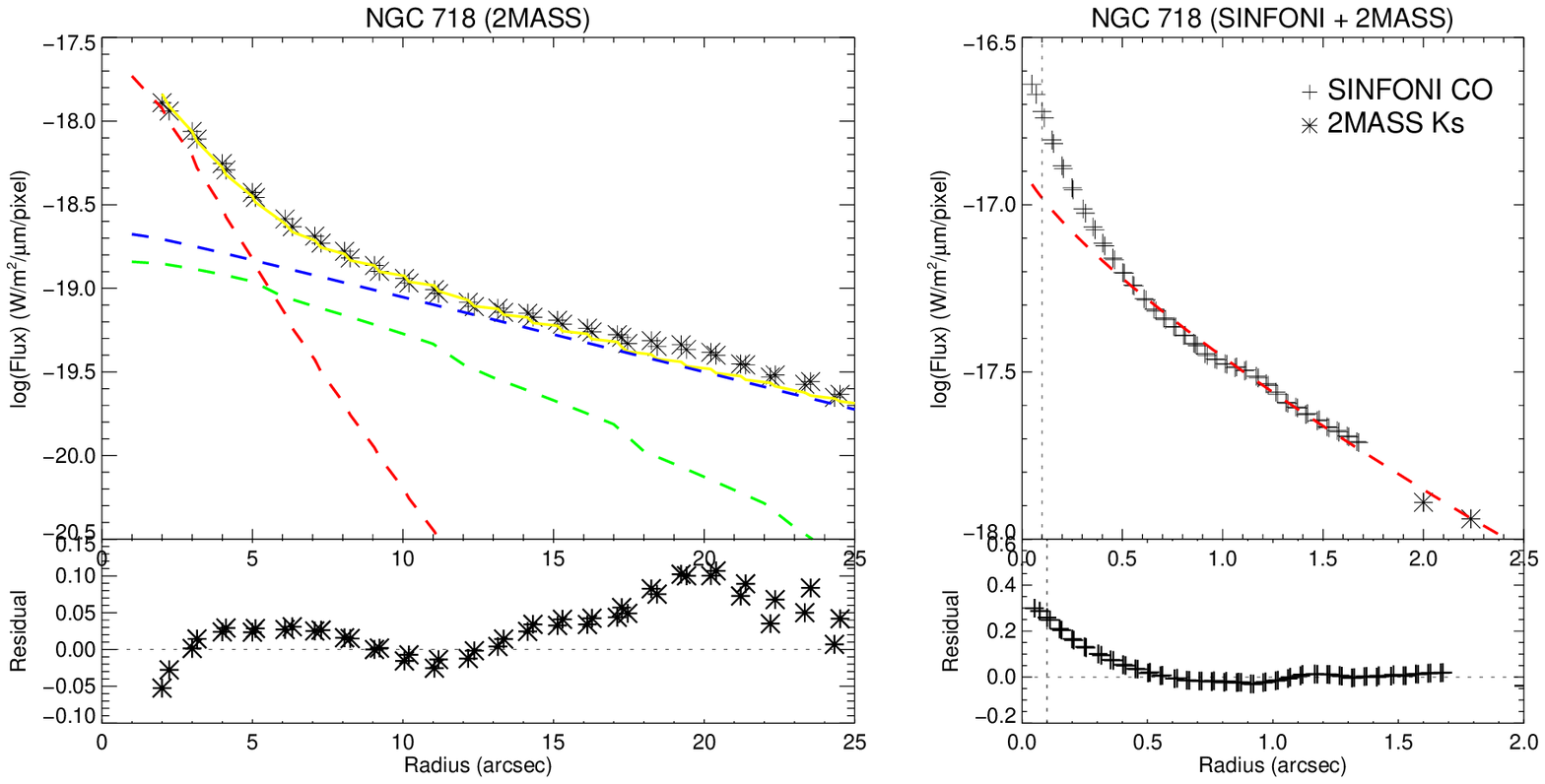}             
       \includegraphics[width=160mm]{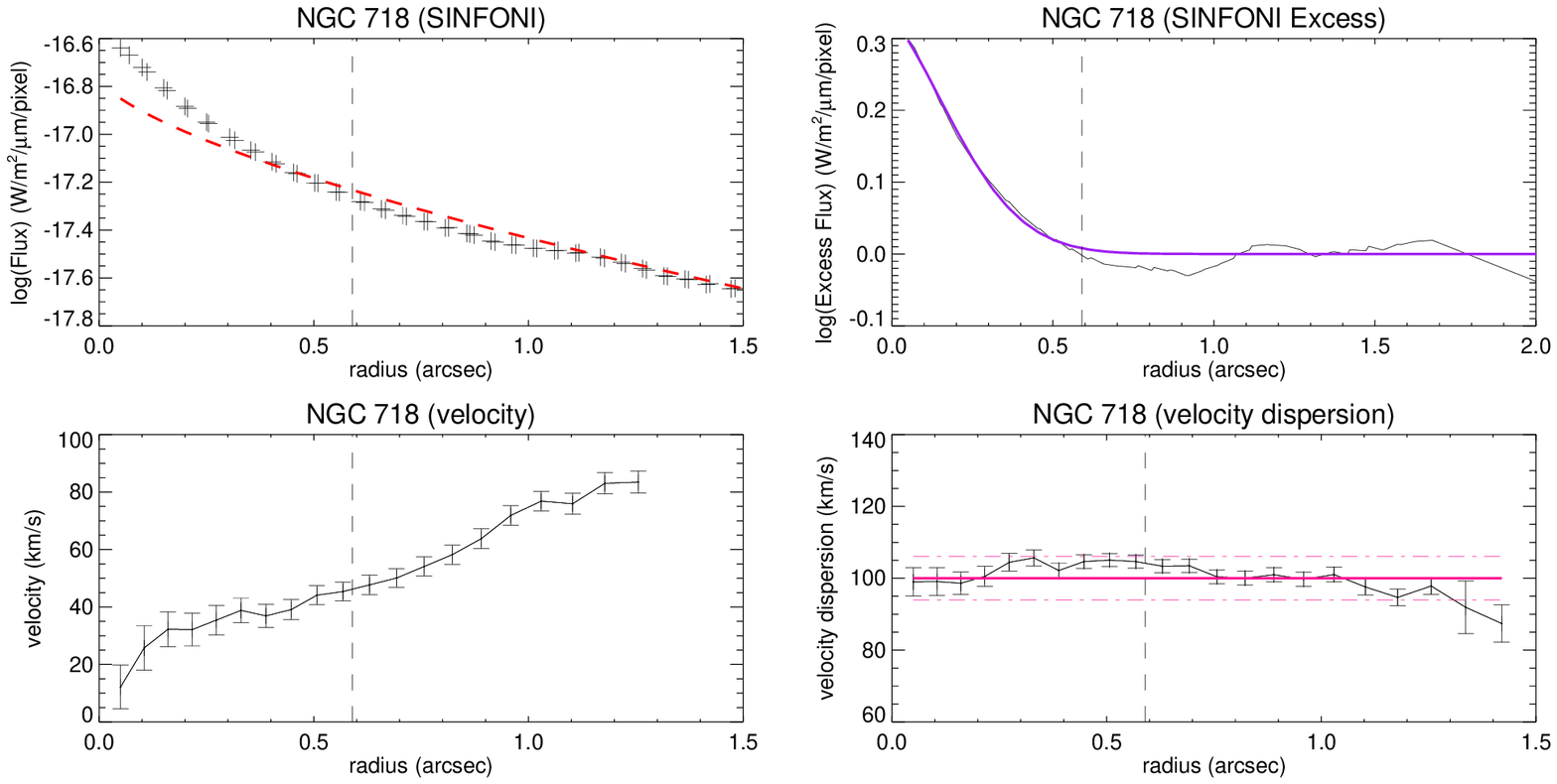}    
         \caption{NGC 718 (Inactive galaxy in Pair 6). See Figure~\ref{fig:appendix-galfit-137-34} for similar descriptions.}
\label{fig:appendix-galfit-718}
\end{center} 
\end{figure*}

\begin{figure*}
\begin{center}
      \includegraphics[width=170mm]{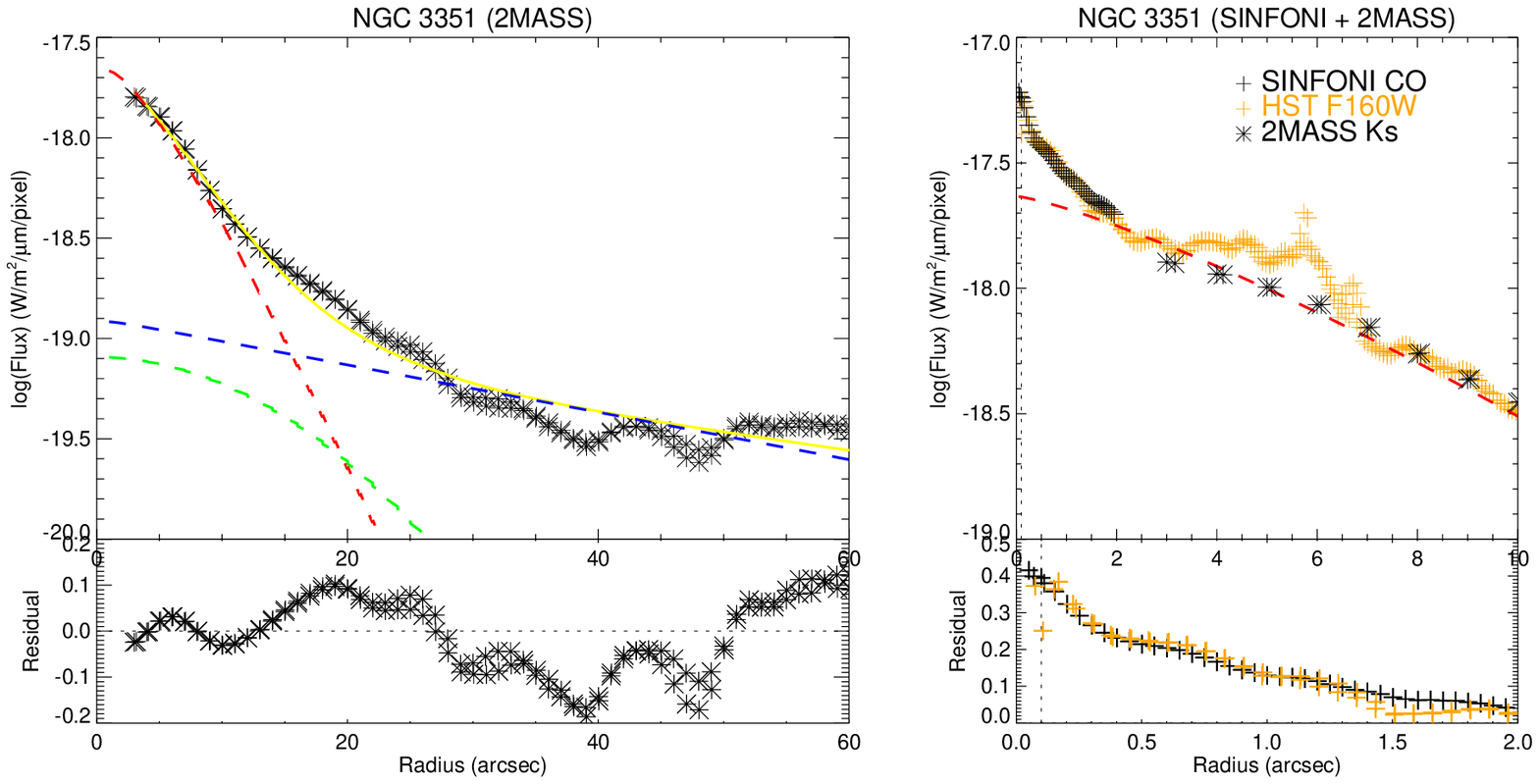}             
       \includegraphics[width=160mm]{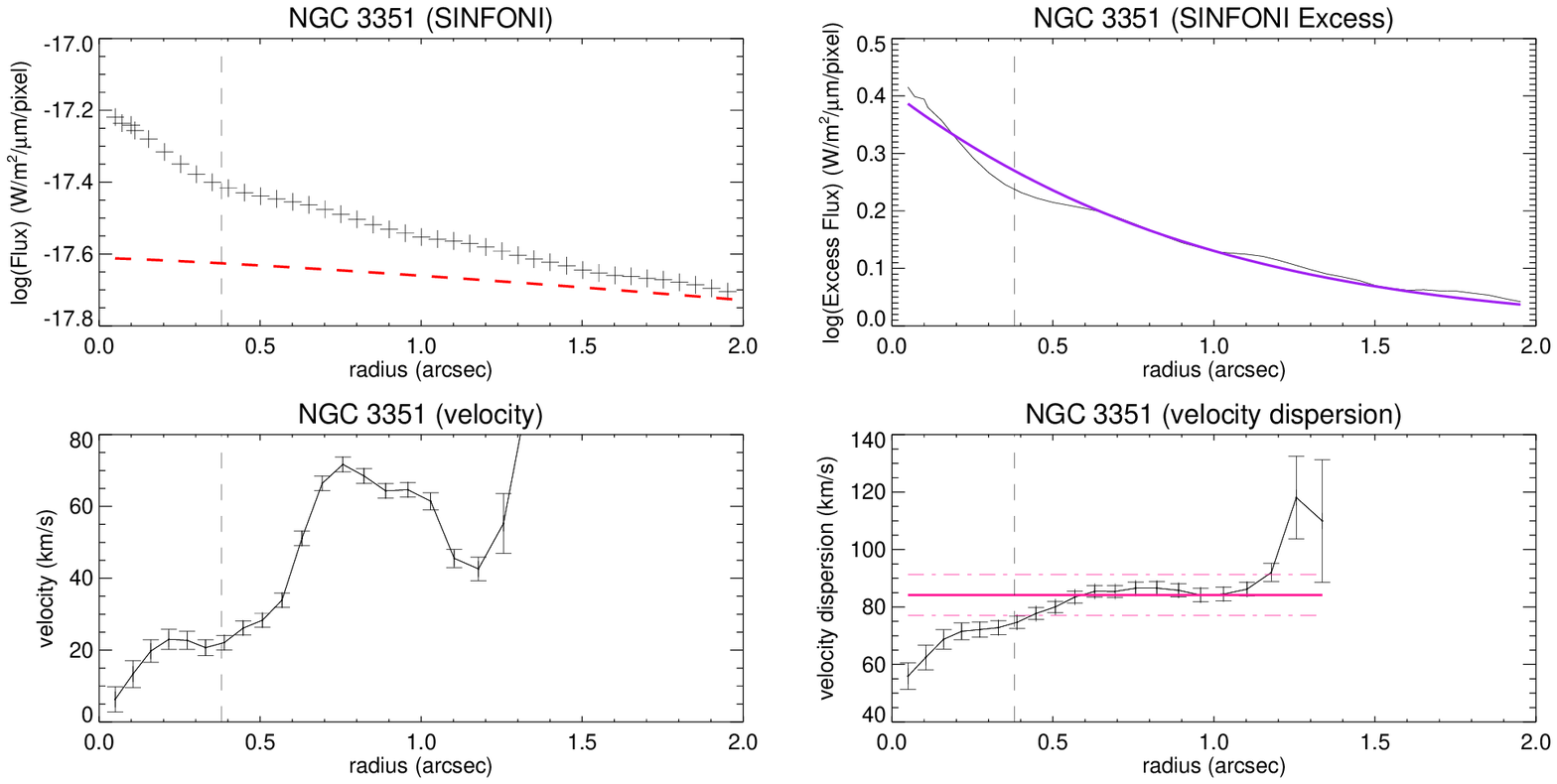}    
         \caption{NGC 3351 (Inactive galaxy in Pair 7). The excess flux distributes entire SINFONI FOV, the black dashed lines in the middle and bottom rows present the size of nuclear cusp. Furthermore, the velocity dispersion drops can be obviously seen inside the nuclear cusp. Other similar descriptions can be seen in Figure~\ref{fig:appendix-galfit-137-34}. }
\label{fig:appendix-galfit-3351}
\end{center} 
\end{figure*}

\begin{figure*}
\begin{center}
      \includegraphics[width=170mm]{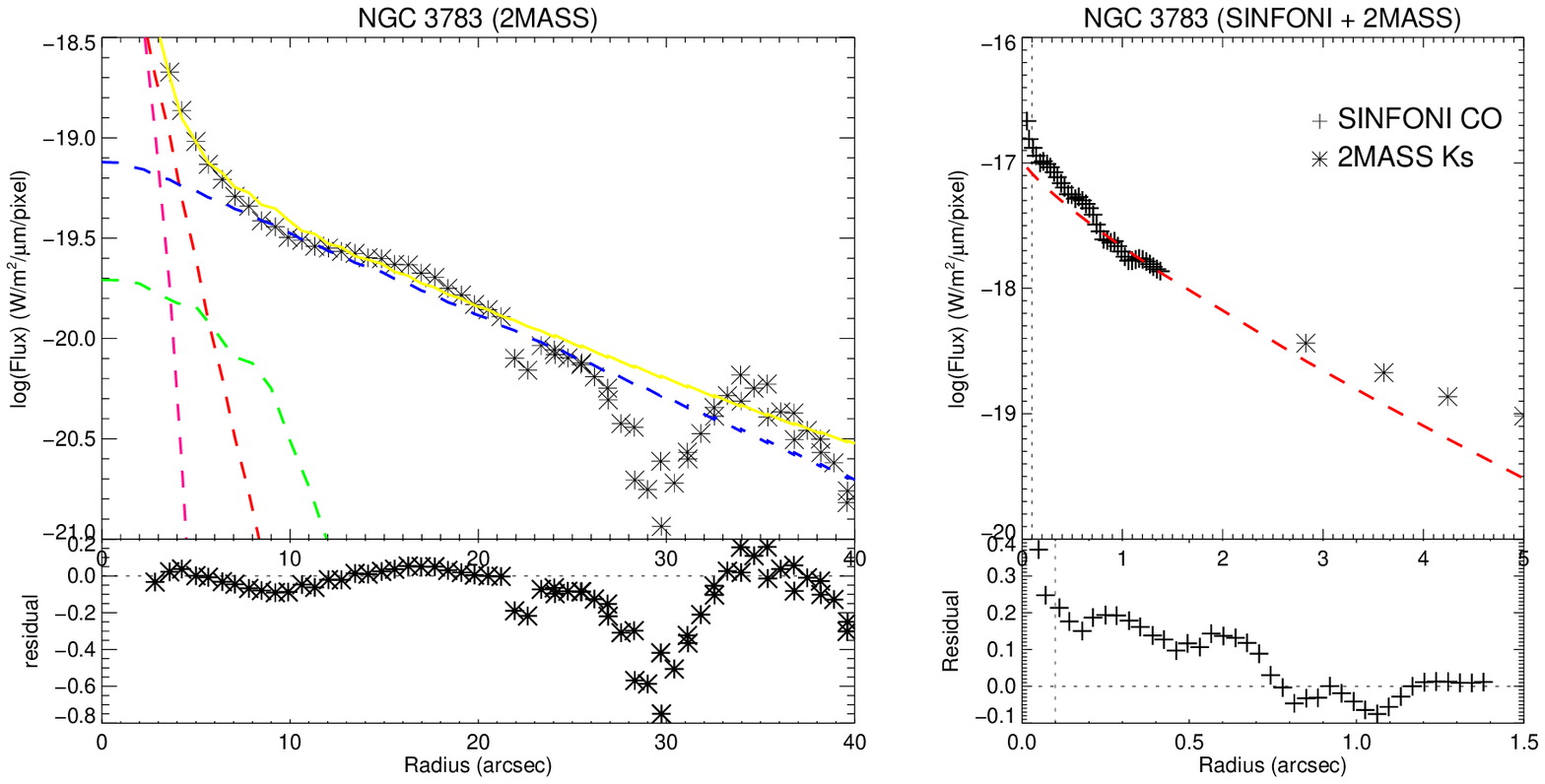}             
       \includegraphics[width=160mm]{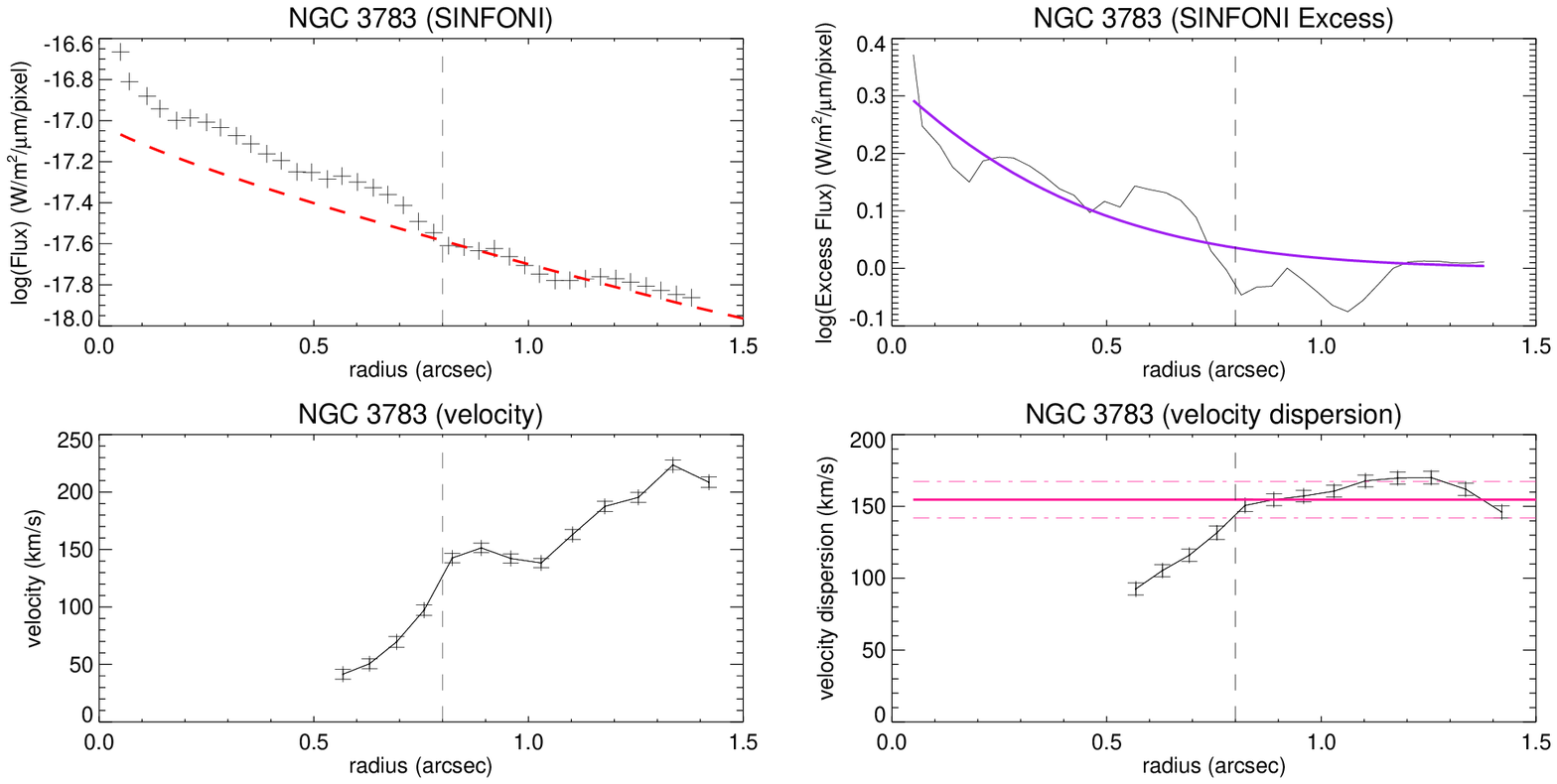}    
         \caption{NGC 3783 (Active galaxy in Pair 6). Because its nucleus is bright in 2MASS Ks image, we add additional PSF during two-dimension fitting and plot it as deep pink dashed line in the top left panel. Other similar descriptions can be seen in Figure~\ref{fig:appendix-galfit-137-34}. }
\label{fig:appendix-galfit-3783}
\end{center} 
\end{figure*}

\begin{figure*}
\begin{center}
      \includegraphics[width=170mm]{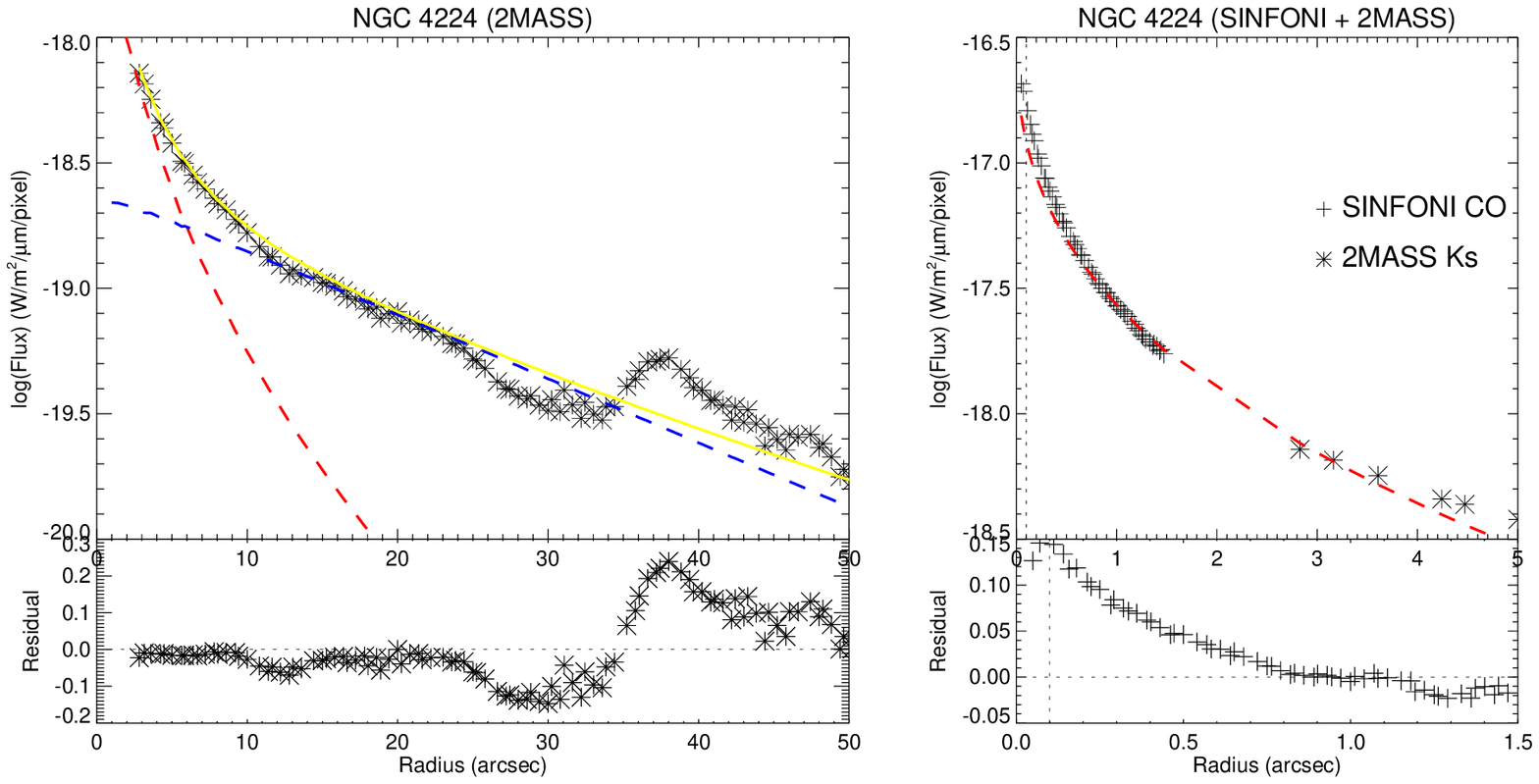}             
       \includegraphics[width=160mm]{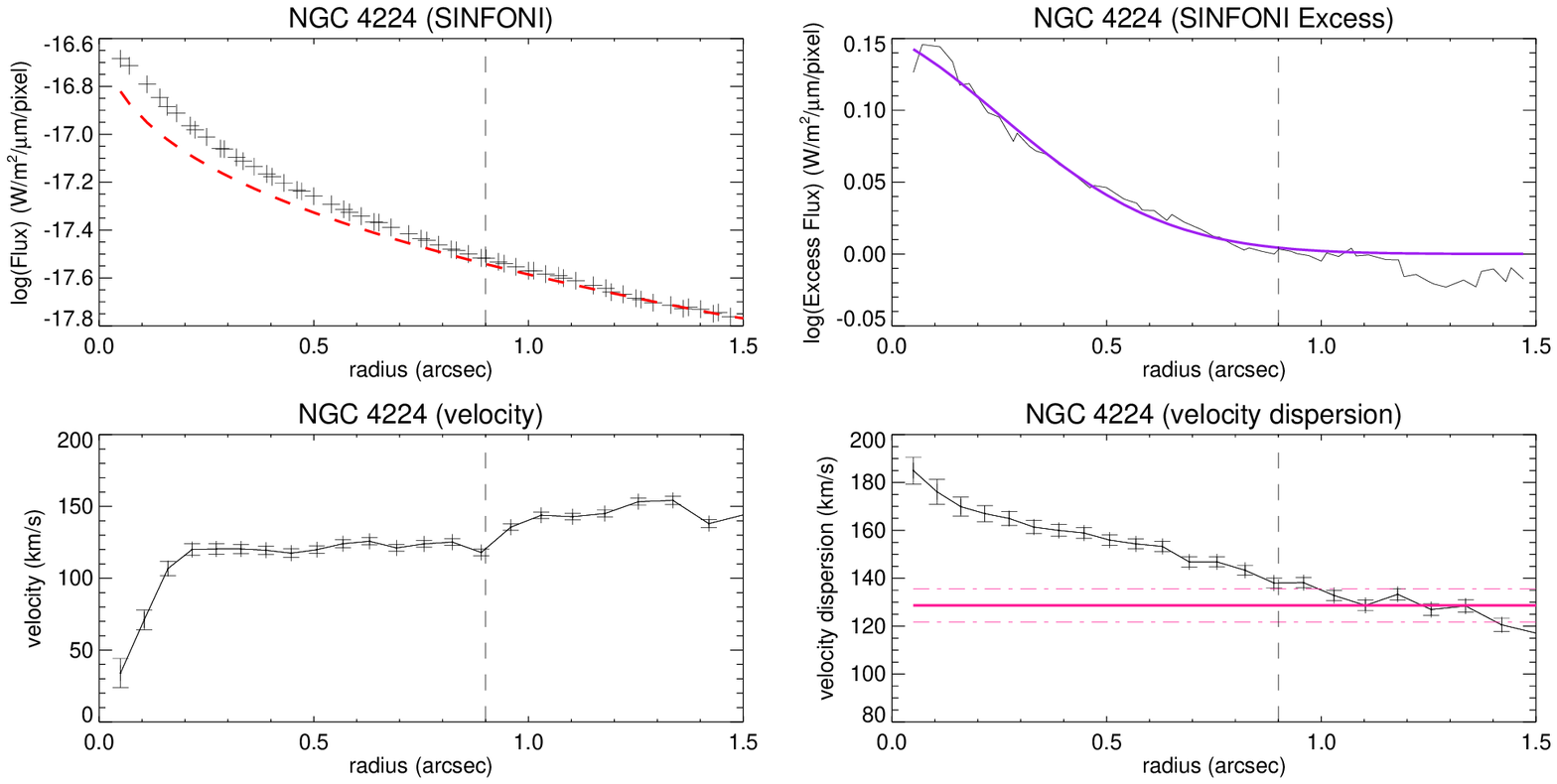}    
         \caption{NGC 4224 (Inactive galaxy in Pair 3 and Pair 5). See Figure~\ref{fig:appendix-galfit-137-34} for similar descriptions.}
\label{fig:appendix-galfit-4224}
\end{center} 
\end{figure*}

\begin{figure*}
\begin{center}
      \includegraphics[width=170mm]{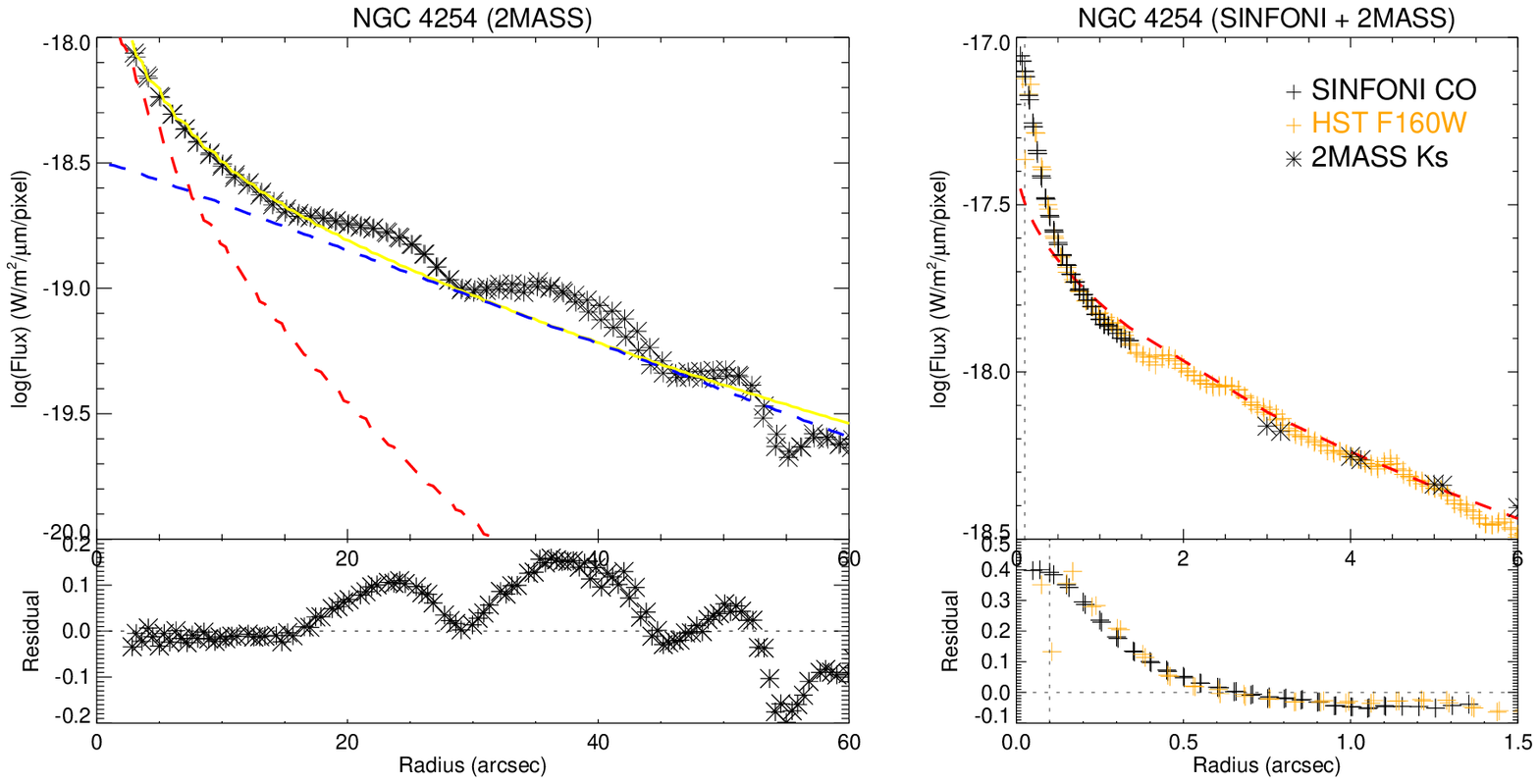}             
       \includegraphics[width=160mm]{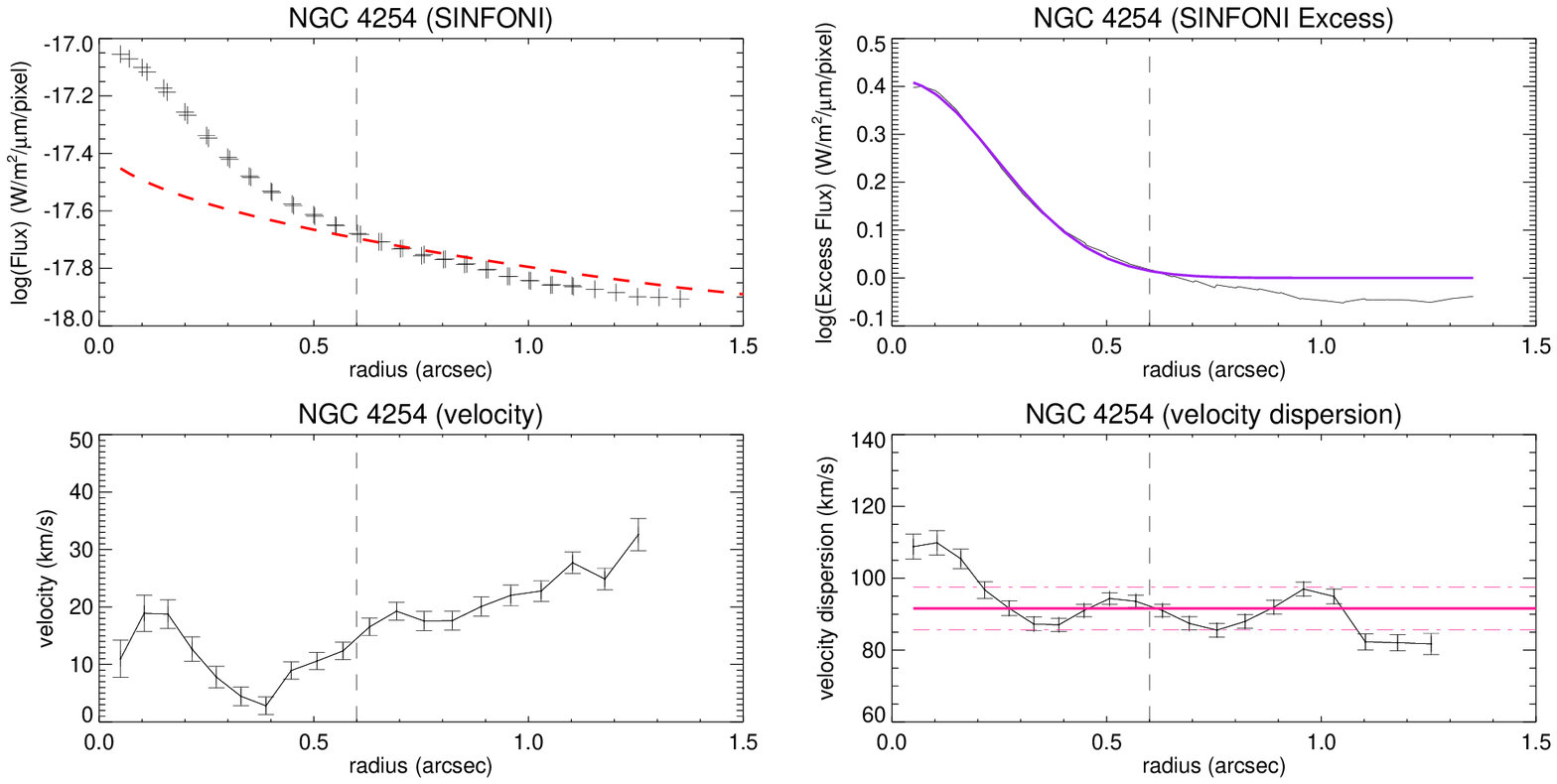}    
         \caption{NGC 4254 (Inactive galaxy in Pair 8). See Figure~\ref{fig:appendix-galfit-137-34} for similar descriptions.}
\label{fig:appendix-galfit-4254}
\end{center} 
\end{figure*}

\begin{figure*}
\begin{center}
      \includegraphics[width=170mm]{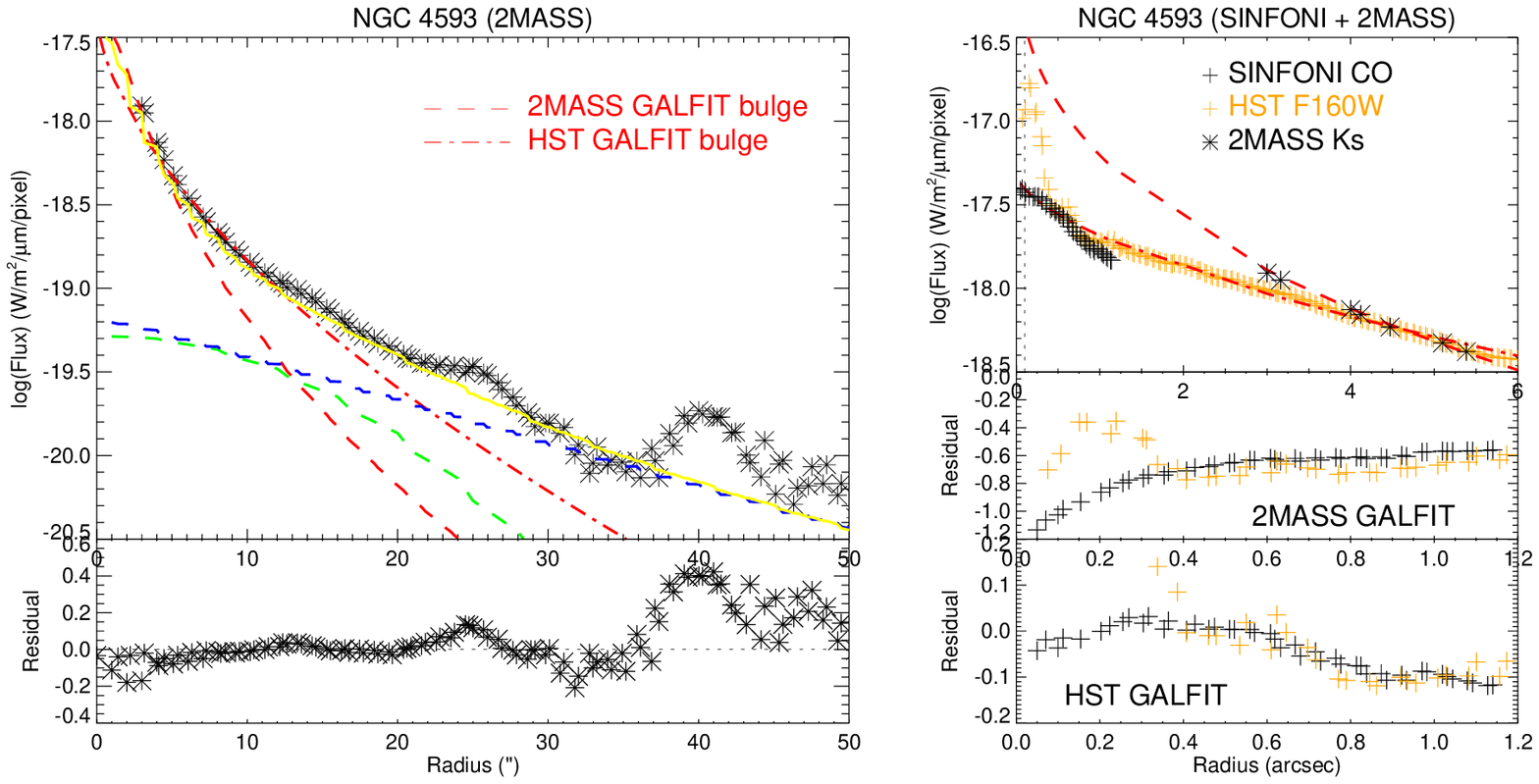}             
       \includegraphics[width=160mm]{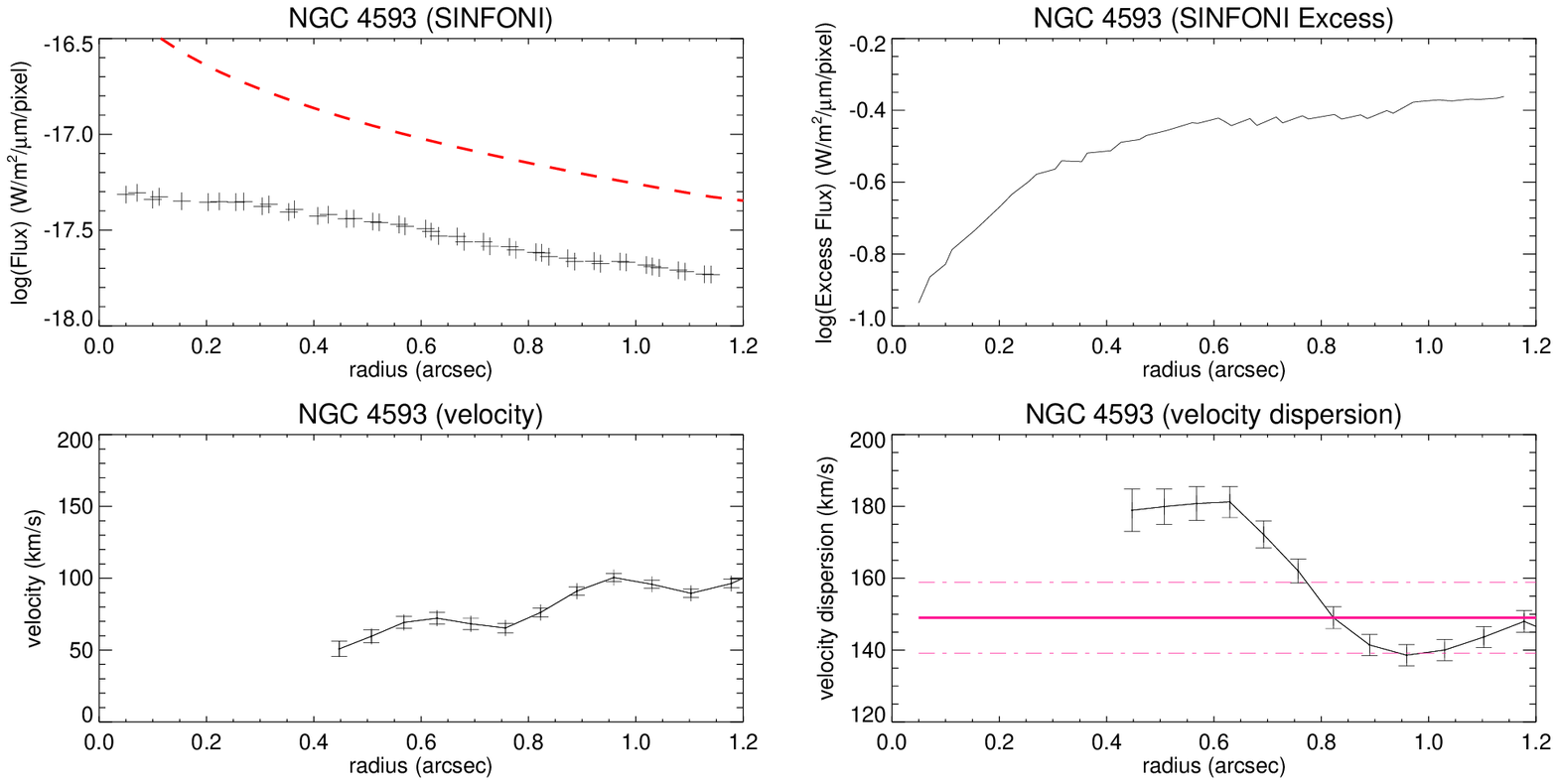}    
         \caption{NGC 4593 (Active galaxy in Pair 7). Top left panel: In addition to measure bulge S\'{e}rsic profile based on 2MASS image, we use GALFIT to fit HST F160W image, which is presented in red dot-dashed line. 
Top right panel: The combined large-scale 2MASS, HST images and small-scale SINFONI image. We do not see any significant nuclear stellar excess toward the centre in both 2MASS and HST residuals. Middle and bottom panels: See Figure~\ref{fig:appendix-galfit-137-34} for similar descriptions.}
\label{fig:appendix-galfit-4593}
\end{center} 
\end{figure*}

\begin{figure*}
\begin{center}
      \includegraphics[width=170mm]{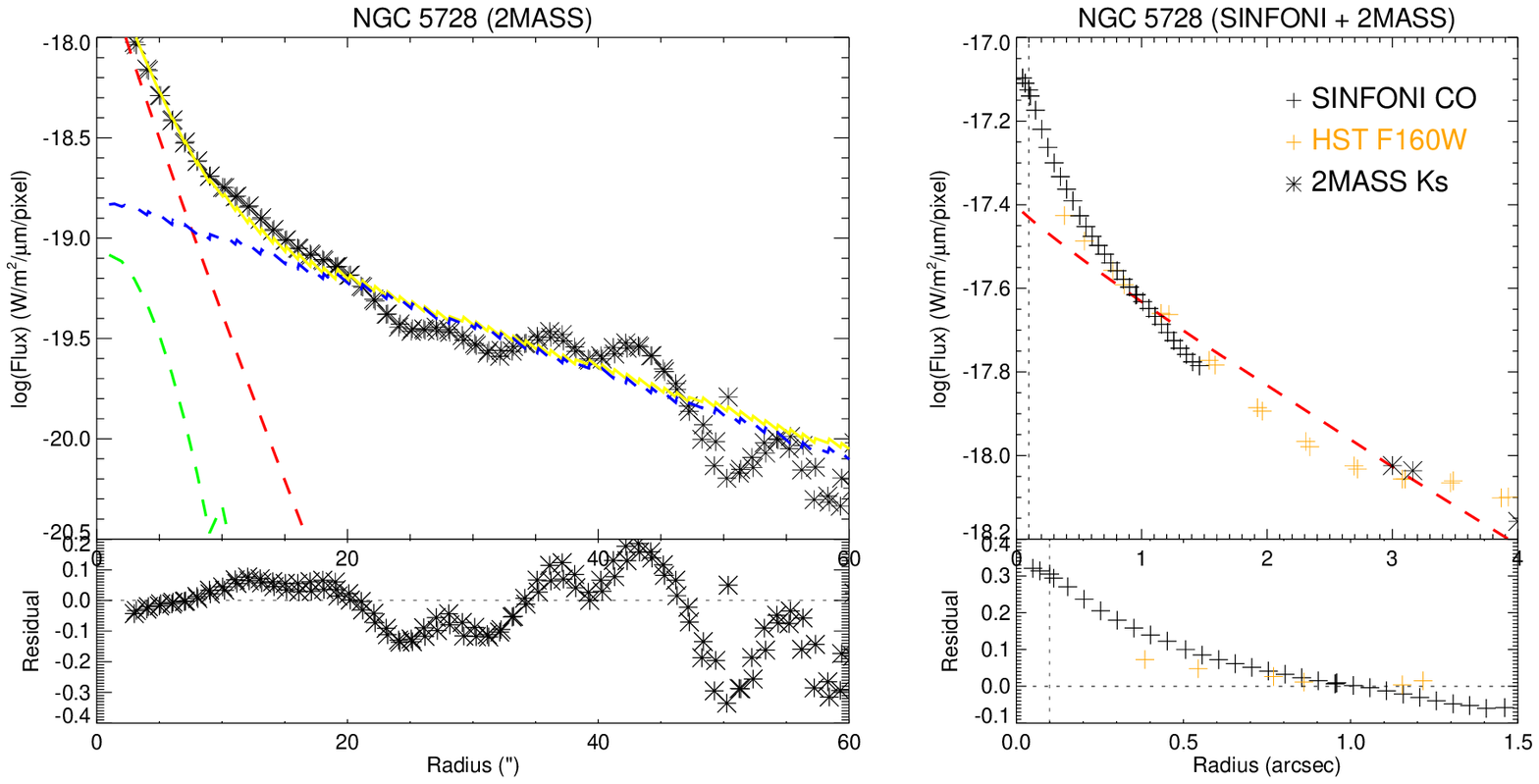}             
       \includegraphics[width=160mm]{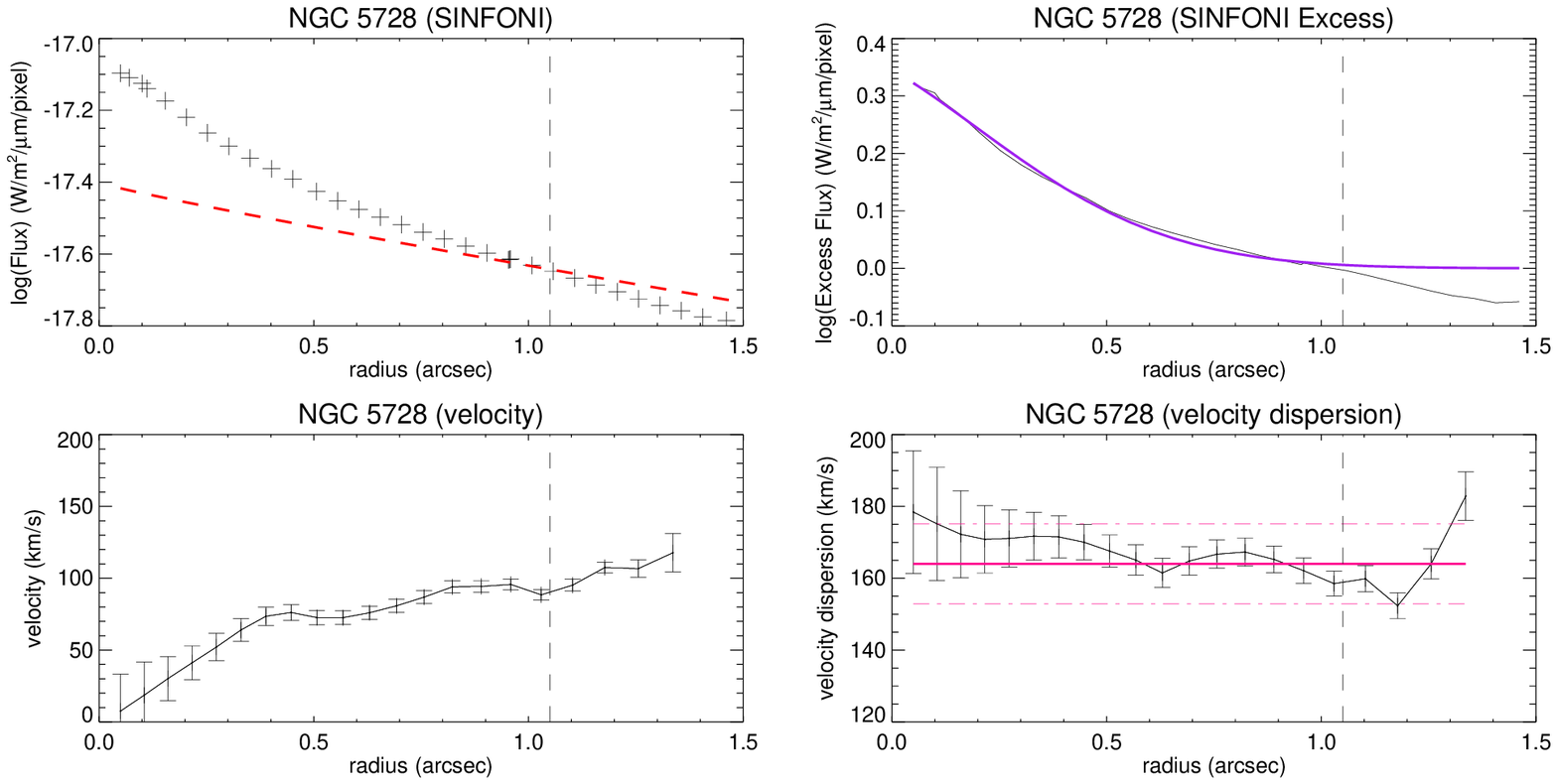}    
         \caption{NGC 5728 (Active galaxy in Pair 4). See Figure~\ref{fig:appendix-galfit-137-34} for similar descriptions.}
\label{fig:appendix-galfit-5728}
\end{center} 
\end{figure*}

\begin{figure*}
\begin{center}
      \includegraphics[width=170mm]{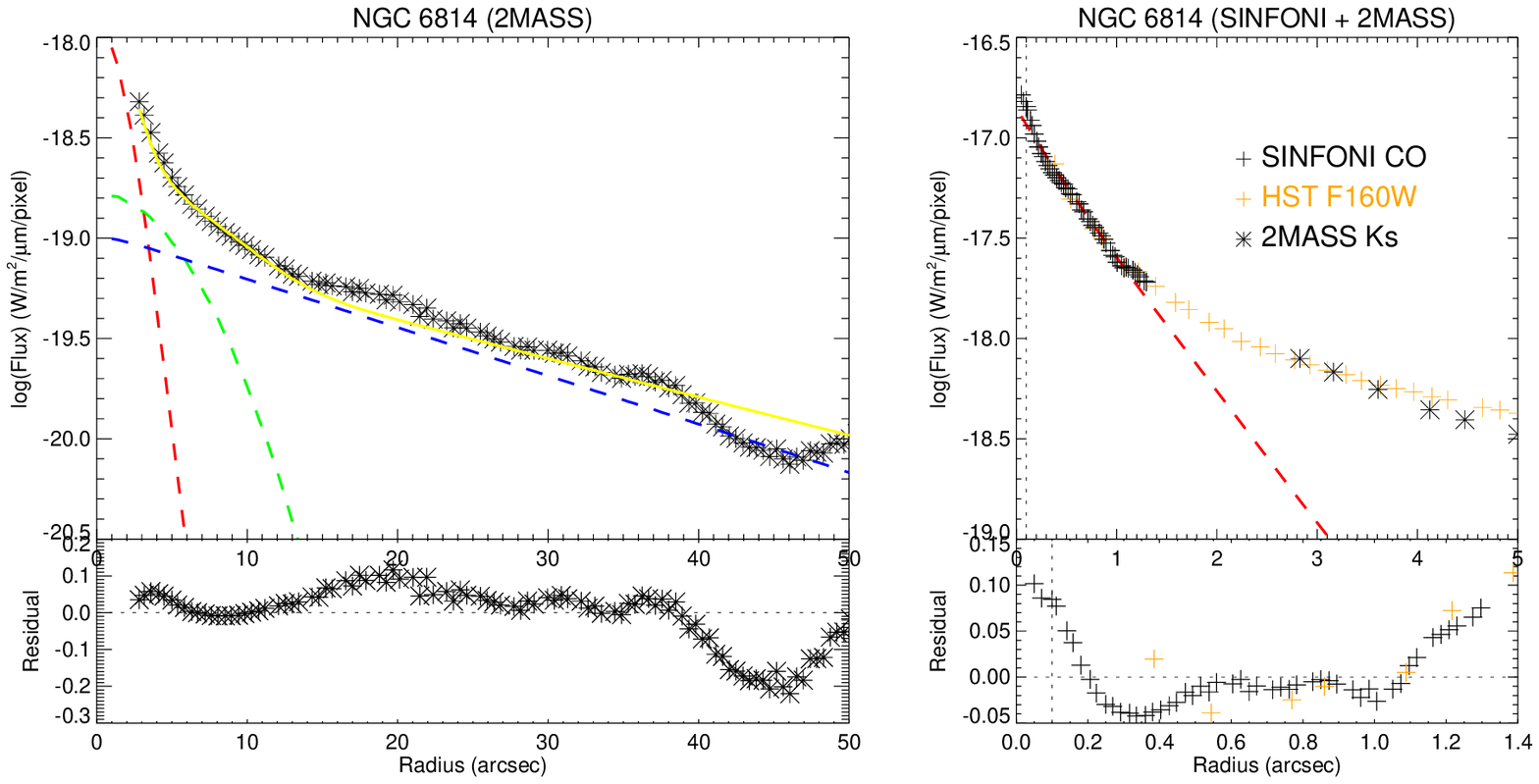}             
       \includegraphics[width=160mm]{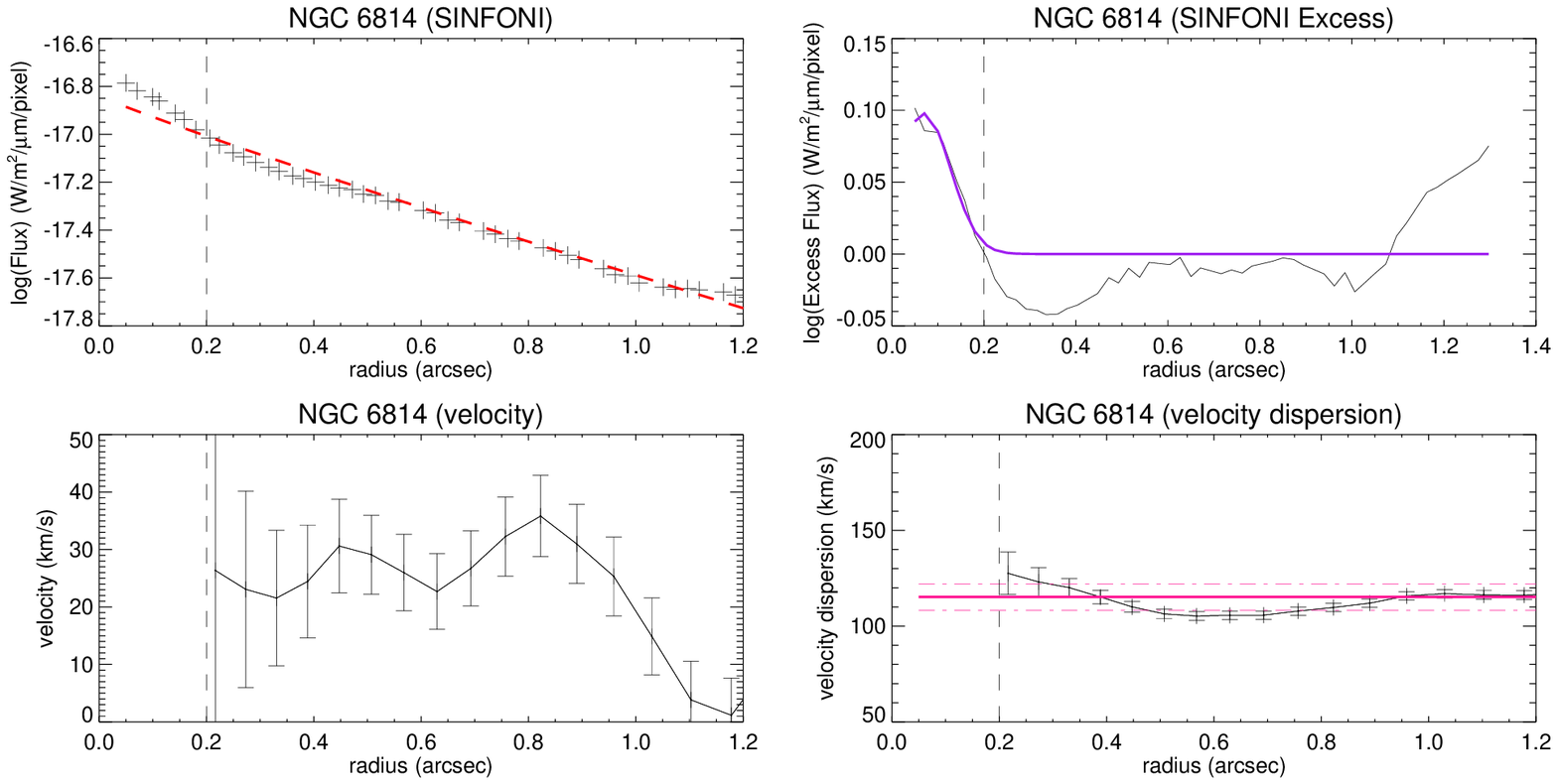}    
         \caption{NGC 6814 (Active galaxy in Pair 8). See Figure~\ref{fig:appendix-galfit-137-34} for similar descriptions.}
\label{fig:appendix-galfit-6814}
\end{center} 
\end{figure*}

\begin{figure*}
\begin{center}
      \includegraphics[width=170mm]{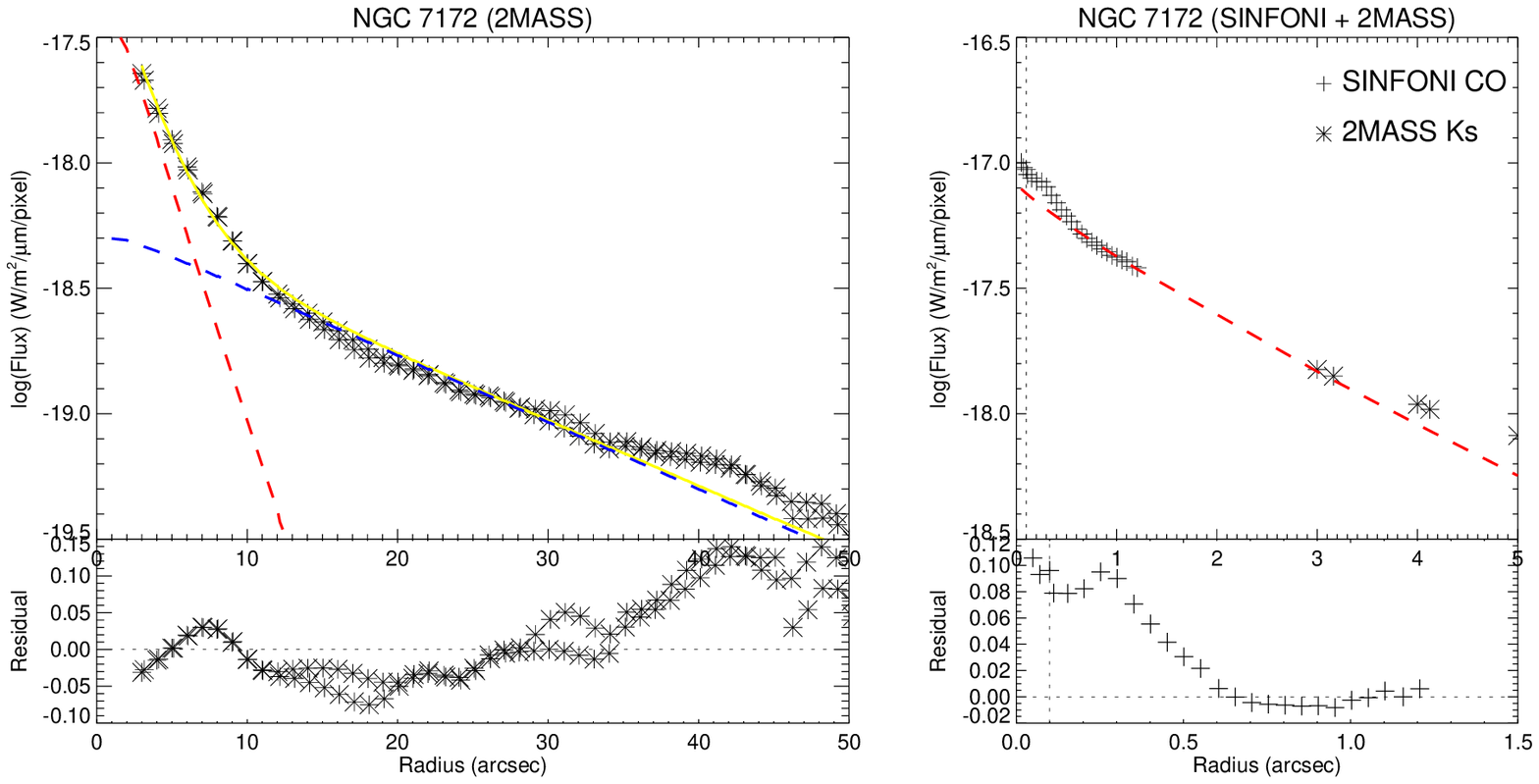}             
       \includegraphics[width=160mm]{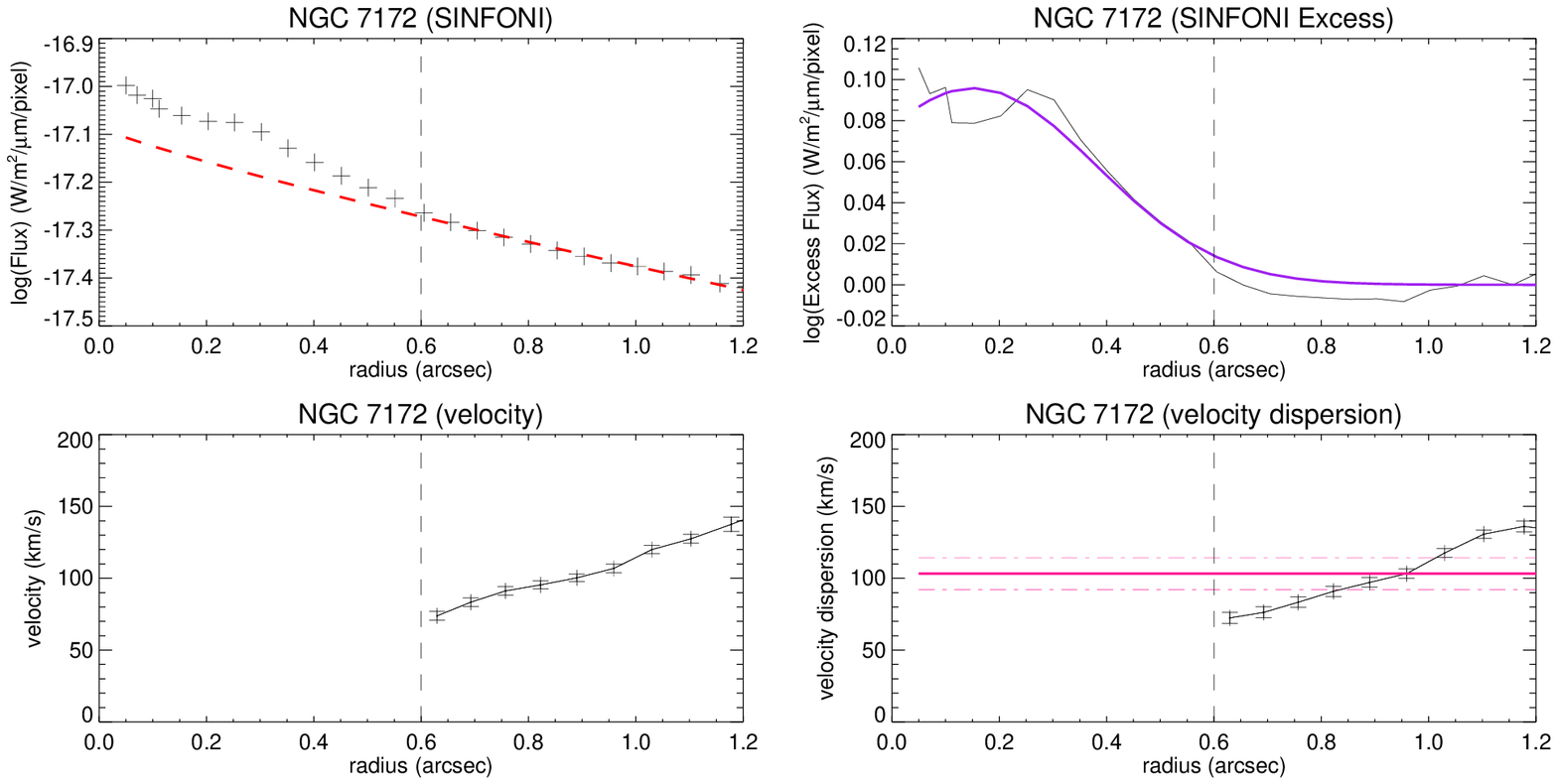}    
         \caption{NGC 7172 (Active galaxy in Pair 3). See Figure~\ref{fig:appendix-galfit-137-34} for similar descriptions.}
\label{fig:appendix-galfit-7172}
\end{center} 
\end{figure*}

\begin{figure*}
\begin{center}
      \includegraphics[width=170mm]{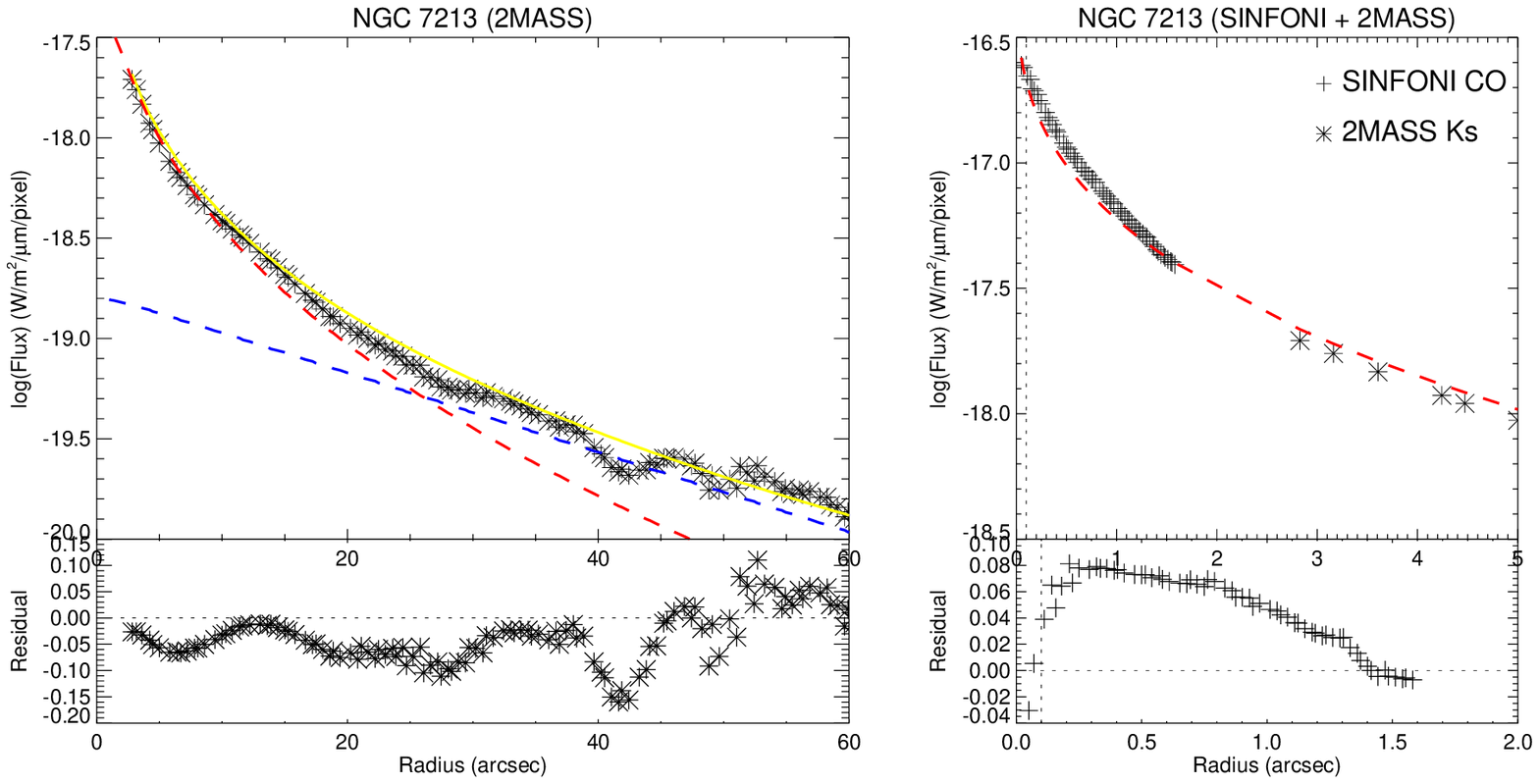}             
       \includegraphics[width=160mm]{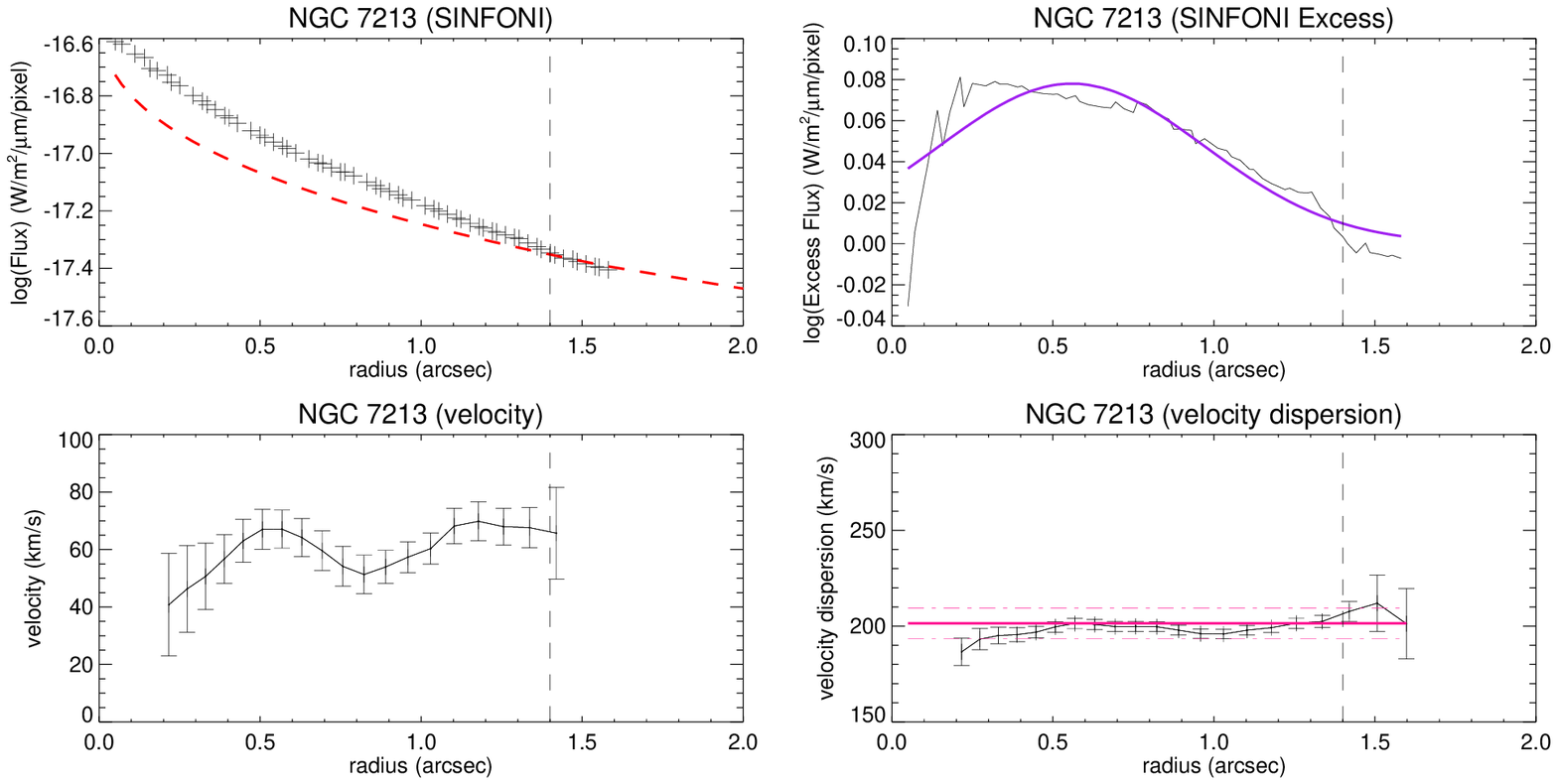}    
         \caption{NGC 7213 (Active galaxy in Pair 2). See Figure~\ref{fig:appendix-galfit-137-34} for similar descriptions.}
\label{fig:appendix-galfit-7213}
\end{center} 
\end{figure*}

\begin{figure*}
\begin{center}
      \includegraphics[width=170mm]{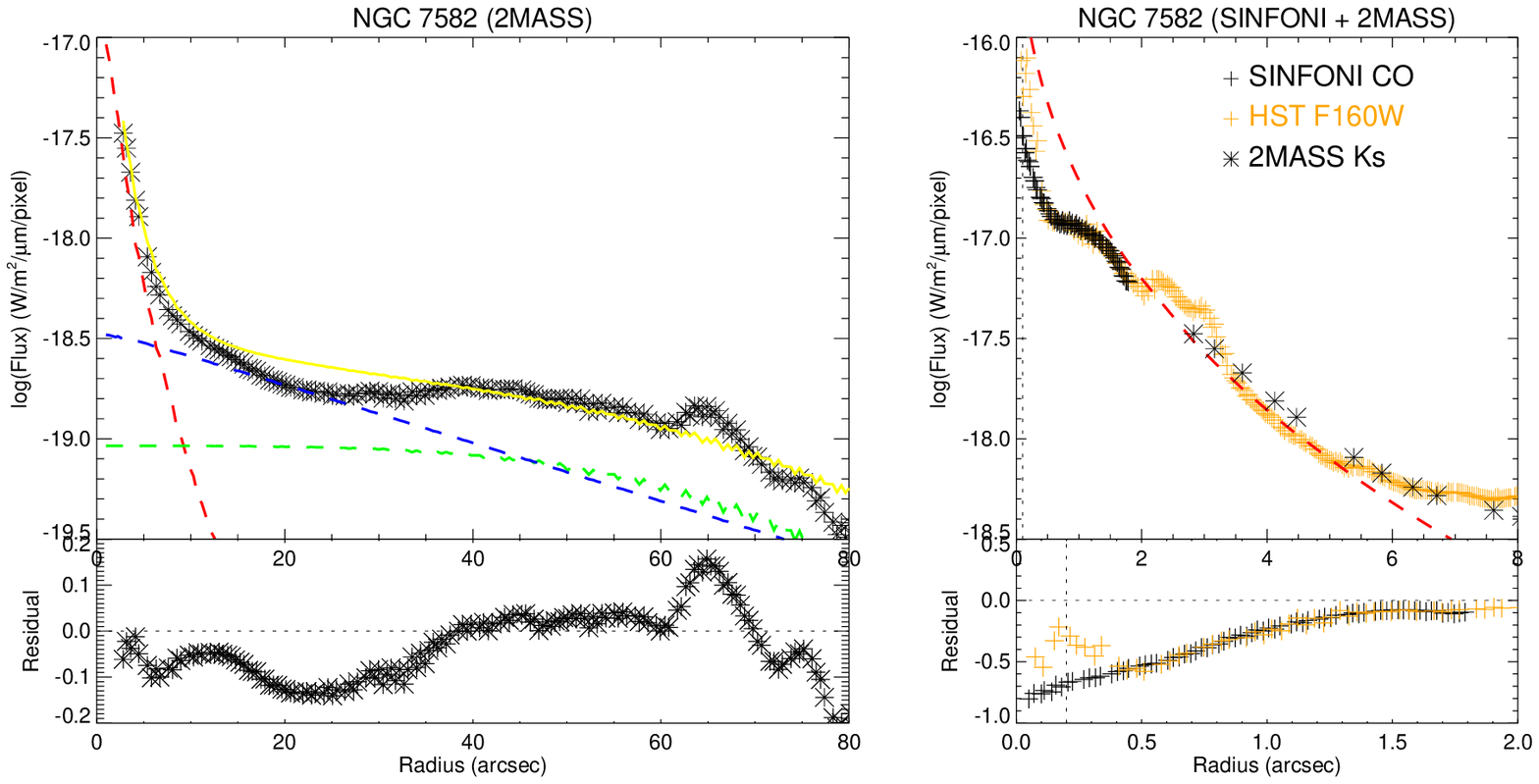}             
       \includegraphics[width=160mm]{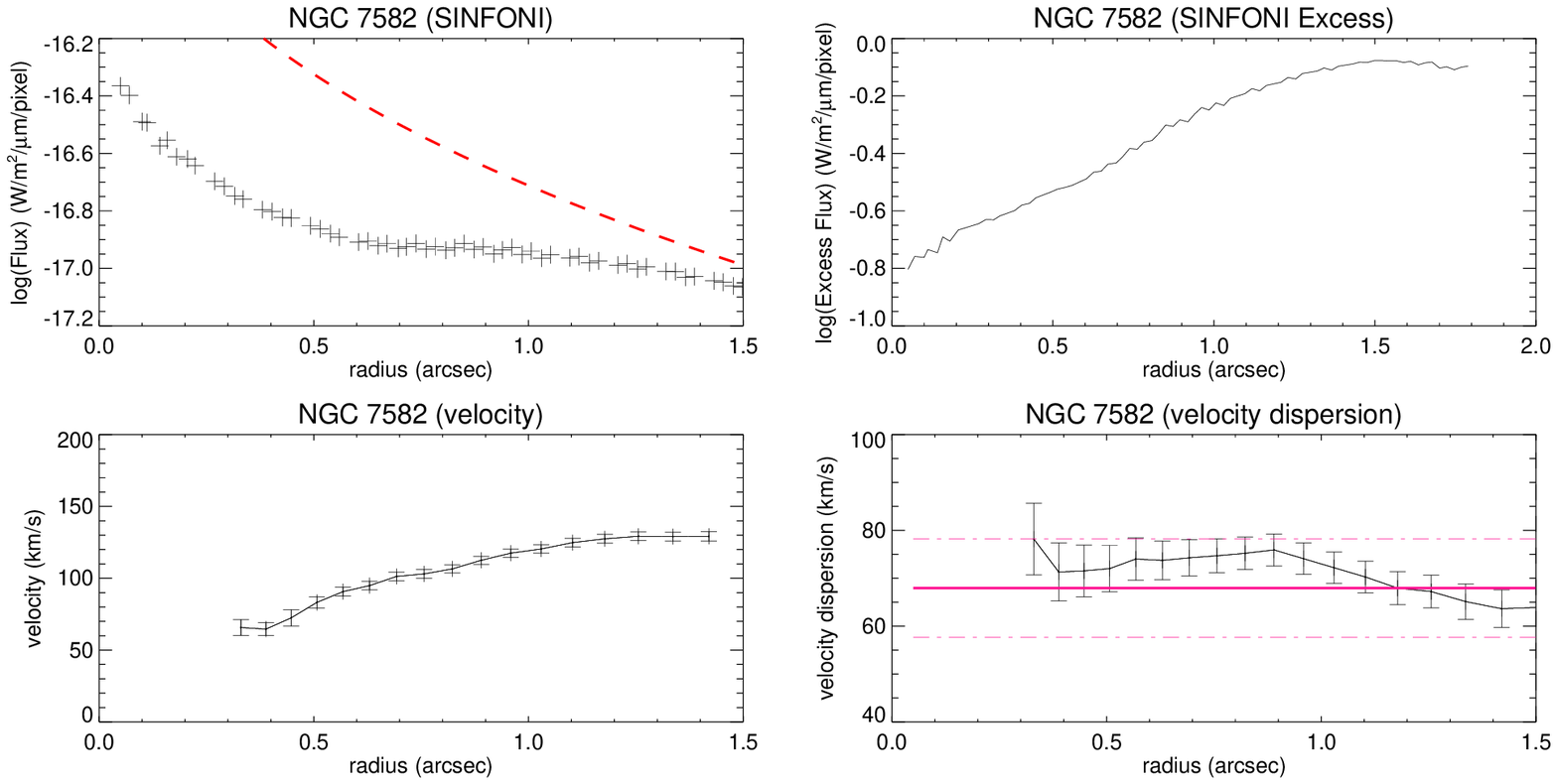}    
         \caption{NGC 7582 (Active galaxy in Pair 5). See Figure~\ref{fig:appendix-galfit-137-34} for similar descriptions.}
\label{fig:appendix-galfit-7582}
\end{center} 
\end{figure*}

\begin{figure*}
\begin{center}
      \includegraphics[width=170mm]{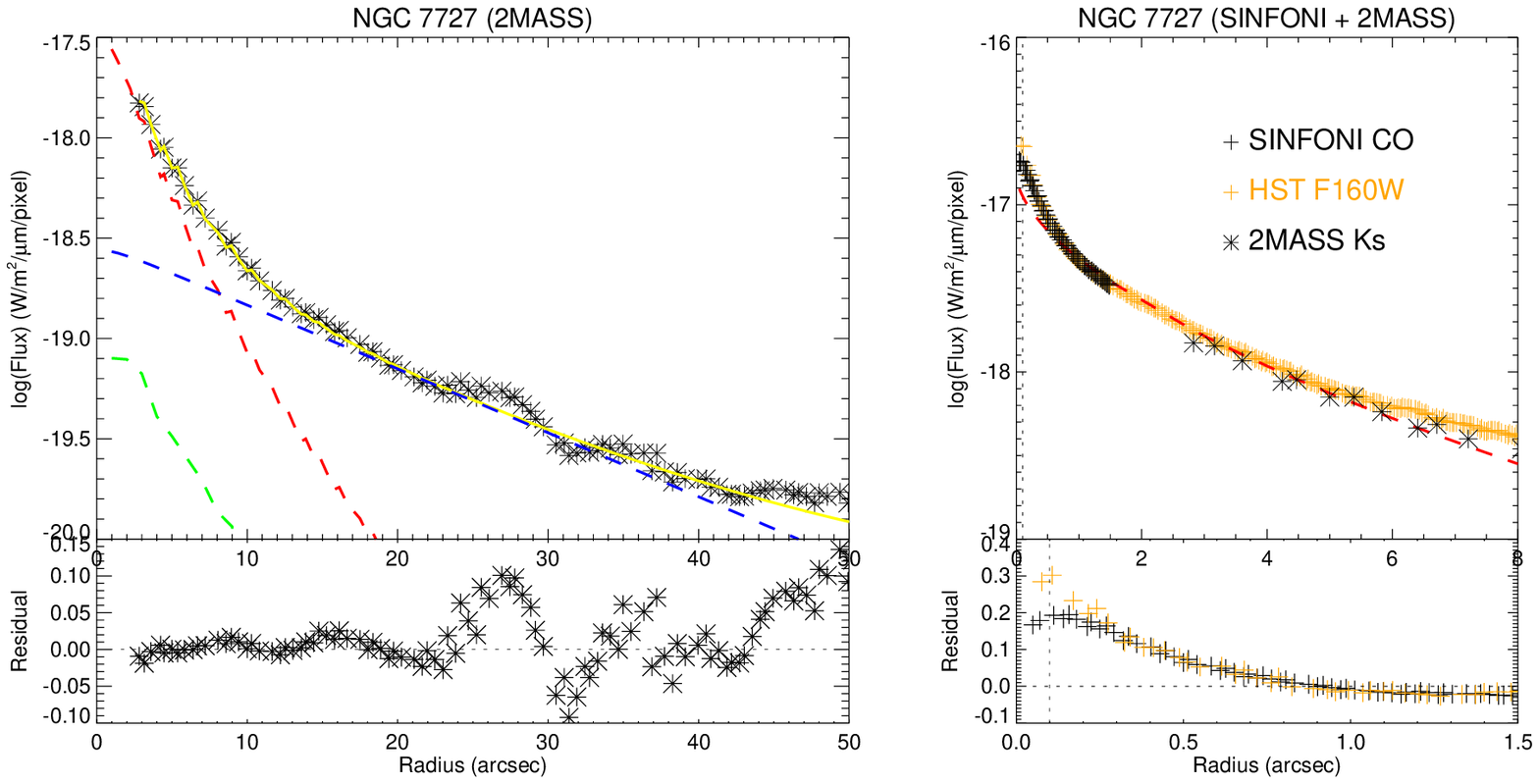}             
       \includegraphics[width=160mm]{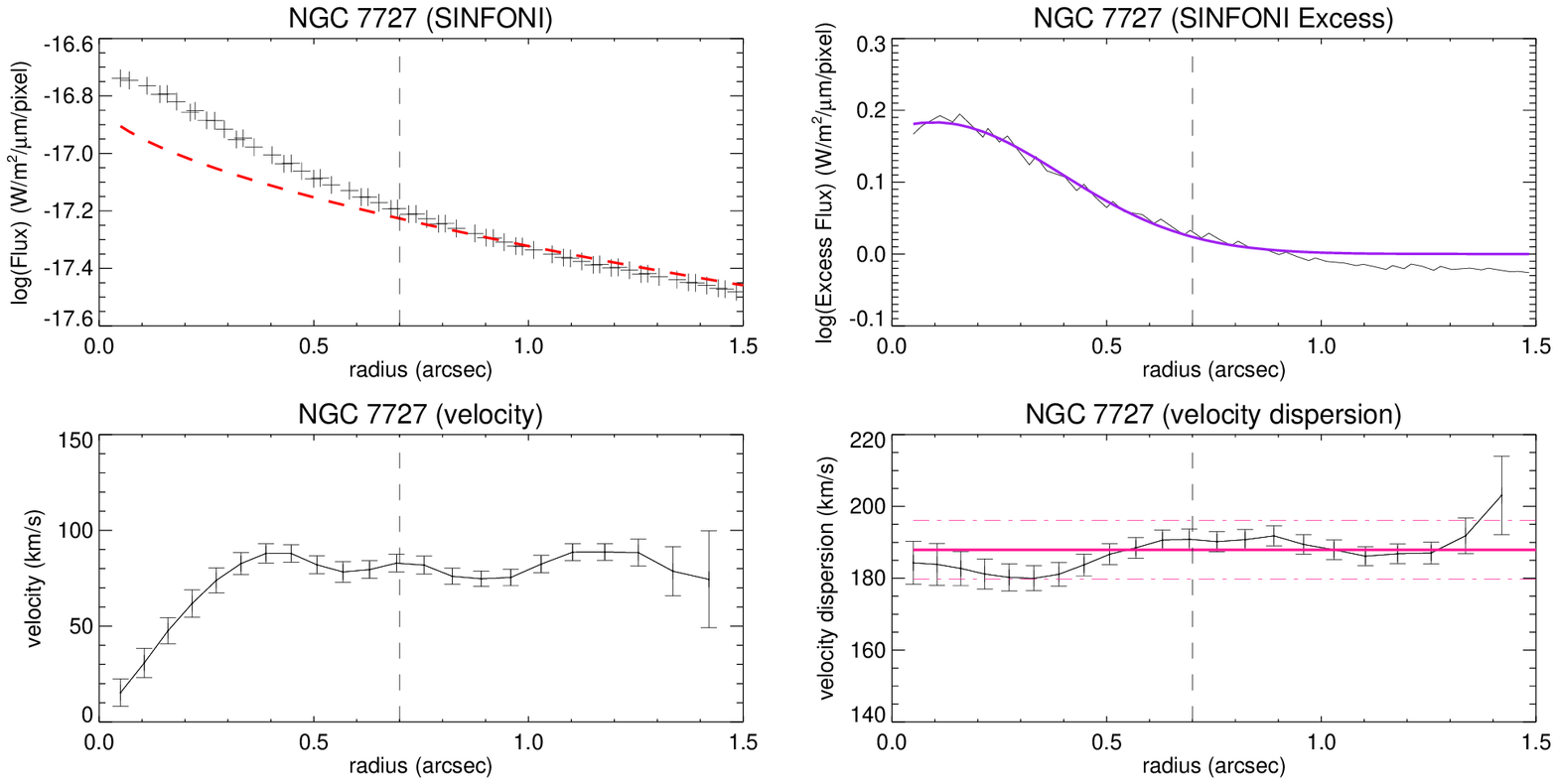}    
         \caption{NGC 7727 (Inactive galaxy in Pair 1, 2, and 4). See Figure~\ref{fig:appendix-galfit-137-34} for similar descriptions.}
\label{fig:appendix-galfit-7727}
\end{center} 
\end{figure*}

\bsp	
\label{lastpage}

\end{document}